\documentclass[longauth]{aa} 
\usepackage{euclid}
\usepackage[colorlinks=true,linkcolor=blue,citecolor=blue]{hyperref}
\usepackage[dvipsnames]{xcolor}
\usepackage{graphicx}
\graphicspath{{graphics}}
\usepackage{txfonts}
\usepackage[separate-uncertainty=true, list-units=repeat]{siunitx}
\usepackage{subcaption}
\usepackage{float}
\usepackage{hyperref}
\usepackage[inline]{enumitem}
\usepackage{threeparttable} 
\usepackage{physics} 	
\usepackage{multirow}
\usepackage{bm}

\newcommand{\lephare}{\texttt{LePHARE}\xspace}
\newcommand{\cigale}{\texttt{CIGALE}\xspace}

\newcommand{\hst}{\textit{HST}}
\newcommand{\JWST}{\textit{JWST}}

\DeclareSIUnit{\Msun}{M_\odot}
\DeclareSIUnit{\year}{yr}
\DeclareSIUnit{\pc}{pc}
\DeclareSIUnit{\mag}{mag}
\DeclareSIUnit{\mas}{mas}
\DeclareSIUnit{\dex}{dex}
\DeclareSIUnit{\jansky}{Jy}
\DeclareSIUnit{\kpc}{kpc}

\newcommand{\logM}{$\log(M_{\star}/{\rm M}_{\odot})$}

\begin{document} 

   \title{The stellar mass function of quiescent and star-forming galaxies and its dependence on morphology in COSMOS-Web}
   \subtitle{}  

   \author{
Marko~Shuntov\inst{\ref{DAWN},\ref{NBI},\ref{Geneva}}\fnmsep\thanks{\email{marko.shuntov@nbi.ku.dk}} \and
Olivier~Ilbert\inst{\ref{LAM}} \and
Claudia~del~P.~Lagos\inst{\ref{ICRAR}, \ref{DAWN}} \and
Sune~Toft\inst{\ref{DAWN}, \ref{NBI}} \and
Francesco~Valentino\inst{\ref{DTU},\ref{DAWN}} \and
Wilfried~Mercier\inst{\ref{LAM}} \and
Hollis~B.~Akins\inst{\ref{UAT}} \and
Nguyen~Binh\inst{\ref{Washington}} \and
Malte~Brinch\inst{\ref{Chile},\ref{Mingal}} \and
Caitlin~M.~Casey\inst{\ref{UAT}, \ref{DAWN}} \and
Maximilien~Franco\inst{\ref{CEA}} \and
Fabrizio~Gentile\inst{\ref{CEA}} \and
Ghassem~Gozaliasl\inst{\ref{Aalto},\ref{Helsinki}} \and
Aryana~Haghjoo\inst{\ref{Riverside}} \and
Santosh~Harish\inst{\ref{Rochester}} \and
Michaela~Hirschmann\inst{\ref{Lausanne},\ref{INAF}} \and
Marc~Huertas-Company\inst{\ref{IAC}, \ref{LERMA}, \ref{Paris-Cite}, \ref{laLaguna}} \and
Shuowen~Jin\inst{\ref{DAWN}, \ref{DTU}} \and
Jeyhan~S.~Kartaltepe\inst{\ref{Rochester}} \and
Anton~M.~Koekemoer\inst{\ref{STScI}} \and
Clotilde~Laigle\inst{\ref{IAP}} \and
Joseph~S.~W.~Lewis\inst{\ref{IAP}} \and
Georgios~E.~Magdis\inst{\ref{DAWN},\ref{DTU}} \and
Henry~Joy~McCracken\inst{\ref{IAP}} \and
Bahram~Mobasher\inst{\ref{Riverside}} \and
Thibaud~Moutard\inst{\ref{ESA}} \and
Pascal~A.~Oesch\inst{\ref{Geneva},\ref{DAWN},\ref{NBI}} \and
Louise~Paquereau\inst{\ref{IAP}} \and
Alvio~Renzini\inst{\ref{Padova}} \and
Michael~R.~Rich\inst{\ref{LA}} \and
David~B.~Sanders\inst{\ref{HawaiiManoa}} \and
Greta~Toni\inst{\ref{Bologna},\ref{INAF-Bologna},\ref{Heidelberg}} \and
Laurence~Tresse\inst{\ref{LAM}} \and
Andrea~Weibel\inst{\ref{Geneva}} \and
John~R.~Weaver\inst{\ref{UMASS}} \and
Lilan~Yang\inst{\ref{Rochester}}
}

\institute{
Cosmic Dawn Center (DAWN), Denmark\label{DAWN} 
\and
Niels Bohr Institute, University of Copenhagen, Jagtvej 128, 2200 Copenhagen, Denmark \label{NBI} 
\and
Department of Astronomy, University of Geneva, Chemin Pegasi 51, 1290 Versoix, Switzerland \label{Geneva} 
\and
Aix Marseille Univ, CNRS, CNES, LAM, Marseille, France \label{LAM} 
\and
International Centre for Radio Astronomy Research (ICRAR), M468, University of Western Australia, 35 Stirling Hwy, Crawley, WA 6009, Australia \label{ICRAR} 
\and
The University of Texas at Austin, 2515 Speedway Blvd Stop C1400, Austin, TX 78712, USA\label{UAT} 
\and
Department of Astronomy, University of Washington Seattle, WA 98105, USA \label{Washington} 
\and
Instituto de Física y Astronomía, Universidad de Valparaíso, Avda. Gran Bretana 1111, Valparaíso, Chile \label{Chile} 
\and
Millennium Nucleus for Galaxies (MINGAL), Chile \label{Mingal} 
\and
Université Paris-Saclay, Université Paris Cité, CEA, CNRS, AIM, 91191 Gif-sur-Yvette, France \label{CEA} 
\and
Department of Computer Science, Aalto University, P.O. Box 15400, FI-00076 Espoo, Finland \label{Aalto} 
\and
Department of Physics, University of, P.O. Box 64, FI-00014 Helsinki, Finland \label{Helsinki} 
\and
Department of Physics and Astronomy, University of California, Riverside, 900 University Ave, Riverside, CA 92521, USA \label{Riverside} 
\and
Laboratory for Multiwavelength Astrophysics, School of Physics and Astronomy, Rochester Institute of Technology, 84 Lomb Memorial Drive, Rochester, NY 14623, USA \label{Rochester} 
\and
Institut de Physique, GalSpec, École Polytechnique Fédérale de Lausanne, Observatoire de Sauverny, Chemin Pegasi 51, 1290 Versoix, Switzerland \label{Lausanne} 
\and
Istituto Nazionale di Astrofisica (INAF), Astronomical Observatory of Trieste, Via Tiepolo 11, 34131 Trieste, Italy \label{INAF} 
\and
Instituto de Astrofísica de Canarias (IAC), La Laguna, E-38205, Spain \label{IAC} 
\and
Observatoire de Paris, LERMA, PSL University, 61 avenue de l'Observatoire, F-75014 Paris, France \label{LERMA} 
\and
Université Paris-Cité, 5 Rue Thomas Mann, 75014 Paris, France \label{Paris-Cite} 
\and
Universidad de La Laguna, Avda. Astrofísico Fco. Sanchez, La Laguna, Tenerife, Spain \label{laLaguna} 
\and
DTU Space, Technical University of Denmark, Elektrovej, Building 328, 2800, Kgs. Lyngby, Denmark \label{DTU} 
\and
Space Telescope Science Institute, 3700 San Martin Drive, Baltimore, MD 21218, USA \label{STScI} 
\and
Institut d'Astrophysique de Paris, UMR 7095, CNRS, Sorbonne Université, 98 bis boulevard Arago, F-75014 Paris, France \label{IAP} 
\and
European Space Agency (ESA), European Space Astronomy Centre (ESAC), Camino Bajo del Castillo s/n, 28692 Villanueva de la Cañada, Madrid, Spain \label{ESA} 
\and
Istituto Nazionale di Astrofisica (INAF), Osservatorio Astronomico di Padova, Vicolo dell'Osservatorio 5, 35122, Padova, Italy \label{Padova}
\and
Department of Physics and Astronomy, UCLA, PAB 430 Portola Plaza, Box 951547, Los Angeles, CA 90095-1547, USA \label{LA} 
\and
Institute for Astronomy, University of Hawai'i at Manoa, 2680 Woodlawn Drive, Honolulu, HI 96822, USA \label{HawaiiManoa} 
\and
University of Bologna - Department of Physics and Astronomy "Augusto Righi" (DIFA), Via Gobetti 93/2, I-40129 Bologna, Italy \label{Bologna} 
\and
INAF–Osservatorio di Astrofisica e Scienza dello Spazio, Via Gobetti 93/3, I-40129, Bologna, Italy \label{INAF-Bologna} 
\and
Zentrum f\"{u}r Astronomie, Universit\"{a}t Heidelberg, Philosophenweg 12, D-69120, Heidelberg, Germany \label{Heidelberg} 
\and
Department of Astronomy, University of Massachusetts, Amherst, MA 01003, USA \label{UMASS} 
}

      \date{Received ; accepted }

  \abstract
  {
We study the stellar mass function (SMF) of quiescent and star-forming galaxies and its dependence on morphology in 10 redshift bins at $0.2<z<5.5$. We used the COSMOS2025 catalog built from the $0.54 \, {\rm deg}^2$ \JWST\ imaging from the COSMOS-Web survey to select galaxies by type using the $NUVrJ$ rest-frame color diagram and classify them morphologically based on their bulge-to-total light ratio ($B/T$). 
\
The SMF of quiescent galaxies shows a rapid early build-up, with the most massive, ${\rm log}(M_{\star}/{\rm M_{\odot}})\gtrsim11$, being assembled by $z\sim1$ and evolving little since. The star-forming SMF evolves more slowly with redshift, following a mass-evolution scenario where galaxies grow in mass via star formation and quench once they reach the characteristic $\log(M^{*}/{\rm M}_{\odot})\sim10.6$. 
\
Bulge systems ($B/T>0.6$) dominate the quiescent SMF at ${\rm log}(M_{\star}/{\rm M_{\odot}})>10$ at all redshifts, while disk systems ($B/T<0.2$) dominate at ${\rm log}(M_{\star}/{\rm M_{\odot}})<9$. However, most bulge-dominated galaxies in the Universe are star-forming with their fraction increasing with redshift and decreasing mass, consistent with them being progenitors of quiescent bulges.
\
We find evidence for the onset of environmental quenching as early as $z\sim3$ from the upturn in the quiescent SMF at ${\rm log}(M_{\star}/{\rm M_{\odot}})<9.5$. This upturn is contributed by disk-dominated galaxies, consistent with environmental quenching scenarios in which satellites are quenched but retain their disk morphologies.
\
The number densities of ${\rm log}(M_{\star}/{\rm M_{\odot}})>10$ quiescent galaxies are lower compared to the recent literature by $0.1-0.7$ dex, but in good agreement with cosmological galaxy formation simulations at $2<z<3$. However, at $z>3$, simulations increasingly underpredict the observations.
\ 
Finally, we build a simple empirical model to describe the redshift evolution of galaxy number densities by parametrizing the quenching rate of all and bulge-dominated galaxies, baryon conversion efficiency and bulge formation function. Our model is consistent with an evolutionary scenario where star-forming galaxies grow a central bulge before permanently quenching in massive halos.
  }
   


\titlerunning{The SMF of quiescent and star-forming galaxies and its dependence on morphology}
   \maketitle


%


\section{Introduction}

Understanding how galaxies assemble their stellar mass and transition from star-forming to quiescent is a central goal of galaxy formation theory. In current models, quenching is driven by a variety of physical processes, including internal mechanisms such as bulge growth and morphological stabilization of disks \citep{Martig2009, DekelBurkert2014, Tacchella2015}, feedback from active galactic nuclei \citep[AGN,][]{Bower2006, croton_many_2006, Lagos2008, dubois_2013}, stellar feedback \citep{Ferrara2000, Gelli2024}, and the exhaustion of gas reservoirs \citep{Lilly2013}. Mergers can also drive quenching by inducing central starbursts and AGN activity that deplete or heat the gas \citep[e.g.,][]{Barnes1996, Hopkins2010}, and are generally considered an internally driven process triggered by external interactions. External mechanisms are driven by the environment \citep{peng_mass_2010} and include stripping \citep{Gunn1972, Moore1996}, strangulation \citep{Larson1980}, and virial shock heating in halos above a mass threshold \citep{birnboim_virial_2003, dekel_galaxy_2006}.
Morphology provides a powerful tracer of these processes: bulge-dominated and disk-dominated systems encode different quenching pathways that eventually result in the population of quiescent ellipticals.  However, the details of these physical processes are still poorly understood, as evidenced by the difficulty of the state-of-the-art cosmological galaxy formation simulations to reproduce the observed numbers of quiescent galaxies, especially at $z\gtrsim3$ \citep[e.g.,][and references therein]{Lagos2025}.

A powerful way to constrain galaxy evolution processes is through the galaxy stellar mass function (SMF), which measures the abundance of galaxies as a function of stellar mass across cosmic time. Furthermore, studying the SMF separated for star-forming and quiescent galaxy as well as by morphological class allows for even better refinement of theoretical models. 
\
Numerous past works have studied the SMF for star-forming and quiescent populations from ground- and space-based observatories out to $z\sim5$ \citep[e.g.,][]{Fontana2004, Arnouts2007, pozzetti_zcosmos_2010, Ilbert2010, Muzzin2013SMF, ilbert_mass_2013, Tomczak2014, Mortlock2015, Moutard2016, davidzon_cosmos2015_2017, McLeod2021, Santini2022, weaver_cosmos2020_2022}. These studies revealed that the quiescent SMF builds up rapidly since $z\sim5$, while the most massive systems appear to be largely in place by $z\sim1$. At the low-mass end, a power-law upturn in the quiescent SMF has been measured out to $z\sim2$, often interpreted as a signature of environmental quenching \citep{peng_mass_2010, Santini2022}. The morphological dependence of the SMF has also been explored \citep[e.g.,][]{Bundy2005, Ilbert2010, Mortlock2015, MHC2016}, showing that bulge-dominated systems shape the knee of the quiescent SMF whereas disk-dominated galaxies shape the low-mass end. 
However, these studies lacked the NIR sensitivity to robustly detect fainter and extremely red quiescent systems as well as the NIR resolution needed to trace their rest-frame optical morphology at $z>3$, limiting constraints on low-mass quiescent galaxies and the morphological dependence of the SMF at early epochs.

The advent of the \textit{James Webb Space Telescope} (\JWST) overcomes these limitations with unprecedented NIR depth and resolution, enabling direct measurements of the quiescent SMF and its morphological dependence into the epoch of early galaxy formation. Early \JWST-based studies have measured the abundances of massive quiescent galaxies at $z>2$ \citep[e.g.,][]{Valentino2023, Carnall2023, Carnall2024, Baker25b, Zhang2025, Stevenson2025}, and have started probing the low-mass end of the SMF with much improved completeness \citep[e.g.,][]{Hamadouche2025}. However, current efforts are still limited by small survey areas ($\lesssim0.2\, {\rm deg}^2$), which restrict the ability to robustly measure the abundance of the rarest and most massive systems, and suffer from cosmic variance. Additionally, most of these works combine \JWST\ and \hst\ with wavelength coverage at $\lambda\gtrsim0.5\, \si{\micron}$ which limits the application at $z<2$ to derive rest-frame colors needed to select quiescent galaxies. In addition, most works have focused on the quiescent population in isolation, yet a complete picture requires treating both the quiescent and star-forming together, since they are evolutionarily linked and a successful model must simultaneously explain the properties of both populations.

Open questions remain regarding the relative importance of different quenching channels and their dependence on stellar mass and cosmic time. In particular, it is unclear whether bulge growth is a universal prerequisite for quenching \citep[e.g.,][]{Martig2009, Barro2013, DekelBurkert2014, Zolotov2015, Tacchella2016} and whether low-mass galaxies can quench without significant morphological transformation \citep[e.g.,][]{Wuyts2011}. Likewise, the onset of environmental quenching remains uncertain: while models predict that dense environments should shape the low-mass end of the quiescent SMF \citep{peng_mass_2010}, the redshift at which these processes first become effective is still debated \citep{DeLucia2024, Lagos2024}. 
Although some observations have revealed systems with evidence of environmental quenching \citep[e.g.,][]{Jin2024, Toni2025}, others do not find such evidence \citep[e.g.,][]{Pan2025} at $z \sim3$, further fueling the debate. Addressing these questions requires simultaneous measurements of the SMF for both star-forming and quiescent populations, separated by morphology, over large areas and spanning the full redshift range where quiescent galaxies are observed.

In this paper, we present the first \JWST-based determination of the SMF for both quiescent and star-forming galaxies along with its dependence on galaxy morphology since the first Gyr -- almost the entire cosmic history in which quiescent galaxies can be found. This is allowed by the unique combination of the deep ($28$ mag AB) \JWST-selected sample (at $\sim~1-5 \, \si{\micron}$), over $\sim0.5 \, {\rm deg^2}$ from the COSMOS-Web \citep{Casey2023CW} survey. Specifically, we used the COSMOS2025 catalog \citep{ShuntovCW2025} that combines dense photometric coverage in over 30 bands spanning $\sim 0.3-8 \, \si{\micron}$, allowing accurate photometric redshifts, rest-frame colors, and stellar masses and quiescent vs. star-forming classification from $z\sim0.2$ to $z\sim5.5$. We leveraged the morphological information from profile fitting of composite bulge and disk models of every source, to carry out classification of disk-, bulge-dominated, and intermediate galaxies and measured the SMF for these populations.

This paper is organized as follows. In Section~\ref{sec:data} we present the data and sample selection; in Section~\ref{sec:measurements} we outline the methodology to measure and fit the SMFs; in Section~\ref{sec:results} we present the results, while in Section~\ref{sec:discussion} we discuss the physical implications of our results. We summarize and conclude in Section~\ref{sec:conclusions}. We adopt a standard $\Lambda$CDM cosmology with $H_0=70$\,km\,s$^{-1}$\,Mpc$^{-1}$ and $\Omega_{\rm m,0}=0.3$, where $\Omega_{\rm b,0}=0.04$, $\Omega_{\Lambda,0}=0.7$, and $\sigma_8 = 0.82$. All magnitudes are expressed in the AB system \citep{1983ApJ...266..713O}. Stellar masses were obtained assuming a \cite{Chabrier03} initial mass function (IMF).

\section{Data and sample selection}\label{sec:data}

To carry out a statistical study of quiescent and star-forming galaxies, we use the COSMOS2025 galaxy catalog, described in detail in \cite{ShuntovCW2025}. Briefly, COSMOS2025 is a catalog of over 700,000 galaxies based on the \JWST\ imaging program COSMOS-Web \citep[GO\#1727, PIs: Casey \& Kartaltepe,][]{Casey2023CW} that provides deep near-infrared imaging in four NIRCam (F115W, F150W, F277W, F444W) and one MIRI (F770W) filter over the central $\sim 0.54 {\, \rm deg}^2$ ($\sim 0.2 {\, \rm deg}^2$ for MIRI) in COSMOS. COSMOS2025 combines these data with ground- and space-based imaging to measure total photometry using a profile-fitting technique in 37 bands spanning \SIrange{0.3}{8}{\micron}. Additionally, it provides morphological measurements for all sources, including Sérsic \citep{Sersic1963} modeling and bulge-disk decomposition. Photometric redshifts and physical parameters from spectral energy distribution (SED) fitting using \lephare{} \citep{arnouts_measuring_2002, ilbert_accurate_2006} as well as non-parametric star formation histories using \cigale{} \citep{Boquien19, Arango-Toro2024}. In this paper, we used the physical parameters derived from \lephare. The catalog is described in detail in \cite{ShuntovCW2025, ShuntovSMF2025}.
 
Since the SMF is one of our principal measurements, we adopted the same sample as in \cite{ShuntovSMF2025}, which studies the SMF of the total population. This includes selection above a limiting magnitude corresponding to $80\%$ completeness, removal of stars, brown dwarf, and AGN/little red dot \citep[LRD,][]{Matthee2024} contamination. We briefly summarize the selection criteria in the following.
\begin{itemize}
    \item $m_{\rm F444W}<m_{\rm lim}$, where $m_{\rm lim} = 27.5$.
    \item stellar mass selection above a completeness limit $M_{\star} > M_{\rm lim}(z)$.
    \item $\chi^2_{\rm gal} < \chi^2_{\rm star}$, coupled with compactness criteria to remove stars, brown dwarfs and bright Type 1 AGN.
    \item Does not satisfy the AGN/LRD criterion 
    (AGN-SED $\lor$ Red) $\land$ Compact, where
    AGN-SED : $\chi^2_{\rm AGN} < \chi^2_{\rm gal}$, 
    Compact : $R_{\rm eff} < \ang{;;0.1} \, \lor \, 0.5<f(\ang{;;0.2})/f(\ang{;;0.5}) < 0.7$,  
    Red : $m_{\rm F277W} - m_{\rm F444W} > 1.5$ \citep{Akins2024}
    \item No X-ray counterpart by cross-matching within $1\, \si{\arcsec}$ with the \cite{Civano_2016} catalog.
    \item Outside of bright star masks defined in Subaru's Hyper-SuprimeCam (HSC).
\end{itemize}
This selection results in a total of 284,002 galaxies from $z=0.2$ to $z=5.5$. The effective area (after applying the HSC star masks) is $\sim 1551\, \si{arcmin}^2$, or $0.431\, \SI{}{\deg}^2$.

We computed the mass completeness limits in the same way as \cite{ShuntovSMF2025}, following the method of \cite{pozzetti_zcosmos_2010}. Briefly, at each $z$-bin, we take the 30$\%$ faintest galaxies in F444W as being the most representative of the population near the limiting stellar mass, and derive their rescaled mass $\text{log}(M_{\text{resc}}) = \text{log} (M_{\star}) + 0.4 ( m_{\text{F444W}} - m_{\rm lim} )$ by scaling the F444W magnitude to the magnitude limit of the survey ($m_{\rm lim}$). Then, we define the limiting stellar mass as the 95th percentile of the $M_{\rm resc}$ distribution.  We compute limiting stellar mass for both quiescent and star-forming galaxies separately.

We separated quiescent galaxies from star-forming ones using the $NUVrJ$ rest-frame colors following \cite{ilbert_mass_2013, davidzon_cosmos2015_2017, Weaver2023}. The $NUV$ is more sensitive to short timescale variation ($\sim10-100$ Myr) of the star formation rate (SFR) than $U$ band ($\sim100-300$ Myr) \citep{arnouts_NRK_2013}
\begin{equation} \label{eq:nuvrj}
    (NUV - r) > 3\times(r - J) + 1 \, {\rm and} \, (NUV - r) > 3.1.
\end{equation}

\begin{figure}[t!]
\centering
\includegraphics[width=1\columnwidth, trim=1.3cm 6cm 2cm 5cm, clip]{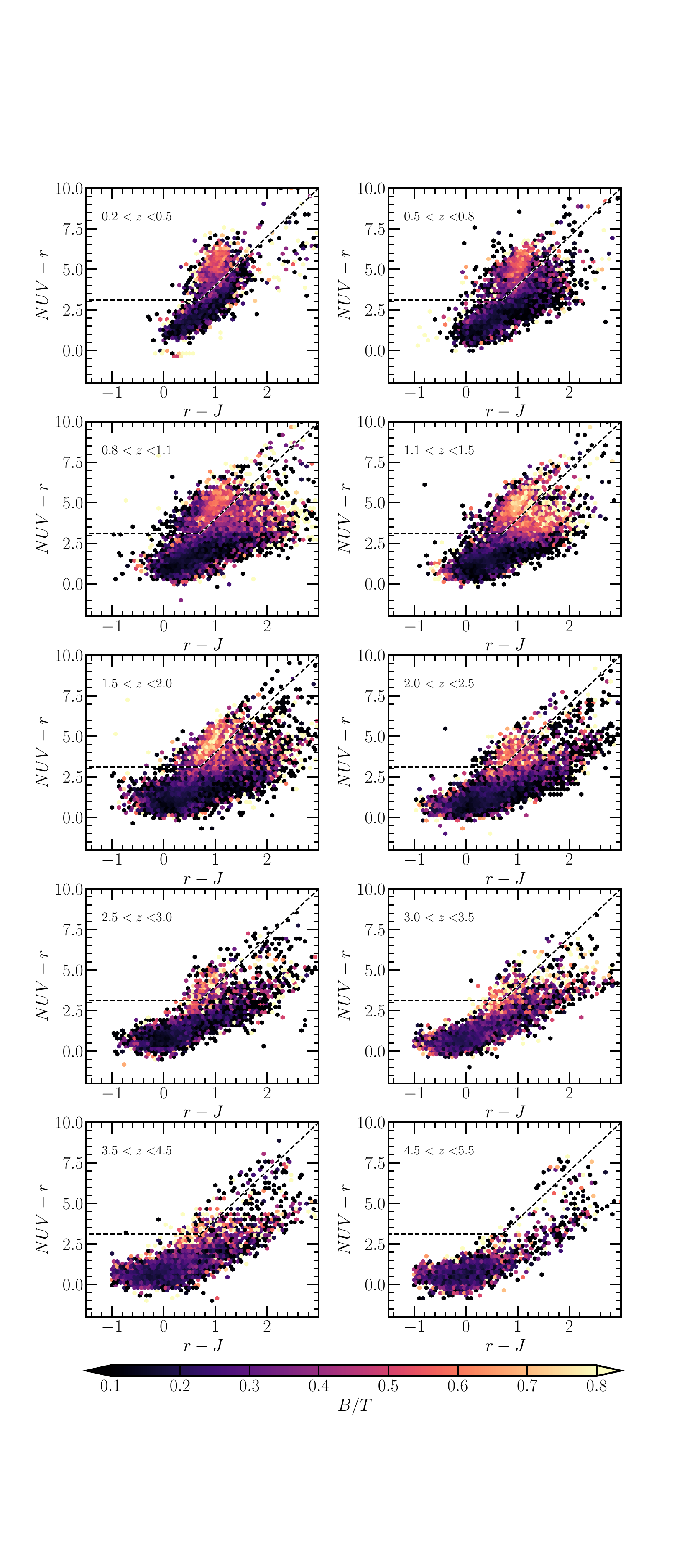}
\caption{Rest-frame $NUVrJ$ color diagram in 10 redshift bins used to select the samples of star-forming and quiescent galaxies separated by the dashed black line (Eq.~\ref{eq:nuvrj}). The histogram is color-coded according to the $B/T$ ratio F115W at $z<2$, F150W at $2<z<3$ and F277W for $z>3$. It includes sources with $\log(M_{\star}/\si{\Msun})>8.5$.}
\label{fig:NUVrJ-distributions}
\end{figure}
This yields a total of 15,608 quiescent and 268,394 star-forming galaxies.
Figure~\ref{fig:NUVrJ-distributions} shows the $NUVrJ$ color diagram in 10 redshift bins that we use to select the star-forming and quiescent samples, marked by the Eq.~\ref{eq:nuvrj} criterion in the dashed black line. We selected our samples in 10 redshift bins from $z=0.2$ to $z=5.5$, identical to those in \cite{ShuntovSMF2025, Weaver2023, davidzon_cosmos2015_2017}, thus facilitating comparison. We note that at $z\gtrsim2.5$ the rest-frame $J$ band moves out of the NIRCam F444W band, and for the majority of the galaxies with no MIRI F770W coverage we rely on the best-fit template extrapolation at longer wavelengths.

We investigated the morphological dependence of the SMF using bulge-disk decomposition measurements from the COSMOS2025 catalog. Briefly, these are done by fitting all sources with a composite $n_{\rm S}=1$ and $n_{\rm S}=4$ model, where the free parameters are the total flux, bulge-to-total ratio ($B/T$), bulge and disk radii and axis ratios \citep[more details in][]{ShuntovCW2025}. To consistently probe rest-frame optical emission ($\sim 0.4-0.6\, \si{\micron}$) across our redshift range, we adopt the $B/T$ measured in F115W for $z<2$, F150W for $2<z<3$, and F277W for $z>3$, following \cite{Yang2025}. 
We divide our samples into three morphological classes based on $B/T$: disk-dominated ($B/T<0.2$), intermediate ($0.2<B/T<0.6$), and bulge-dominated ($B/T>0.6$). These bins are motivated by the empirical correlation between $B/T$ and Sérsic index ($n_{\rm S}$), where $B/T<0.2$ and $B/T>0.6$ correspond to $n_{\rm S}\lesssim1.5$ and $n_{\rm S}\gtrsim3$, respectively, i.e., exponential disks and de Vaucouleurs profiles (Yang et al., in prep.).
The $NUVrJ$ diagram in Fig.~\ref{fig:NUVrJ-distributions} is color coded by the $B/T$ and shows a gradient of increasing $B/T$ values towards the quiescent region. Additionally, there is a gradient in the quiescent region going from more disk-dominated blue $r-J$ towards bulge-dominated red $r-J$ systems.

We note here several caveats regarding the $B/T$ decomposition. Even though single component fits are independently done for all sources, we used the double component fits, regardless of which one provides a better fit. However, given the fact that $B/T$ can also be 0 or 1, means that single component ($n_{\rm S}=1$ or $n_{\rm S}=4$) are effectively allowed. In reality, disks and bulges exhibit a wide range of $n_{\rm S}$ values, however the $n_{\rm S}=1$ or $n_{\rm S}=4$ approximations are acceptible for the purpose of estimating the $B/T$. Furthermore, there could be a possible bias arising due to the surface brightness limitation of the survey -- for fainter and higher redshift sources, even though they intrinsically have both a disk and a bulge component, the disk can be near or fainter than the surface brightness limit, which would bias high the $B/T$ measurement.

\begin{figure*}[th!]
\centering
\includegraphics[width=0.85\textwidth, trim=0cm 0.4cm 0cm 0cm, clip]{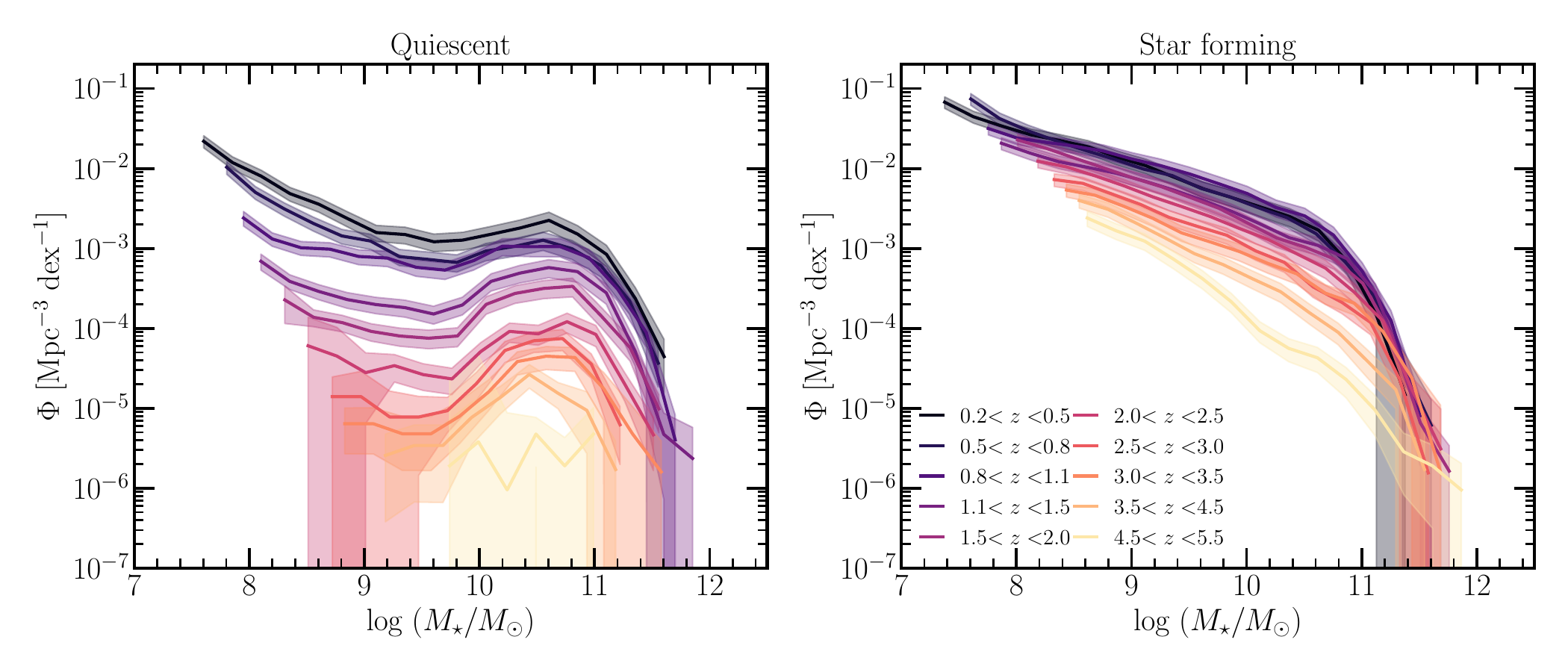}
\caption{Stellar mass function for quiescent (left) and star-forming (right) galaxies in 10 redshift bins from $z=0.2$ to $z=5.5$. The solid lines mark the measurements, while the filled areas envelop the $1\,\sigma$ confidence interval including Poisson, cosmic variance and SED-fitting errors. The SMFs are measured with a bin size of $\Delta \log M_{\star}=0.25$.}
\label{fig:SMF-Q-SF-twopanels}
\end{figure*}

\section{Measurements}\label{sec:measurements}

\subsection{$1/V_{\rm max}$ estimator for the SMF and associated uncertainties}

We measured SMFs in the same way as described in \cite{ShuntovSMF2025}. Briefly, this consists of using the $1/V_{\rm max}$ estimator \citep{schmidt_space_1968} to correct for the \citet[][]{Malmquist1922} bias that affects the number densities of faint galaxies. The $1/V_{\rm max}$ weights each galaxy by the maximum volume in which it would be observed given the redshift range of the sample and magnitude limit of the survey. 
For each redshift bin, the SMF is computed starting from the mass completeness limit of the quiescent and star-forming samples accordingly, using a uniform bin with size $\Delta \log M_{\star} = 0.25$.

The uncertainties in the SMF measurements include contributions from three components: 1) Poisson noise due to counting galaxies in bins of stellar mass, $\sigma_{\rm Pois}$; 2) SED fitting uncertainties, $\sigma_{\rm SED}$, in the $M_{\star}$ estimates due to photometric uncertainty and SED fitting degeneracies. These are estimated by drawing 1000 random samples from ${\rm PDF}(M_{\star})$, computing $\Phi(M_{\star})$ for each, and taking the standard deviation for each stellar mass bin. 3) Cosmic variance $\sigma_{\rm CV}$, arising from the fact that our survey might be probing a biased (under- or over-dense) volume of the Universe, computed following \cite{Jespersen2025, ShuntovSMF2025}. The final uncertainty is $\sigma_{\Phi}^2 = \sigma_{\rm Pois}^2 + \sigma_{\rm SED}^2 + \sigma_{\rm CV}^2$.

The SMF is also known to be affected by the \cite{Eddington1913} bias. This is due to the exponential cutoff of the SMF, which means that lower mass galaxies can be upscattered, thus inflating the number densities at the high-mass end. We account for the Eddington bias in the same way as \cite{ShuntovSMF2025} when fitting the functional forms of the SMF (\S\ref{sec:functional-forms-smf}) by convolving them with a kernel, $\mathcal{D}(M_{\star})$, that describes the stellar mass uncertainty in bins of mass and redshift. We built $\mathcal{D}(M_{\star})$ for each $z$-bin and each sample (e.g., star-forming, quiescent) by stacking the PDF($M_{\star}$), centered at the median of the distribution, of all galaxies in the given sample. This means that the best-fit function represents the intrinsic SMF, and all quantities that are inferred from it account for the Eddington bias.

\subsection{Functional forms to describe the SMF and MCMC fitting}\label{sec:functional-forms-smf}
The SMF of star-forming galaxies is typically well described by a parametric function which is a composite of a power law and an exponential cut-off function that describes the low- and high-mass ends respectively, often referred to as the \cite{Schechter76} function. The SMF of quiescent galaxies is commonly parametrized as a two-component Schechter function \citep[at least out to $z\sim2$, e.g.,][]{ilbert_mass_2013, Weaver2023, McLeod2021, Santini2022}, which can theoretically be explained as the result of two different quenching mechanisms \citep[e.g., mass- and environment-driven,][]{peng_mass_2010}. The single Schechter function is written in terms of the logarithm of the stellar mass as
\begin{equation} \label{eq:schechter}
\begin{split}
    & \Phi\, \dd({\rm log}M_{\star}) = \\
    & = {\rm ln}(10) \, \Phi^*\, e^{-10^{{\rm log}M_{\star} - {\rm log}M^*}}  \left(10^{{\rm log}M_{\star} - {\rm log} M^*}\right)^{\alpha+1} \dd ( {\rm log}M_{\star}),
    \end{split}
\end{equation}
where $M^*$ is the characteristic stellar mass that marks the so-called `knee', $\alpha$ is the low-mass slope, $\Phi^*$ is the overall normalization. The double Schechter function is given by
\begin{equation} \label{eq:double-schechter}
\begin{split}
    & \Phi\, \dd({\rm log}M_{\star}) = {\rm ln}(10) \,  e^{-10^{{\rm log}M_{\star} - {\rm log}M^*}} \times \\
    & \left[ \Phi^*_1  \left(10^{{\rm log}M_{\star} - {\rm log} M^*}\right)^{\alpha_1+1} + \Phi^*_2  \left(10^{{\rm log}M_{\star} - {\rm log} M^*}\right)^{\alpha_2+1} \right] \dd ( {\rm log}M_{\star}),
    \end{split}
\end{equation}
where the components share the same characteristic stellar mass $M^*$, but have different normalization $\Phi_1^*$ and $\Phi_2^*$ and low-mass slopes $\alpha_1$ and $\alpha_2$.

We fit the star-forming SMF with the single Schechter form. For the quiescent SMF we adopt a double Schechter out to $z=3.5$ and single Schechter in the last two bins at $z>3.5$. Similarly, for the quiescent SMF separated by $B/T$ we use the double Schechter out to $z=3.5$ and the single Schechter beyond, that we chose empirically.

We used the Markov Chain Monte Carlo (MCMC) method to fit the functional forms of the SMF via the affine-invariant ensemble sampler implemented in the \texttt{emcee} code \citep{foreman-mackey_emcee_2013}. We defined a Gaussian likelihood, used $6\times n_{\rm param}$ walkers, and defined a chain convergence criteria using the auto-correlation time $\tau$ with the requirement that $\tau > 60$ times the length of the chain and that the change in $\tau$ is less than $5\%$. We discarded the first $2\times\text{max}(\tau)$ points of the chain as the burn-in phase and thinned the resulting chain by $0.5\times\text{min}(\tau)$. We imposed flat priors on all parameters. For the double Schechter function fits, we required $\alpha_1 < \alpha_2$ and log$(\Phi^*_2/\Phi^*_1) > 0$. When quoting best-fit parameter values and uncertainties, we take the median and $1\,\sigma$ percentiles of the posterior distribution. When displaying the SMF and quantities derived from it, the median and $1\,\sigma$ envelopes are computed by randomly sampling the posterior 500 times. We provide the fitting results in Appendix~\ref{appdx:fitting-results}.

\section{Results}\label{sec:results}

\subsection{The stellar mass function of quiescent and star-forming galaxies}\label{sec:smf-q-sf}

\begin{figure*}[t!]
\centering
\includegraphics[width=1\textwidth]{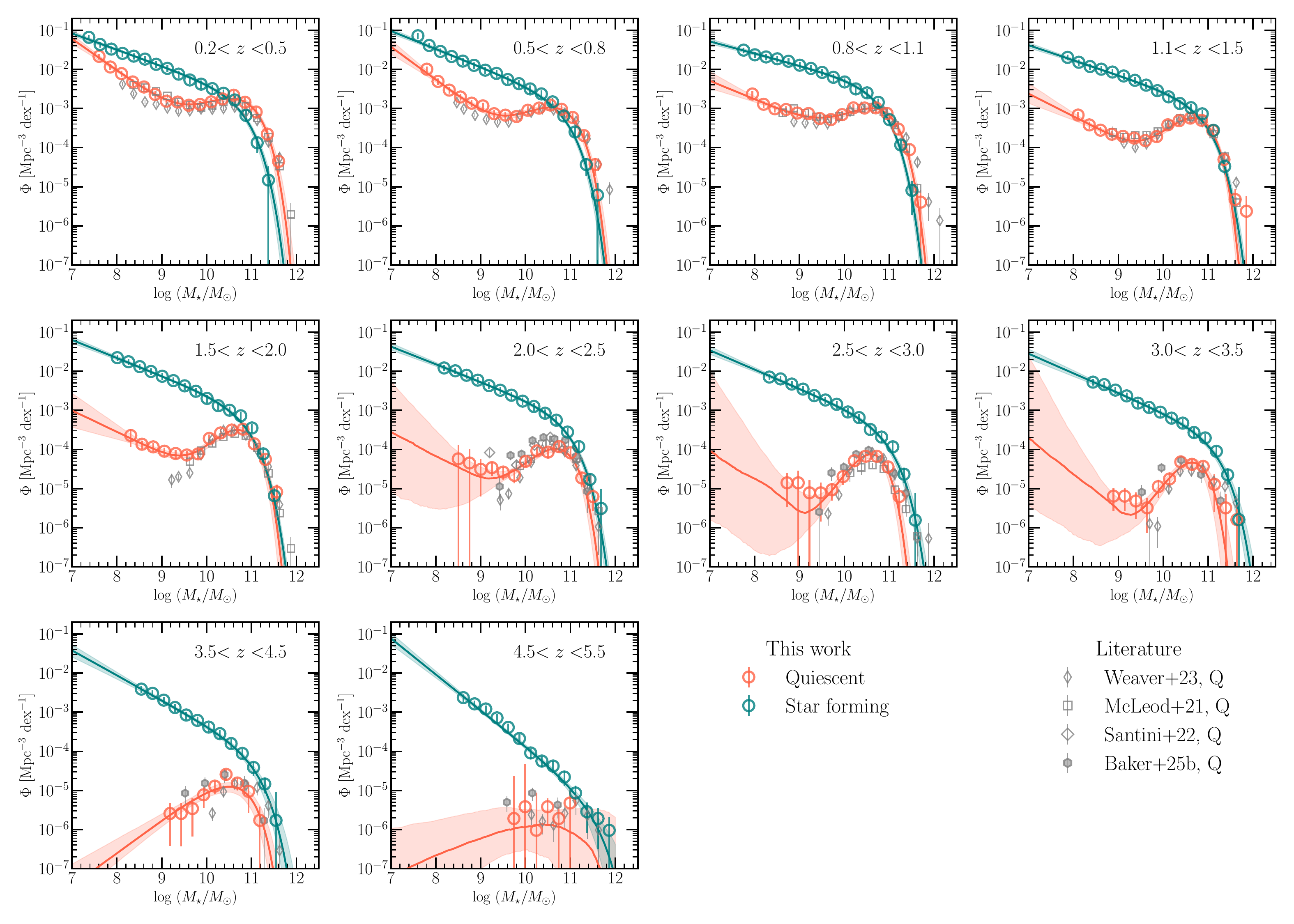}
\caption{Stellar mass function for quiescent (red) and star-forming (teal) galaxies. Each panel, corresponding to a different $z$-bin, shows the measurements (symbols) and best-fit intrinsic functions (solid lines and envelopes). The models are also shown extrapolated beyond the measured mass range. Comparison with recent literature works includes \cite{Weaver2023, McLeod2021, Santini2022, Baker25b} in gray symbols. This work indicates, for the first time, an upturn in the quiescent SMF at \logM~$\lesssim9.5$ out to $z\sim 3.25$.}
\label{fig:SMF-Q-SF}
\end{figure*}

We present the measurements of the SMF of quiescent and star-forming galaxies in 10 redshift bins in the $0.2<z<5.5$ range in Fig.~\ref{fig:SMF-Q-SF-twopanels}. In Fig.~\ref{fig:SMF-Q-SF} we show them in different panels for each $z$-bin along with the best-fit functions and the $1\,\sigma$ confidence interval. 

The quiescent SMF shows a clear mass-dependent evolution across the entire redshift range, revealed by the improved mass completeness at the low-mass end. The number densities of \logM~$\sim11$ galaxies at $0.2<z<4.5$ increase by about 2 dex, while those of \logM~$\sim9.5$ increase by over 2.5 dex. The evolution of even lower-mass quiescent galaxies \logM~$<9.5$ appears even faster, as evidenced by a flattening of the low-mass slope with increasing redshift. There is also a notable difference in the number density evolution in different epochs at the low- and high-mass end. The density of \logM~$>11.2$ galaxies shows almost no evolution at $0.2<z<1.1$, while it decreases by about $1-2$ dex at $z>1.1$, showing that the majority of massive galaxies have been quenched by $z\sim1$. The low-mass end, \logM~$<10$, although showing a slower evolution at $0.2<z<1.1$, has a considerable evolution at all redshifts, indicating that low-mass galaxies are continuously quenched throughout cosmic history.

Our measurements reveal for the first time an upturn of the quiescent SMF at \logM~$<9.5$ out to $z\sim 3.25$. Previously, this was established out to $z\sim2.5$ from studies based on both \hst\ \citep[e.g.,][]{Santini2022} and \JWST\ \citep{Hamadouche2025}. This drives the double Schechter form of the quiescent SMF out to $z\sim3$. However, despite the upturn of the SMF at lower masses, the number densities remain relatively low and near the volume limit of the COSMOS-Web survey; therefore the uncertainties are relatively high. Additionally, the exact classification of quiescent galaxies can also have a considerable effect on the inferred number densities at the low-mass end as we discuss in \S\ref{sec:quiescent-classification}.

The star-forming SMF shows a considerably slower change across the $0.2<z<5.5$ range, with strong mass-dependent evolution. This is evidenced by unchanged number densities of \logM~$>11.2$ galaxies, at least since $z\sim3$, while e.g., \logM~$<10.5$ galaxies increase by about 1 dex since $z\sim3$.
The star-forming SMF is expected to follow a mass evolution scenario -- galaxies grow in mass via star formation, which results in a horizontal shift in the SMF. For example, a growth of 1 dex in mass is required to match the SMFs at $z\sim2.25$ and $z\sim0.3$. This is further evidenced by the low-mass slope remaining roughly constant with redshift (c.f., \S\ref{sec:params-evolution}, Fig.~\ref{fig:Parameters-vs-redshift}). The lack of evolution in number densities of massive galaxies beyond the `knee' indicates that a quenching process that is mass dependent kicks in after galaxies grow beyond $M^*$, ceases star formation and moves galaxies to the high-mass end of the quiescent SMF \citep[][c.f., \S\ref{sec:discussion}]{peng_mass_2010}.

We compared our measurements with the recent literature for the case of the quiescent SMF in Fig.~\ref{fig:SMF-Q-SF}. These include \cite{Weaver2023, McLeod2021, Santini2022} from pre-\JWST studies and \cite{Baker25b} based on \JWST\ samples, shown in gray symbols. We note that the quiescent classification is different in these studies: \cite{Weaver2023} use the $NUVrJ$, \cite{McLeod2021} and \cite{Santini2022} use the $UVJ$ rest-frame colors, while \cite{Baker25b} use a sSFR-based classification. In general, there is very good agreement at all redshifts. Notable difference is the upturn at \logM~$<9.5$ that our work shows out to $z\sim 3.25$, which is not seen in \cite{Weaver2023}, likely because of our deeper detection at $>1 \,\si{\micron}$. \cite{Baker25b} analyze 745 quiescent galaxies at \logM~$>9.5$ and $2<z<7$ in $\sim 800$ arcmin$^2$ of public JWST fields and their SMFs are generally in agreement with ours. However, they find slightly increased densities of \logM~$\sim9.5-10.2$ galaxies at $z\sim2.25$, $z\sim3.25$ and $z\sim4$, as well as in the massive end \logM~$>11$ at $z\sim2.75$; these differences remain within $\sim 2\,\sigma$ uncertainties.

\subsection{Morphological dependence of stellar mass function}\label{sec:smf-q-sf-morpho}

\begin{figure*}[t!]
\centering
\includegraphics[width=1\textwidth]{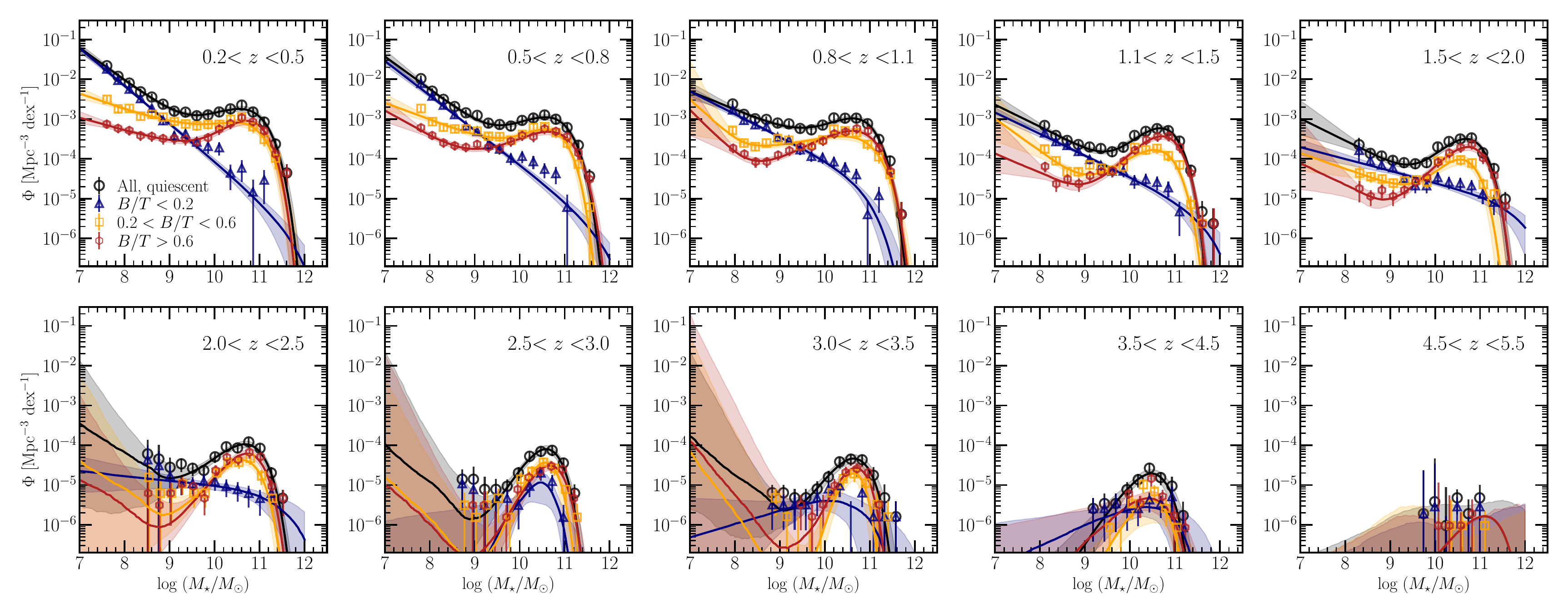}
\caption{Stellar mass function for quiescent galaxies separated by the bulge-to-total ratio. Each panel, corresponding to a different $z$-bin, shows the measurements (symbols) and best-fit functions (solid lines and envelopes) for all quiescent (black), disk-dominated ($B/T<0.2$, blue), intermediate ($0.2<B/T<0.6$, orange) and bulge-dominated ($B/T>0.6$, dark red). The models are also shown extrapolated beyond the measured mass range.}
\label{fig:SMF-Q-SF-BT}
\end{figure*}

\begin{figure*}[t!]
\centering
\includegraphics[width=1\textwidth]{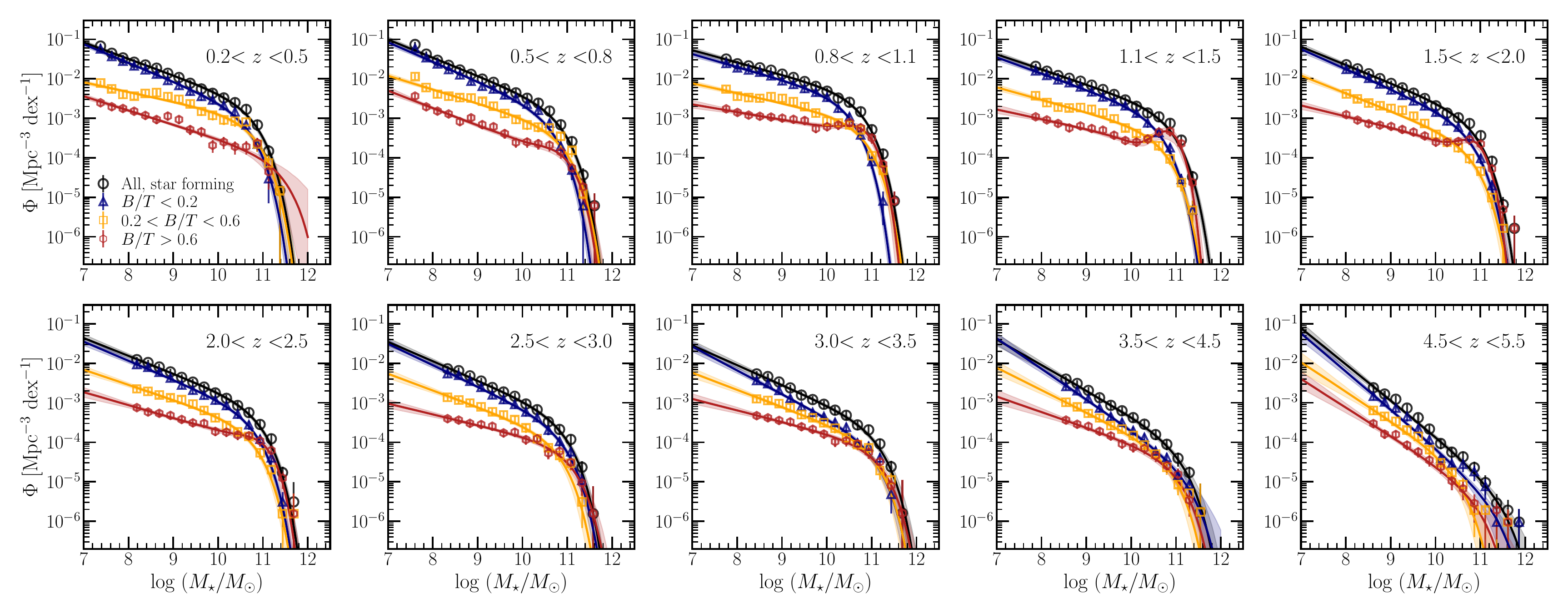}
\caption{Stellar mass function for star-forming galaxies separated by the bulge-to-total ratio. Each panel, corresponding to a different $z$-bin, shows the measurements (symbols) and best-fit functions (solid lines and envelopes) for all quiescent (black), disk-dominated ($B/T<0.2$, blue), intermediate ($0.2<B/T<0.6$, orange) and bulge-dominated ($B/T>0.6$, dark red). The models are also shown extrapolated beyond the measured mass range.}
\label{fig:SMF-SF-BT}
\end{figure*}

We investigated the morphological dependence of the quiescent and star-forming SMF by measuring it for three samples selected by their bulge-to-total ratio. This unveils which morphological populations dominate the number densities at different stellar masses. In Fig.~\ref{fig:SMF-Q-SF-BT} and Fig.~\ref{fig:SMF-SF-BT} we show the SMF for all (black), disk-dominated ($B/T<0.2$, blue), intermediate ($0.2<B/T<0.6$, orange) and bulge-dominated ($B/T>0.6$, dark red), for both quiescent and star-forming galaxies. In this section we first describe the quiescent SMF for the three classes and then the star-forming one.

Quiescent bulge systems ($B/T>0.6$) dominate the number densities of quiescent galaxies at \logM~$\gtrsim10$ at virtually all redshifts, and they shape the `knee' of the SMF of the total quiescent population. This can also be seen in Fig.~\ref{fig:Parameters-vs-redshift} where the values of $M^*$ are similar for the total and $B/T>0.6$ best fit SMFs. Their number densities decrease with decreasing mass, and by \logM~$\lesssim9$ quiescent bulge systems are the least abundant at all redshifts. The high-mass Schechter component that shapes the knee is sharper at high redshifts but becomes flatter with decreasing redshift (also indicated by flattening of the slope $\alpha_2$ with decreasing redshift, Fig~\ref{fig:Parameters-vs-redshift}). This indicates that high-mass bulge-dominated quiescent systems form the fastest, while lower-mass ones form more gradually with time and are almost as abundant as their massive counterparts by $z\sim0.2$, an incarnation of the downsizing scenario \citep{1996AJ....112..839C, de_lucia_formation_2006, thomas_environment_2010}.

Quiescent disk systems ($B/T<0.2$) dominate the quiescent SMF at masses below \logM~$\lesssim9$ because they have significantly steeper $\alpha_1$ slopes, at least out to $z\sim1.75$. Their contribution at higher redshifts becomes difficult to probe because of the limited survey depth, including the difficulty in robustly measuring morphology via profile fitting of lower $S/N$ sources. This population shows a single Schechter form, whose normalization does not appear to evolve with redshift (Fig.~\ref{fig:Parameters-vs-redshift}). Combined with the slope becoming shallower with increasing redshift, this results in a roughly constant abundance of massive quiescent disks with redshift (at least out to $z\sim2.5)$, which have been observed \citep[e.g.,][]{Toft2017, DEugenio2024, Slob2025}.

Intermediate systems have similar contributions from both the bulge and disk components ($0.2<B/T<0.6$) and their SMF lies between that of the bulges and disks. At lower redshifts ($z<1$) they contribute to the high mass end ($\gtrsim M^*$) equally as the bulge-dominated systems to the quiescent SMF, but their number densities decrease faster with redshift out to $z\sim2$. However, at $z>2$ it closely resembles the bulge-dominated SMF. It is unclear whether this result is physical or an artifact, due to the difficulty in robustly measuring the $B/T$ at these redshifts where sources are dimmer. 

Star-forming bulge systems (Fig.~\ref{fig:SMF-SF-BT}) are about $1-1.5$ dex less abundant than star-forming disks at \logM~$\lesssim10$, but are equally abundant at \logM~$>11$. Regardless, the abundance of star-forming bulges is relatively high at all redshifts, with their relative abundance increasing with redshift (discussed further in \S\ref{sec:quiescent-fractions}, Fig.~\ref{fig:Quiescent-fractions}).

\begin{figure*}[t!]
\setlength{\abovecaptionskip}{-0.2cm}
  \centering
  \begin{subfigure}{0.49\textwidth}
    \includegraphics[width=\linewidth]{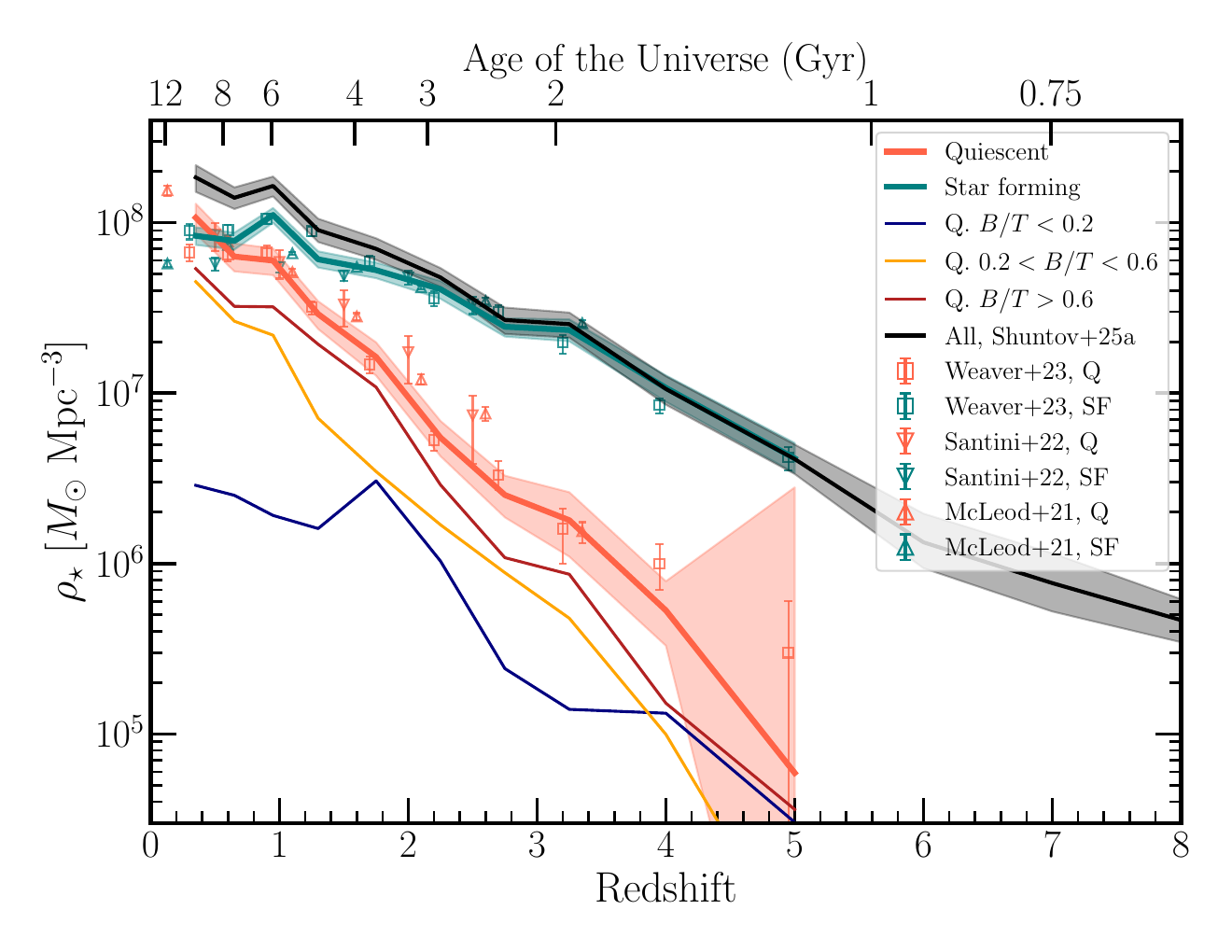}
    \label{fig:SMD}
  \end{subfigure}
  \hfill
  \begin{subfigure}{0.49\textwidth}
    \includegraphics[width=\linewidth]{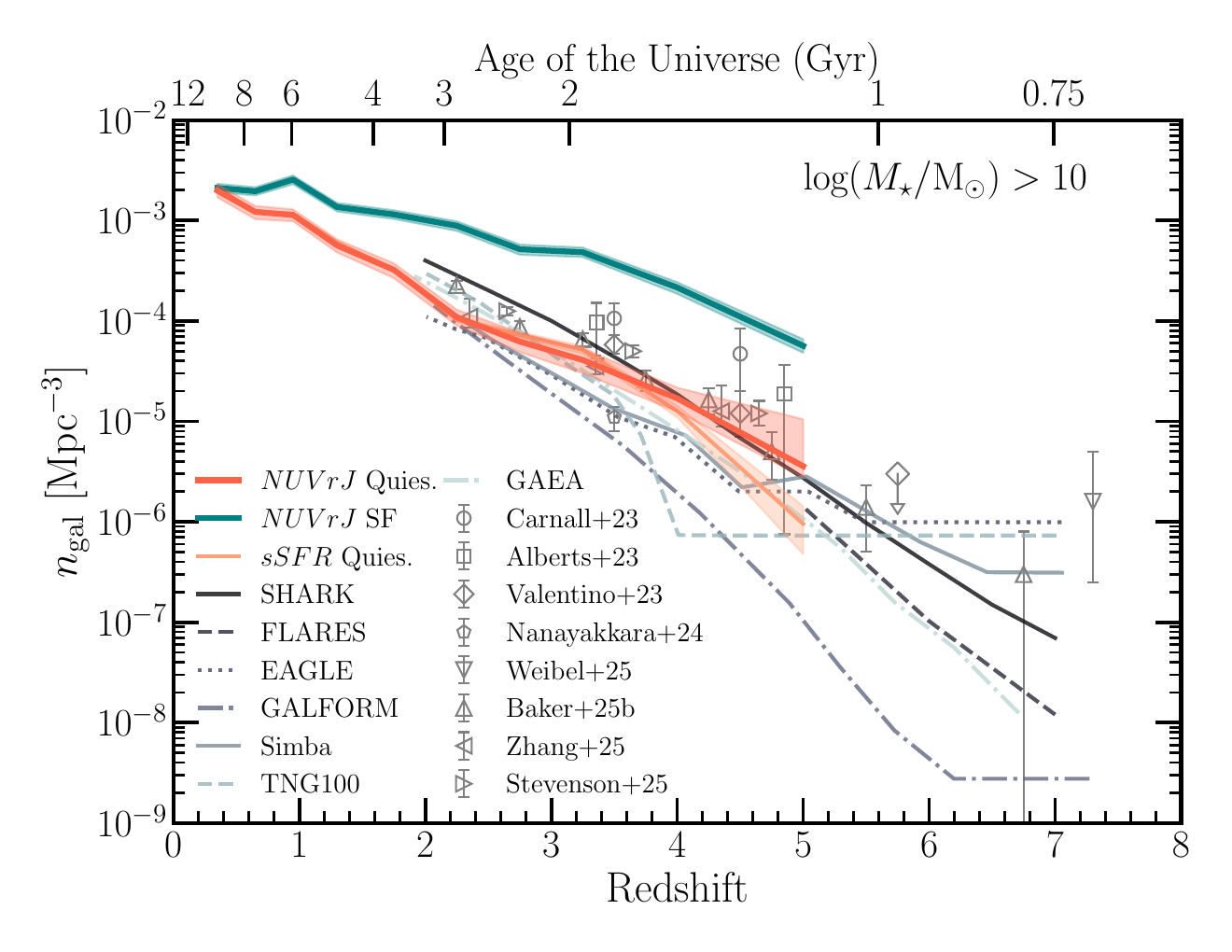}
    \label{fig:ND}
  \end{subfigure}
  \caption{Stellar mass and number density of quiescent and star-forming galaxies. The left panel shows the stellar mass density for total quiescent and star-forming population, as well as for the $B/T$-selected quiescent samples for \logM~$>8$. It also includes literature comparison from recent observational studies. The right panel shows the number densities of galaxies with \logM~$>10$ selected using the $NUVrJ$ rest-frame colors, as well as a ${\rm sSFR}/{\rm yr }^{-1} < 0.2/t_{\rm H}(z)$ selection for the quiescent from our work. We compare with a compilation of simulations (lines) and observations (symbols with error bars). All are selected to have \logM~$>10$, but the quiescent classification varies for the different observational works. In the simulations, quiescent galaxies are classified as sSFR$< 5\times 10^{-11}$ yr$^{-1}$ and include uncertainties by convolving he stellar masses with Gaussian-distributed errors with a width of 0.3 dex. In our, work the sSFR selection yields $n_{\rm gal}$ lower by $\sim0.2-0.5$ dex at $z\gtrsim4$ compared to the $NUVrJ$ selection.
  }
  \label{fig:NumberDensities}
\end{figure*}

Disk systems dominate the star-forming SMF at all redshifts and masses of \logM~$\lesssim11$, making it the most abundant galaxy population at low and intermediate masses. This reflects the fact that star formation preferentially occurs in rotationally-supported gas-rich disks where cold gas can efficiently cool and fragment \citep{FallEfstathiou1980, Kennicutt2012}.
\
SMF of intermediate star-forming systems lies between the two, and since this population is a mix of both, it is difficult to interpret.

Several studies in the literature have investigated the dependence of galaxy abundances on their morphology, stellar mass and star formation activity. Here we qualitatively compare our measurements to similar ones from the literature. \cite{Ilbert2010} measured the SMF of quiescent and star-forming galaxies for elliptical and spiral/irregular classes in S-COSMOS out to $z=2$. Our results are qualitatively consistent in finding that elliptical galaxies dominate the massive end of the quiescent SMF and that low-mass star-forming galaxies are predominantly disks. \cite{MHC2016} measured the SMF for several morphological classes based on \hst\ observations using machine learning and obtained qualitatively consistent results. Similarly, \cite{MHC2025} uses the same COSMOS-Web galaxy catalog coupled with machine learning-based morphological classification of essentially the same galaxies as our sample. They also find that the massive end of the quiescent SMF is dominated by bulge systems, but reveal that the low-mass end is dominated by peculiar galaxies. As shown in Fig.~A.1 of \cite{MHC2025}, the peculiar class displays a mixture of disk and clumpy/irregular structures, which are fitted with a low $B/T$ profile in our work. \cite{Lang2014} performed a bulge+disk decomposition using profile fitting in $H$-band from CANDELS \citep{Grogin2011,Koekemoer2011} and found that the quiescent fraction increases with both $B/T$ ratio and stellar mass, consistent with our trends.

\subsection{Number and stellar mass densities of quiescent galaxies}\label{sec:smd-numdensities}

The cosmic stellar mass density (SMD, $\rho_{\star}$) measures how much stellar content there is in a volume of the Universe and can inform us about the contributions from different galaxy populations. Along with the overall number density of quiescent galaxies, they provide powerful observational benchmarks for theoretical models and simulations. Using our compilation of quiescent and star-forming galaxies along with the measured SMFs, we computed their cosmic SMD and the number density, shown in Fig.~\ref{fig:NumberDensities}.

The SMD of quiescent and star-forming galaxies, computed by integrating the SMFs from a lower limit of \logM~$>8$, is shown in the left panel of Fig.~\ref{fig:NumberDensities}. We computed these by integrating the fitted intrinsic SMFs, corrected for the Eddington bias. We also show the SMD of the $B/T$-selected quiescent samples, as well as the SMD of the total population, for which we take the results from \cite{ShuntovSMF2025}. Star-forming galaxies dominate the cosmic SMD at $z\gtrsim0.65$, with $\Delta \rho_{\star}\gtrsim1.5$ dex at $z>4$ compared to the quiescent one. However, the SMD of quiescent galaxies shows a steeper increase with time by about a factor of 1,000 since $z\sim5$, while the star-forming one increased by only about a factor of 30 over the same time. The quiescent SMD equals the star-forming one at $z\sim0.65$, and dominates the cosmic SMD at $z\lesssim0.65$. This means that most of the stellar mass in the Universe today is locked in quiescent galaxies. The SMD for $B/T$-selected quiescent galaxies shows that most of the SMD from quiescent galaxies is locked in bulge-dominated systems. 

We compared our SMD measurements with several pre-\JWST\ works from the literature \citep{Weaver2023, McLeod2021, Santini2022}. There is very good consistency with \cite{Weaver2023} which bases their analysis in COSMOS and uses similar selection methods (e.g., $NUVrJ$ colors). \cite{McLeod2021} and \cite{Santini2022} show consistent trends, but find higher SMD of quiescent galaxies, which could be due to the treatment and effect of the Eddington bias, given the fact that the SMFs are in better agreement than $\rho_{\star}$.

The number density of quiescent and star-forming galaxies with masses \logM~$>10$ as a function of redshift is shown in the right panel of Fig.~\ref{fig:NumberDensities}. These are computed by simply counting the number of galaxies and dividing by the total volume. The uncertainties include Poisson, SED fitting and cosmic variance, incorporated using the $1\,\sigma$ percentiles of the fitted SMF functions. Quiescent galaxies show a steady increase in density by about a factor of 1,000 since $z\sim5$, reaching $\sim 2\times 10^{-3} \, {\rm Mpc}^{-3}$ by $z\sim0.35$. Massive star-forming galaxies, on the other hand, increase in number density only by about a factor of 30 (1.5 dex) over the same timespan.

The number densities of quiescent galaxies have been studied extensively in the literature \citep[e.g.,][]{Girelli2019, Carnall2020, Gould2023, Valentino2023, Carnall2023, Long2024, Baker25b, Zhang2025, Stevenson2025, YangQG2025}. However, different surveys and studies have shown a high degree of dispersion ($\sim1-2$ dex). This can be due to quiescent classification methods, but also due to cosmic variance that can significantly impact the observed number densities in smaller fields \citep[e.g., by a factor of $2-3$,][]{Valentino2023}.
We compared our measurements conducted in the largest area probed by \JWST\ to date with the recent literature, namely \cite{Carnall2023, Valentino2023, Alberts2024, Nanayakkara2024, Weibel2025, Baker25b, Zhang2025, Stevenson2025}. In general, we found lower number densities of quiescent galaxies by about $0.1-0.7$ dex. The closest agreement is with \cite{Baker25b} who also analyzes a large area ($\sim 800$ arcmin$^2$, about half of the area in our work) but selects the quiescent sample with a combination of $UVJ$ colors and sSFR cuts. At $2\lesssim z \lesssim 3$ they find higher number densities by about $0.1-0.2$ dex at $1-3 \, \sigma$, compared to our work, but are consistent at $z\gtrsim3.5$. The consistency between these two \JWST-based works that analyze two of the largest areas is encouraging and shows that they are likely converging towards real number of density of quiescent galaxies, with the effect of cosmic variance minimized.
The remaining discrepancy is likely due to the different photometric bands available in both works and the different SED fitting codes and the configuration of their physical prescriptions that can affect the estimated redshifts and physical properties.

\subsection{Fraction of quiescent galaxies and its morphology dependence}\label{sec:quiescent-fractions}

\begin{figure}[t!]
\centering
\includegraphics[width=0.85\columnwidth]{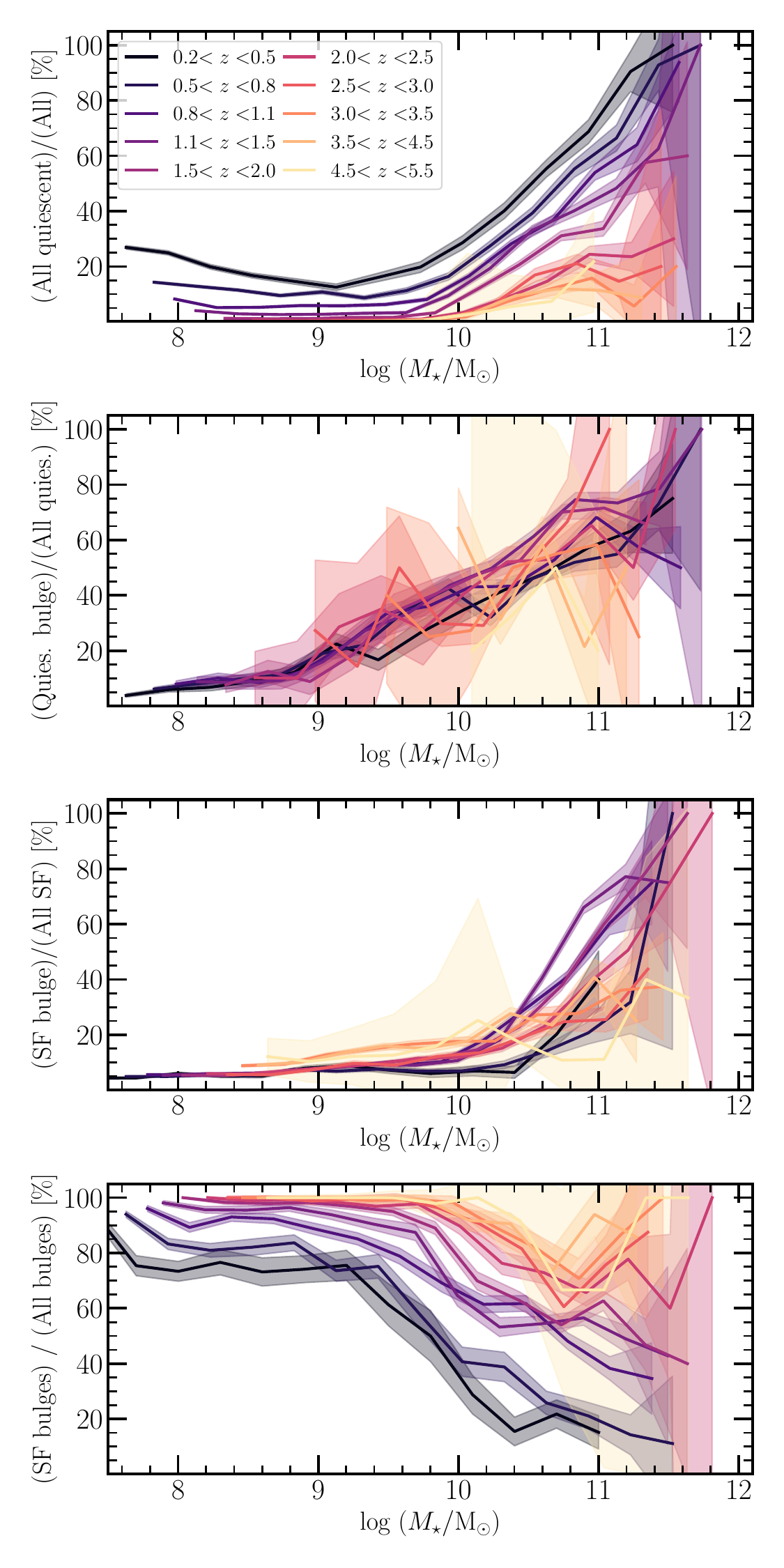}
\caption{Fraction of quiescent and bulge-dominated galaxies as a function of stellar mass in the 10 redshift bins from $z=0.2$ to $z=5.5$. The first panel shows the fraction of quiescent to all galaxies, the second panels shows the fraction of quiescent bulges to all quiescent, the third panel shows the fraction of star-forming bulges to all star-forming galaxies, and the fourth panel shows the fraction of star-forming bulges compared to the total bulge population.}
\label{fig:Quiescent-fractions}
\end{figure}

We computed the fraction of quiescent and bulge-dominated galaxies as a function of stellar mass, using the number of quiescent and star-forming galaxies per mass bin. We note that by not using the best-fit Schechter functions for this, we do not account for the potential effect of the Eddington bias that might impact the measurements at the high-mass end. We propagated the Poisson, cosmic variance and SED fitting uncertainties to the total error bar. The results are shown in Fig.~\ref{fig:Quiescent-fractions}. 

The fraction of quiescent galaxies (compared to the total population), $f_q$ is shown in the first (top) panel of Fig.~\ref{fig:Quiescent-fractions}. In general, $f_q$ increases with stellar mass at all redshifts. Out to $z\sim1.5$ quiescent galaxies account for up to $\sim100\%$ of the \logM~$\sim11.5$ population. There is also a clear trend with redshift --- at a given stellar mass, there are always higher $f_q$ at lower redshifts. At the highest masses that we can probe, at $z\gtrsim3$, quiescent galaxies make less than $\sim 30\%$ of the galaxy population. 

The fraction of low-mass quiescent galaxies increases with decreasing stellar mass at \logM~$<9$, at least out to $z\sim1$. This potentially indicates a mass-dependent quenching mechanism of low-mass galaxies, or an additional mechanism on top of the mass-independent environmental quenching under the \cite{peng_mass_2010} paradigm (such as stellar feedback \citealt{Ferrara2000}, discussed further in \S\ref{sec:discussion}). These fractions indicate that at most $\sim 20 \%$ of low-mass galaxies have been affected by a low-mass quenching mechanism over entire cosmic history.

Next, we investigated the fraction of quiescent and star-forming galaxies with bulge-dominated morphology ($B/T>0.6$) compared to the total quiescent and star-forming population accordingly, shown in the second and third panels of Fig.~\ref{fig:Quiescent-fractions}. The fraction of quiescent bulges increases almost linearly with stellar mass at all redshifts and bulges make the majority of massive quiescent galaxies ($\gtrsim60 \%$) at all redshifts (albeit with large uncertainties). There is no significant trend with redshift in this case, suggesting that the quenching mechanisms of bulge-dominated galaxies are mass dependent but constant over time.
On the other hand, the fraction of star-forming bulges remains low ($\lesssim10\%$) out to \logM~$\sim10$ at all redshifts, after which it increases with mass. At \logM~$\gtrsim11.4$, bulges dominate the SF population with fractions of up to $\sim80-100 \%$.

\begin{figure*}[t!]
\centering
\includegraphics[width=0.95\textwidth]{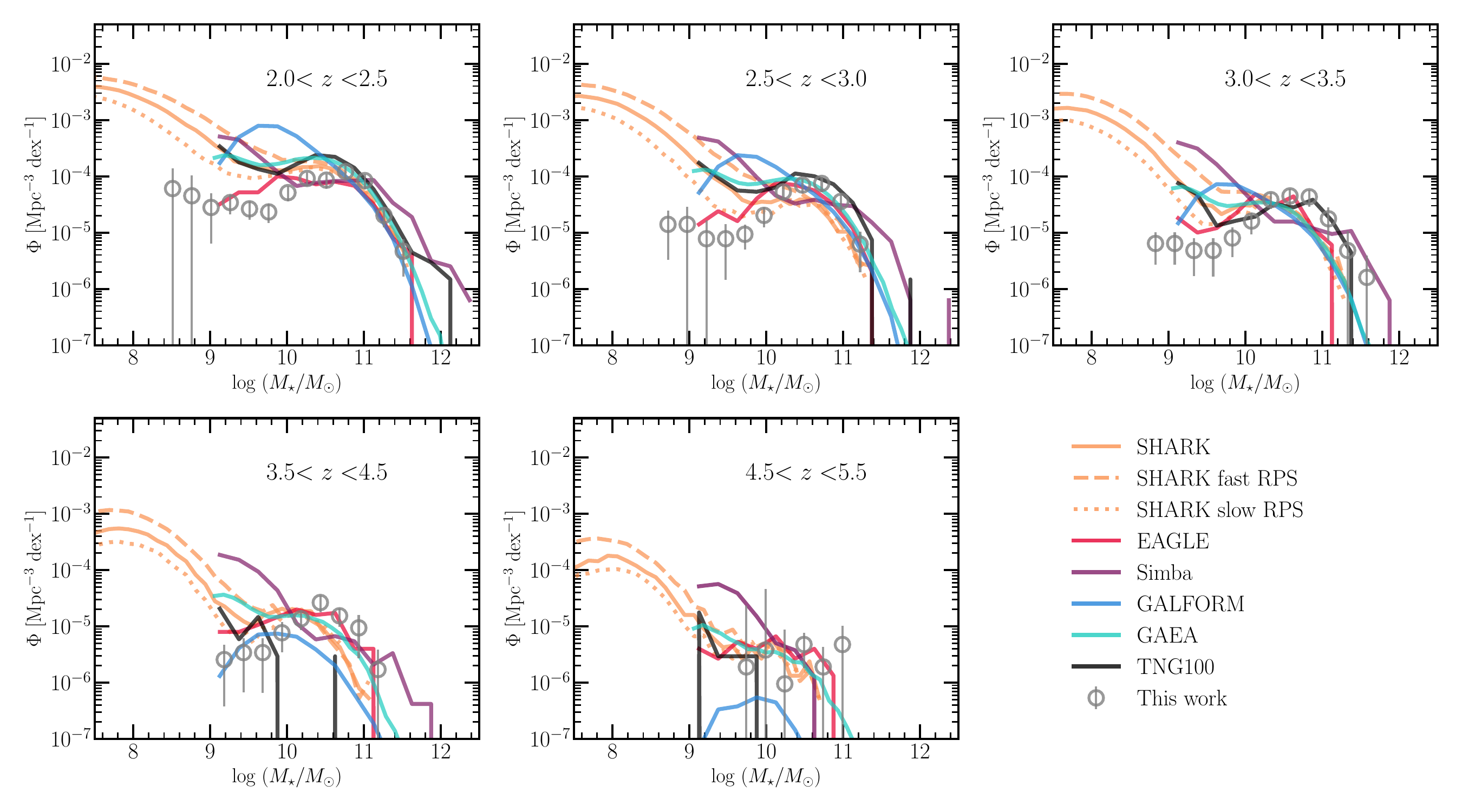}
\caption{Comparison of the observed quiescent SMF with semi-analytical models and hydrodynamical simulations. These include \textsc{SHARK} \citep{Lagos2018, Lagos2024} run with two alternative runs with slow and fast ram-pressure stripping, \textsc{GALFORM} \citep{GALFORM_Lacey2016}, \textsc{GAEA} \citep{DeLucia2024}, \textsc{Simba} \citep{SIMBADave2019}, \textsc{EAGLE} \citep{Schaye2015, Crain2015}, \textsc{TNG100} \citep{springel_first_2018, pillepich_first_2018}. These are computed using a $UVJ$ color selection and include uncertainties by convolving the stellar masses with a Gaussian kernel of 0.3 dex.}
\label{fig:QSMF-vs-sims}
\end{figure*}

Finally, we computed the fraction of star-forming galaxies that are bulge-dominated compared to the total bulge-dominated population (both star-forming and quiescent, shown in the bottom panel). There is a clear trend with both redshift and mass, where star-forming galaxies make most of the bulge-dominated population at high redshift.  
At $z>2$ more than $60\%$ of massive (\logM~$>10.5$) bulges were star-forming, but at later epochs this population dwindles down to $\lesssim 20\%$ by $z\sim0.5$. This coincides with the epoch in which massive quiescent ellipticals grow rapidly \citep[e.g.,][]{Bell2004, Cimatti2006}.
At $z>3$, $\sim 100 \%$ of bulge-dominated galaxies with \logM~$<10$ are star-forming, while this fraction decreases down to $\sim 70\%$ with increasing mass. At a fixed stellar mass, the fraction of star-forming bulges decreases with redshift. At $z<1$, star-forming bulges make about $80\%$ at \logM~$<9$, which decreases with mass to less than $20\%$ at \logM~$\sim 11$. These results confirm that there is a significant population of star-forming bulges, especially at lower masses and earlier epochs \citep[e.g.,][]{Barro2013, Barro2014, Barro2017, Tacchella2015, Gomez-Guijarro2019}. The prevalence of this population decreases with both increasing mass and cosmic time, indicating an evolutionary link between star formation activity and morphology which we discuss further in \S\ref{sec:morpho-transform-quenching}.

\section{Discussion} \label{sec:discussion}

\subsection{Comparison with models and simulations} \label{sec:model-sims-comparison}

Observations of the SMF, stellar and number densities of quiescent galaxies serve as a crucial benchmark against which the physics in semi-analytical models (SAM) and hydrodynamical simulations is calibrated. Our work aims to provide a comprehensive set of such benchmark observables from photometrically selected quiescent galaxies from \JWST\ spanning $z=0.2 - 5.5$. In this section, we discuss the comparison with a compilation of SAMs and hydrodynamical simulations.

In Fig.~\ref{fig:NumberDensities} (right panel) we compare the number densities of quiescent galaxies with SAMs: \textsc{Shark} \citep{Lagos2018, Lagos2024} \textsc{GALFORM} \citep{GALFORM_Lacey2016} and \textsc{GAEA} \citep{DeLucia2024}, and hydrodynamical simulations: \textsc{Simba} \citep{SIMBADave2019}, \textsc{EAGLE} \citep{Schaye2015, Crain2015}, \textsc{TNG100} \citep{springel_first_2018, pillepich_first_2018} and \textsc{Flares} \citep{Lovell2021, Vijayan2022, Wilkins2023}. These use the same mass selection of \logM~$>10$, but classify quiescent using ${\rm sSFR} < 5\times 10^{-11} \, {\rm yr}^{-1}$. For a more just comparison, we also computed the $n_{\rm gal}$ using a ${\rm sSFR}/{\rm yr }^{-1} < 0.2/t_{\rm H}(z)$, where  $t_{\rm H}$ is the Hubble time, shown in the light red color in Fig.~\ref{fig:NumberDensities}. 
The stellar masses in the simulations are convolved with Gaussian-distributed errors with a width of 0.3 dex, for a more just comparison with observations. This is the same compilation as in \cite{Lagos2025}, where the effect of different $M_{\star}$ and sSFR selections are discussed. In the redshift range of $2\lesssim z \lesssim3$ there is a reasonable agreement where our measurements lie in the scatter of the different simulations. However, there is a notable trend with redshift --- at $z>3$ almost all simulations systematically under-produce the observed abundances of $NUVrJ$-quiescent galaxies, with differences increasing with redshift. The exception is \textsc{Shark} that retains good agreement at $z\gtrsim3.5$ but overpredicts the abundances at $z<3$. Our sSFR selection yields quiescent abundances lower by $\sim0.2-0.5$ dex at $z\gtrsim4$ compared to the $NUVrJ$ selection, and are in closer agreement with models like \textsc{Simba}, \textsc{EAGLE}, and \textsc{GAEA}.

We compared our quiescent SMFs in Fig.~\ref{fig:QSMF-vs-sims} with several simulations including \textsc{Shark}, \textsc{GALFORM}, \textsc{GAEA}, \textsc{Simba}, \textsc{EAGLE} and \textsc{TNG100}, computed using a $UVJ$ selection. There are varying degrees of agreement depending on the simulation, redshift and mass range. In general, the simulations come closest to our measurement at $z\lesssim 2.5$, but at higher redshifts the high-mass end (\logM~$\gtrsim10.5$) becomes increasingly difficult to match to our observations. The scatter between the simulations and the observations increases with redshift -- at $z\sim3.25$, all but TNG100 disagree with our SMF. While showing good agreement at lower redshifts at the high-mass end, at $z>3.5$ the TNG100 SMF drops precipitously and almost no (\logM~$\gtrsim10$) quiescent galaxies are found. At these redshifts \textsc{EAGLE} appears to come closest to our observations, while the rest of the simulations systematically underpredict the SMF at \logM~$\gtrsim10.2$. The exception is the $4.5<z<5.5$ bin where \textsc{Shark} and \textsc{Simba} are also within our observational errorbars (which are rather large). Quenching at the high-mass end is typically regulated by the AGN feedback and black hole physics implemented in the simulations \citep[][for a recent review on this topic]{Lagos2025}. However, the details of the implemented AGN physics impact the quenching of intermediate- and lower-mass galaxies too \citep[e.g., via the hot-halo mode that can drive environmental quenching][]{gabor_hot_2015}. Therefore, our SMF measurements over the largest redshift and mass range to date serve as an important benchmark.

At the intermediate- and low-mass end (\logM~$\lesssim10$ ) the discrepancies are the highest among both the simulations and our measurements. The low-mass end of the SMF is shaped by the prescriptions about the physics of environmental quenching and stellar feedback (including sub-grid physics in hydrodynamical simulations) and different models produce largely different results. In general, there is disagreement between the simulations and our measurements. At $z>2$, \textsc{Eagle} and \textsc{GALFORM} do not show the power law increase of the low-mass SMF, while this can be seen in \textsc{Shark}, \textsc{Simba} and \textsc{TNG100}. To demonstrate the effect of the environmental ram pressure stripping (RPS), we show in Fig~\ref{fig:QSMF-vs-sims} the SMF from \textsc{Shark} run with a fast and slow RPS. Fast RPS produces more low-mass quiescent galaxies, however the slow RPS captures better the pronounced dip in the observed SMF at \logM~$\sim10$. However, \textsc{Shark}, in general, produces $\sim 0.5-1$ dex higher abundances in this regime compared to our observations, even though the observed trends between galaxy group/cluster mass and satellite galaxies in the local Universe is well reproduced (Oxland et al., submitted). We discuss further in \S\ref{sec:low-mass-end-smf} where we interpret our results within the context of environmental quenching.

\subsection{Schechter parameter evolution with redshift and physical implications}\label{sec:params-evolution}

\begin{figure}[t!]
\centering
\includegraphics[width=1\columnwidth, trim=0cm 1.5cm 1cm 0cm, clip]{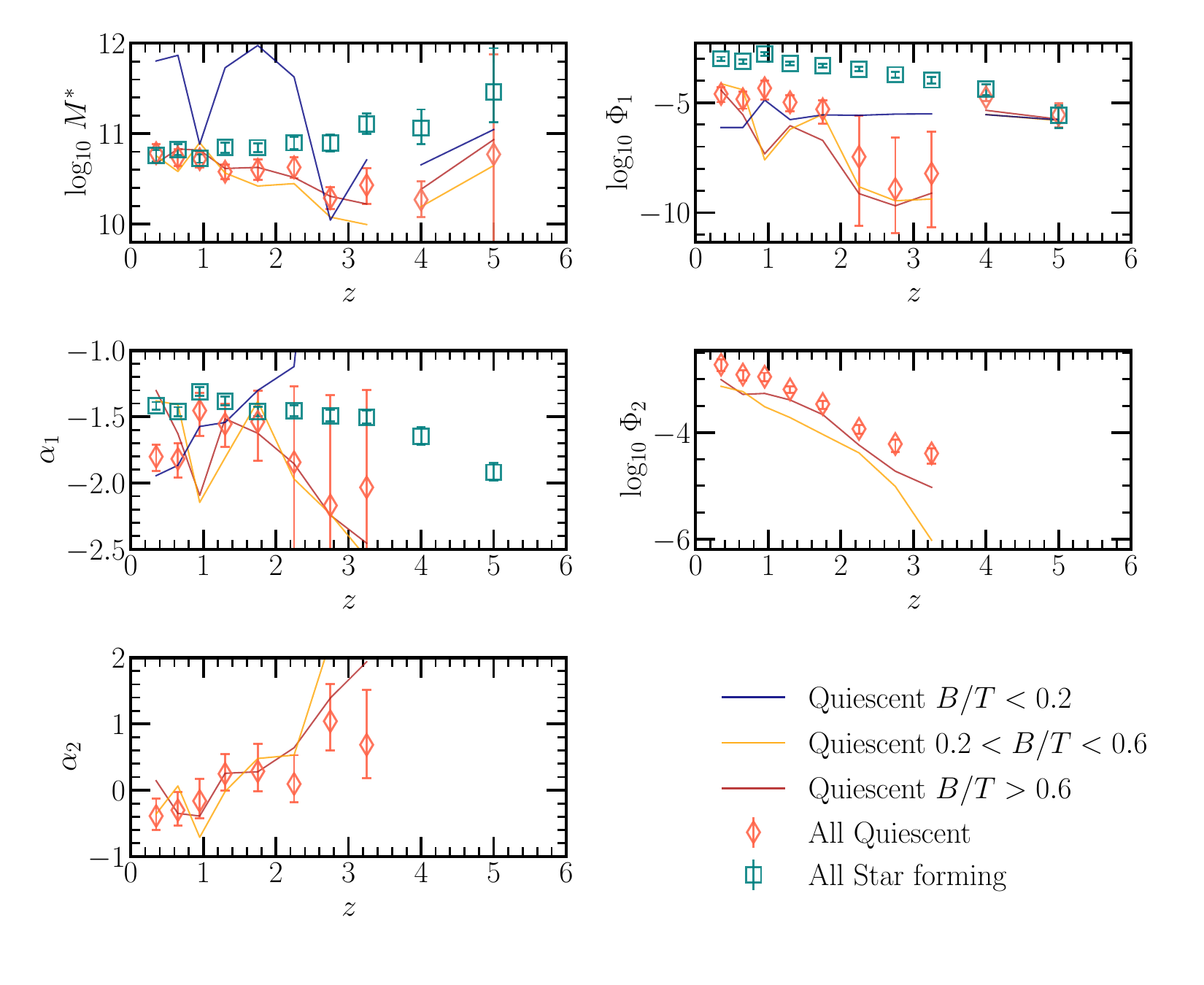}
\caption{Best-fit Schecher function parameters as a function of redshift. The symbols with errorbars show the best-fit and $1\,\sigma$ uncertainties for the total quiescent and star-forming galaxies. The lines show the best-fit values for the disk-dominated ($B/T<0.2$, blue), intermediate ($0.2<B/T<0.6$, orange) and bulge-dominated ($B/T>0.6$, dark red) SMFs.}
\label{fig:Parameters-vs-redshift}
\end{figure}

The redshift evolution of the Schechter parameters that  describe the observed SMF holds important physical insights. We show this in Fig.~\ref{fig:Parameters-vs-redshift} for the total quiescent and star-forming populations (symbols with $1\,\sigma$ uncertainty) and for the three $B/T$-selected quiescent samples (lines with no uncertainties for clarity). 

The characteristic stellar mass is $\log(M^*/\si{\Msun}) \sim 10.6$, roughly consistent with no evolution with redshift. Out to $z\sim1$, both the quiescent and star-forming SMF have the same values for $M^*$. However at $z>1$, $M^*$ shows mild decrease/increase with redshift for quiescent/star-forming galaxies (except the last two bins for the quiescent). $M^*$ for the $B/T$-selected quiescent sample has similar values with the bulge-dominated SMF having almost identical values, as we discussed in \S\ref{sec:smf-q-sf}. One of the predictions of the \cite{peng_mass_2010} model is that due to mass quenching, when star-forming galaxies grow beyond $M^*$ they quench and move to the quiescent SMF, which would have the same $M^*$. Our results show that this is the case only out to $z\sim1$. However, it remains uncertain if this trend is significant --- stellar mass uncertainties manifesting in the Eddington bias that might not be fully accounted for in our method can inflate the inferred $M^*$. Additionally, the relatively small volume of our survey limits more robust measurements of the number densities of more massive galaxies. Robustly probing the high-mass end will be crucial in testing the mass quenching scenario at $z>1$.

The normalizations of the single and double Schechter functions $\Phi_1$ and $\Phi_2$ decrease with redshift, considerably more steeply for the quiescent than for the star-forming sample, reflecting the results that we discussed in \S\ref{sec:smf-q-sf}. $\Phi_1$ for bulge-dominated and intermediate galaxies shows similar decrease as for the total population. Interestingly, $\Phi_1$ for the disk-dominated quiescent SMF remains roughly constant with redshift. This indicates a constant (in terms of number density) population of $\sim M^*$ disk-dominated quiescent galaxies.

The slope $\alpha_1$ for star-forming galaxies remains roughly constant with redshift with values $\alpha \sim1.4$ and starts to steepen at $z>4$. Along with the little-to-no evolution of $M^*$, it means that the shape of the star-forming SMF does not change significantly since $z\sim4$. $\alpha_1$ for quiescent galaxies is slightly steeper by $\sim 0.1-0.5$, with higher uncertainties. Within the \cite{peng_mass_2010} model, the low-mass galaxies are quenched by environmental processes whose efficiency is independent of mass. This means that the low-mass Schechter component of the quiescent SMF closely resembles that of the star-forming one, i.e., equal $\alpha_1$ values. The slight deviations towards steeper slopes for the quiescent SMF could be an indication that there are deviations from the mass independence of environmental quenching over larger mass ranges ($>2$ dex). Our results indicate that there is an additional steepening of the very low-mass quiescent SMF in excess of the power-law slope of the Schechter component, which could indicate a third quenching mechanism acting in the lowest mass galaxies (e.g., \logM~$\lesssim8.5$). This could be stellar feedback that is strong enough to expel cold gas from the gravitational potential well of low-mass galaxies \citep[e.g.,][]{Ferrara2000, Gelli2024}. However, the relatively low $S/N$ photometry of this population and the higher uncertainty in their colors and classification, prevents us from drawing more robust conclusions.

The slope $\alpha_2$ of the high-mass Schechter component for quiescent galaxies shows an increase with redshift, which makes the `knee' of the SMF more pronounced. Since the massive component of the quiescent SMF is built from that of the star-forming galaxies, the two must be connected. The \cite{peng_mass_2010} model predicts that the quenching probability is simply proportional to its SFR, or more precisely SFR/$M^*$. This results in the quiescent and star-forming SMF that have the same $M^*$ and slopes related as $\alpha_2 = \alpha_{1, \rm{SF}} + 1+\beta$, where $\beta \approx 0$ \cite[e.g.,][]{Schreiber2015} is the slope of the sSFR$-M_{\star}$ relation; this assumes that $\alpha_{1, \rm{SF}}$ and $M^*$, which is the same for the star-forming and quiescent populations, are constant with redshift. Our independent fits for the two populations show that $\alpha_2 - \alpha_{1,{\rm SF}} \approx 1$ only out to $z\sim1$ but increases with redshift. 
However, \cite{Porras-Valverde2025} demonstrate using semi-analytical models that the slope $\alpha_2$ is related to the scatter between black hole mass and stellar mass ($\sigma_{\rm BH}$). Assuming that the quenching probability is instead proportional to the black hole mass\footnote{Therefore, the empirical relation between the quenching probability and SFR in the \cite{peng_mass_2010} model is indirect, thanks to the relation between black hole accretion rate and SFR \citep{Silverman2009, Daddi2007b}}, a larger scatter means a wider stellar mass range over which quenching can occur, resulting in a shallower slope. Our results of steepening slope with increasing redshift qualitatively imply a decreasing $\sigma_{\rm BH}$ with redshift.

\subsection{Quenching rates, baryon conversion efficiency and bulge creation: empirical modeling of the galaxy populations} \label{sec:empirical-models}

We built an empirical model to describe the redshift evolution of the number densities of star-forming and quiescent galaxies $n_{\mathrm{SF}}(z)$ and $n_{\mathrm{Q}}(z)$ with \logM~$>10$, that we presented in \S\ref{sec:smd-numdensities}. Our starting point is the abundance matching principle, where the assumption is that the number densities of galaxies above a stellar mass threshold is equal to that of halos above a halo mass threshold that is scaled by the baryonic fraction $f_{\rm b }\approx 0.16$ and baryon conversion efficiency $\epsilon_{\star}$ \citep{behroozi_most_2018}.

The total galaxy number density above a threshold of $M_{\star}$ is related to the halo number density $n_{\mathrm{h}}(z)$ through the baryon conversion efficiency dependent on redshift $\epsilon_{\star}(z)$
\begin{equation}
\begin{split}
    n_{\mathrm{tot}}(z, >M_{\star}) & =  n_{\mathrm{SF}}(z, >M_{\star}) + n_{\mathrm{Q}}(z, >M_{\star})  \\ & = n_{\rm h}\left(z, >M_{\star}\,f_{\rm b}^{-1}\,\epsilon_{\star}^{-1} \right),
    \label{eq:ntot_z}
\end{split}
\end{equation}
where $M_{\star}\,f_{\rm b}^{-1}\,\epsilon_{\star}^{-1}$ is simply the halo mass $M_{\rm h}$. We model the number density of quiescent galaxies at a given redshift to be related to that of star-forming galaxies times a quenching rate $Q(z)$
\begin{equation}
    \frac{d n_{\mathrm{Q}}}{dz} = \frac{dt}{dz} \, Q(z)\, n_{\mathrm{SF}}(z).
    \label{eq:dndzQ}
\end{equation}
Since we want to express this function in units of Gyr$^{-1}$ we used $dt/dz = -{1}/{[(1+z) H(z)]}$ which is given by the cosmology and $H(z)$ is expressed in Gyr$^{-1}$. Using Eq.~\eqref{eq:ntot_z}, the star-forming density can be written as
\begin{equation}
    n_{\mathrm{SF}}(z)  = n_{\rm h}\left(z, \epsilon_{\star} \right) - n_{\mathrm{Q}}(z),
    \label{eq:nsf_z}
\end{equation}
where we omit the mass threshold dependence in writing the equations for brevity. Substituting Eq.~\eqref{eq:nsf_z} into Eq.~\eqref{eq:dndzQ}, we obtain:
\begin{equation}
    \frac{d n_{\mathrm{Q}}}{dz} = \frac{dt}{dz}\, Q(z) \left[ \, n_{\mathrm{h}}(z, \epsilon_{\star}) - n_{\mathrm{Q}}(z) \right].
    \label{eq:finalODE_z}
\end{equation}
Finally, the star-forming population follows from Eq.~\eqref{eq:nsf_z}. For the quenching rate and the stellar fractions we adopt power-law functions
\begin{equation}
    \label{eq:emod-funcs}
    Q(z) = Q_{0}\, (1+z)^{\alpha}; \ \
    \epsilon_{\star}(z) = \epsilon_{1} (1+z)^{\beta_1} + \epsilon_{2} (1+z)^{\beta_2}.
\end{equation}
We chose these forms empirically by testing several different ones and choosing the simplest forms that provide a good fit. We take $n_{\rm h}$ as known, which we obtain from the halo mass function (HMF) calculator \texttt{colossus} \citep{Diemer2018}, using the \cite{diemer2020} HMF. We used the numerical differential equation solver \texttt{scipy.integrate.solve\_ivp} to integrate Eqs.~\ref{eq:finalODE_z}\&\ref{eq:nsf_z}, fit our measurements of $n_{\rm Q}$ and $n_{\rm SF}$ and constrain the parameters of Eq.~\ref{eq:emod-funcs} using \texttt{emcee}.

We then built a similar model to describe the bulge-dominated star-forming and quiescent populations. In this case, we assume that bulge-dominated star-forming galaxies are created at a rate $S_{\rm B}(z) \, [{\rm Mpc}^{-3}\, {\rm Gyr^{-1}}]$, which are then quenched at a rate $Q_{\rm B}(z)$ to create the quiescent bulge population; this quenching then removes galaxies from the star-forming population at a rate $-Q_{\rm B}(z)$. This can be written as
\begin{align}
    \label{eq:nq-B}
    \frac{d n_{\mathrm{Q,B}}}{dz} & = \frac{dt}{dz}\, Q_{\rm B}(z) \, n_{\rm SF,B}\\
    \label{eq:nsf-B}
    \frac{d n_{\mathrm{SF,B}}}{dz} &= \frac{dt}{dz}\, \left[ S_{\rm B}(z) - Q_{\rm B}(z) \, n_{\rm SF,B} \right].
\end{align}
This simple model postulates that bulge-dominated star-forming galaxies are progenitors of bulge-dominated quiescent galaxies, and that other processes such as mergers do not contribute significantly to the quiescent bulge numbers. 
For $Q_{\rm B}(z)$ and $S_{\rm B}(z)$ we take a power-law and a Gaussian forms, that we chose empirically as the simplest form that fit the data.
\begin{equation}
    \label{eq:emod-funcs2}
    Q_{\rm B}(z) = Q_{\rm B,0}\, (1+z)^{\gamma}; \ \
    S_{\rm B}(z) = S_{0} \exp\!\left[ -\frac{1}{2} \left( \frac{z - z_c}{\sigma_{\rm B}} \right)^{2} \right],
\end{equation}
where $Q_{\rm B,0}$ and $\gamma$ are the scaling and exponent of the power law, while $S_0$, $z_c$ and $\sigma_{\rm B}$ are the normalization, mean redshift and standard deviation of the Gaussian. We fitted the parameters using \texttt{emcee} by numerically solving Eqs.~\ref{eq:nq-B}\&\ref{eq:nsf-B} on the observed number densities (\S\ref{sec:smd-numdensities}).

\begin{figure}[t!]
\centering
\includegraphics[width=0.99\columnwidth, trim=0.0cm 0cm 0cm 0cm, clip]{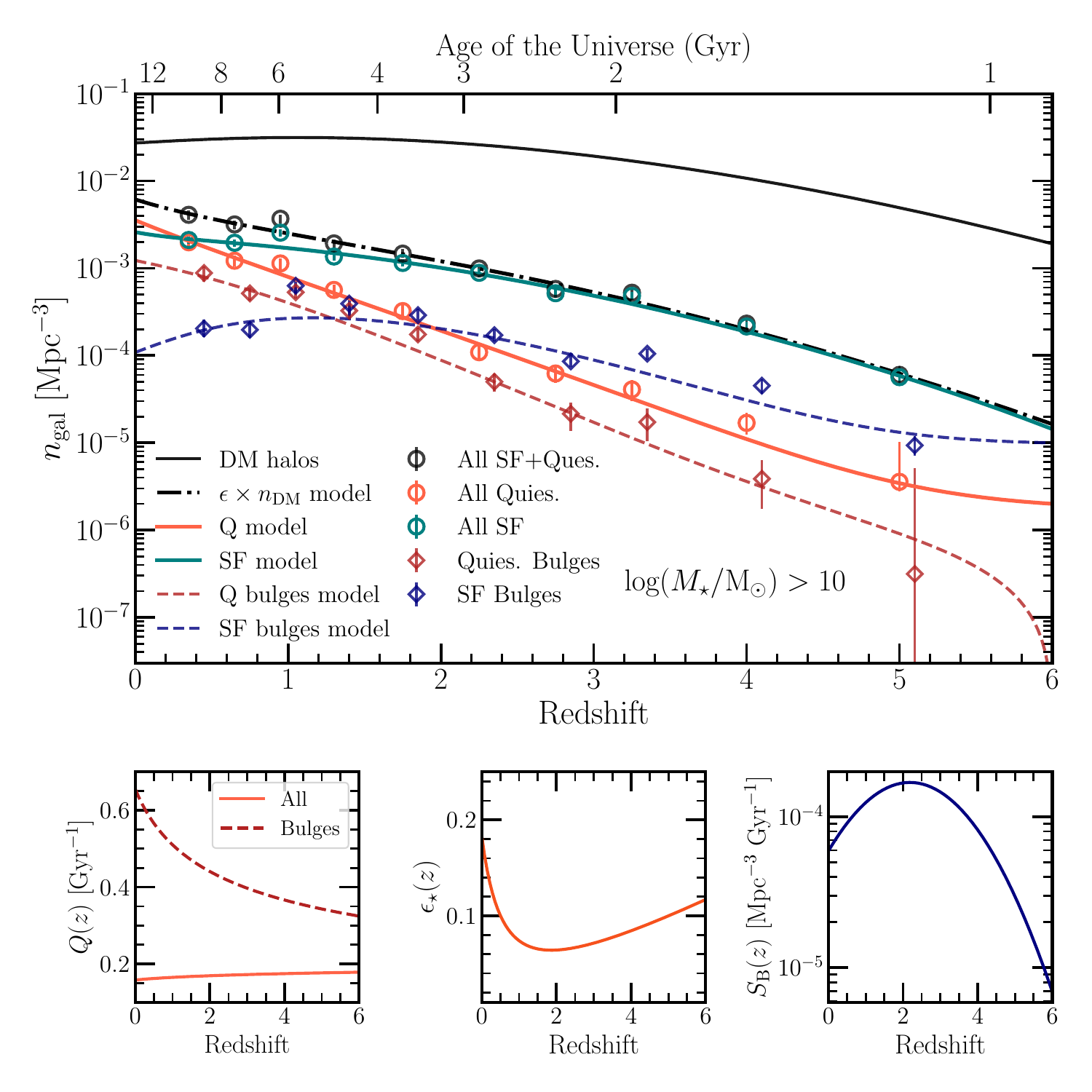}
\caption{Empirical modeling of the number densities of star-forming and quiescent galaxies as well as their bulge-dominated sub-populations. The symbols correspond to the measurements, while the curves to the best fit models as described by the set of differential equations. The bottom panels show the best-fit parametric forms of the quenching rates, integrated SFE and bulge formation function.}
\label{fig:num-dens-modeling}
\end{figure}

\begin{table}[t!]
\caption{Median and $1\sigma$ confidence intervals of the fitted model parameters of Eq.~\ref{eq:emod-funcs} and  Eq.~\ref{eq:emod-funcs2}.}
\centering
\begin{tabular}{cc|cc}
\hline
\hline
Parameter &  Median$^{+1\sigma}_{-1\sigma}$ & Parameter &  Median$^{+1\sigma}_{-1\sigma}$ \\
\hline
$Q_0$        & $0.16^{+0.03}_{-0.03}$ & $Q_{\rm B,0}$ & $0.66^{+0.27}_{-0.12}$ \\
$\alpha$     & $0.06^{+0.16}_{-0.15}$ & $\gamma$  & $-0.36^{+0.18}_{-0.33}$ \\
$\epsilon_1$ & $0.011^{+0.003}_{-0.003}$ & $S_{0}$  & $(1.69^{+0.18}_{-0.24})\times 10^{-4}$ \\
$\beta_1$    & $1.18^{+0.16}_{-0.13}$ & $z_{c}$ & $2.18^{+0.11}_{-0.25}$ \\
$\epsilon_2$ & $0.17^{+0.02}_{-0.02}$ & $\sigma_{\rm B}$ & $1.51^{+0.28}_{-0.14}$ \\
$\beta_2$    & $-1.82^{+0.24}_{-0.14}$ & \\
\hline
\hline
\end{tabular}
\label{tab:emp-model-params}
\end{table}

We present the results in Fig.~\ref{fig:num-dens-modeling}, where the top panel shows the measurements and best fit models of the number densities of \logM~$>10$ galaxies. The black solid line shows the theoretical limit of $n_{\rm h} (10^{10}/f_b)$, while the black dash-dotted line shows the best fit $n_{\rm h} (10^{10}/f_b/\epsilon_{\star})$. The number density measurements are shown in open circles with errorbars and are well fitted by our model shown in solid (dashed) orange (dark red) and teal (dark blue) lines for the quiescent (bulge-dominated quiescent) and star-forming (bulge-dominated star-forming) populations respectively. The bottom panels show the best-fit functions for the quenching rates of all and bulge-dominated galaxies, the baryonic conversion efficiency and bulge formation function. The best fit model parameters are given in Table~\ref{tab:emp-model-params}.

Our empirical model shows that the baryon conversion efficiency has non-monotonic form that decreases with cosmic time from $\sim 10\%$ at $z\sim6$ to about $6\%$ at $z\sim1.5$ and increases out to $\sim20 \%$ by $z\sim0$. These values are consistent with studies focused on the stellar-to-halo mass relationship \citep[e.g.,][]{Shuntov2022, Paquereau2025}.

The quenching rate of massive (\logM~$>10$) galaxies is about $0.17$ Gyr$^{-1}$, with mild decrease with cosmic time. Given the definition of Eq.~\ref{eq:dndzQ}, this means that $\sim17\%$ of star-forming galaxies quench per 1 Gyr. 

The bulge-dominated population is predominantely star-forming out to $z\sim1.5$ after which $n_{\rm SF,B}$ declines while quiescent bulges become more abundant, showing a constant increase with time. The star-forming bulge galaxy formation function, $S_{\rm B}$, grows by about 1 dex from $z \sim 5$ to $z \sim 2$, where it peaks at $\sim 2\times 10^{-4} \, {\rm Mpc^{-3}\, Gyr^{-1}}$, and subsequently decreases by $\sim 0.5$ dex toward $z \sim 0$. Within our framework (Eqs.~\ref{eq:nq-B}\&\ref{eq:nsf-B}), this function determines the reservoir of galaxies available to quench into the bulge-dominated quiescent population.

Star-forming bulge galaxies quench at higher rates than the overall population, with $Q_{\rm B}(z)$ rising from $\sim 30\%\,{\rm Gyr}^{-1}$ at $z \sim 6$ to $\sim 60\%\,{\rm Gyr}^{-1}$ at $z \sim 0$. This trend can be explained by the idea that permanent quenching requires the host halo to cross the critical mass scale $M_{\rm h, crit} \sim 10^{12}\, M_\odot$, where virial shock heating and AGN-driven hot halo feedback operate \citep{birnboim_virial_2003, Keres2005, DekelBurkert2014}. Since $n_{\rm h}(>10^{12} M_\odot)$ increases with time, more halos cross this threshold at lower redshifts, raising the probability that a bulge galaxy will be quenched. In addition, the decline in cosmological gas accretion and cold inflow at late times further reduces the fuel supply, making it easier to quench at lower redshifts. Finally, once a bulge has formed, the buildup of additional stellar mass proceeds more slowly with time: a massive central bulge stabilizes the disk against fragmentation \citep[morphological quenching;][]{Martig2009}, thereby promoting the transition to quiescence. These results are qualitatively consistent with theoretical predictions in which bulge growth, compaction events, and subsequent stabilization \citep[e.g.,][]{Zolotov2015, Tacchella2016} play a central role in driving the quenching of massive galaxies.

We note that this is a simple empirical model that comes with caveats, such as the assumption that quiescent bulges are exclusively descendants of star-forming bulges. In reality, some fraction of quiescent disks can grow a bulge at later times, an incarnation of the so-called progenitor bias \citep{vanDokkum2001, Carollo2013}. However, the fraction of quiescent disks is low, therefore this should not be a dominant channel.
The next step is to extend this framework to be mass dependent and fit the stellar mass functions, as well as to adopt a more physically motivated model parametrization, in order to provide more detailed insight into different quenching mechanisms. We leave this for future work.

\subsection{Morphological transformations as prerequisite for quenching} \label{sec:morpho-transform-quenching}

Our results are qualitatively consistent with with several theoretical frameworks that link morphological transformations to the cessation of star formation. In the compaction scenario \citep[e.g.,][]{DekelBurkert2014, Zolotov2015, Tacchella2016}, bulge growth precedes quiescence: cold gas streams at high redshifts induce violent disk instabilities and dissipative inflows that trigger central starbursts and the rapid buildup of dense stellar bulges. This is in line with our finding that the majority of bulge-dominated galaxies are star forming with this fraction increasing at earlier times and lower stellar masses (Fig.~\ref{fig:Quiescent-fractions}, bottom panel).

Quenching then can proceed in several ways. The formation of a central bulge can stabilize the disk and suppress further central inflows. Once a bulge is in place, morphological quenching \citep{Martig2009} becomes effective, as the deeper central potential prevents efficient fragmentation of the gas disk. The dissipative inflows and the formation of a central bulge also create conditions for rapid growth of the central black hole and triggering an AGN whose feedback quenches the galaxy. AGNs are thought to be one of the dominant mechanisms to keep the galaxy quenched by heating the halo gas \citep[hot halo maintainance mode, ][]{gabor_hot_2015}. In parallel, a decline in the cold gas reservoir -- either because of a drop in accretion rate with cosmic time, or because of shock heating of inflows in massive halos and/or strong outflows -- leaves the stellar component dominant, whose velocity dispersion stabilizes the system and leads to a quiescent elliptical. These scenarios are also consistent with our finding that the quenching rate of bulges increases with cosmic time.

\subsection{The low-mass end of the SMF as evidence of environmental quenching} \label{sec:low-mass-end-smf}

The low-mass end of the quiescent SMF is thought to be shaped by environmental quenching mechanisms such as strangulation \citep{Larson1980, Moran2007}, ram pressure stripping \citep{Gunn1972}, and galaxy harassment \citep{Moore1996}, which operate in dense environments \citep{Boselli2006, peng_mass_2010, Cortese2021}. 
This is also demonstrated in Fig.~\ref{fig:QSMF-vs-sims} by the \textsc{Shark} quiescent SMF, run with different ram-pressure stripping (RPS) parameters. These models predominantly impact the low-mass end, with more aggressive RPS leading to an enhanced abundance of quenched low-mass galaxies. We refer to Oxland et al., (2025, submitted) for a thorough discussion of these different RPS models and how they impact quenching in group and cluster galaxies in \textsc{Shark}.

Environmental quenching is also not expected to strongly transform galaxy morphology. As galaxies are accreted as disks into more massive halos, the denser environment, including the hot halo mode, quenches them while largely preserving their disk structure. Our results are consistent with this picture: disk systems ($B/T<0.2$) dominate the low-mass quiescent SMF at all redshifts (Figs.~\ref{fig:SMF-Q-SF-BT}\&\ref{fig:Quiescent-fractions}), albeit with large uncertainties at $z>2.5$. Taken together, both the upturn of the quiescent SMF at the low-mass end and the prevalence of disk-dominated morphologies provide strong evidence for environmental quenching.

Intriguingly, our results suggest that environmental quenching is already established by $z\sim3$, as indicated by the upturn in the low-mass SMF. This may appear surprising, since group- and cluster-scale halos -- which provide the virialized hot gas environments needed to heat or strip satellites -- are not expected to become abundant until $z\sim2$ \citep[e.g.,][]{Chiang2013}. However, SAMs such as \textsc{Shark} and \textsc{GAEA} \citep{DeLucia2024} predict that environmental quenching can shape the low-mass end of the SMF at all redshifts out to $z\sim5$. Our measurements qualitatively agree with this picture out to $z\sim3$, but lack the depth ($\log M_\star < 9$) at higher redshifts to confirm or refute it. Deeper observations will be needed to robustly trace this regime, but these results indicate that environment (RPS and tidal stripping) can quench low-mass galaxies at all redshifts and early times.

We caution against over-interpreting the low-mass results, which as shown in Appendix~\ref{sec:quiescent-classification} are sensitive to the definition of quiescence \citep[also,][]{Lagos2024}. The nature of low-mass, potentially quiescent galaxies therefore remains uncertain. Crucially, \JWST\ has the capability to directly test the SAM predictions: if the quiescent SMF indeed continues to rise at the low-mass end, this should be detectable with surveys that reach completeness limits of $\log (M_\star/{\rm M_{\odot}}) \sim 8.5-9$ for quiescent galaxies. Spectroscopic follow-up of this population, while observationally expensive, will be essential to confirm their quiescent nature and to provide further insight into the mechanisms driving environmental quenching.

\section{Conclusions}\label{sec:conclusions}

In this work, we presented the SMF measurements for quiescent and star-forming galaxies, classified by their $NUVrJ$ rest-frame colors, in 10 redshift bins at $0.2<z<5.5$ selected from COSMOS2025 catalog. We also investigated the morphological dependence of the SMF for both populations by separating them based on their bulge-to-total ratio in disk-dominated ($B/T<0.2$), intermediate ($0.2<B/T<0.6$), and bulge-dominated ($B/T>0.6$). We summarize our results as follows. 
\begin{itemize}
    \item The SMFs of quiescent and star-forming galaxies shows a strong mass-dependent evolution: quiescent galaxies build up rapidly at early times until $z\sim1$. The most massive (\logM~$\gtrsim 11$) quiescent systems show little evolution at $0.2<z<1.1$ but decline by $1$–$2$ dex at $z>1.1$, implying most massive galaxies quenched by $z\sim1$. In contrast, star-forming SMFs evolve more slowly, consistent with steady mass growth until quenching above $M^*$ transfers galaxies to the quiescent population.

    \item As a function of morphology, quiescent bulge systems dominate the number densities at \logM~$>10$ at all redshift, shaping the `knee' of the SMF. Quiescent disk systems dominate the SMF of quiescent galaxies at \logM~$\lesssim9$. For the star-forming population, disk systems dominate the SMF at all redshifts and masses of \logM~$\lesssim11$, while bulge-dominated star-forming systems contribute significantly at \logM~$\gtrsim11$. However, most bulge-dominated galaxies in the Universe are star-forming with their fraction increasing with redshift and decreasing mass, consistent with them being progenitors of quiescent bulges.

    \item We measure an upturn in the quiescent SMF at the low-mass end (\logM~$<9.5$) as early as $z\sim 3$. This upturn is mostly contributed by disk-dominated galaxies, consistent with environmental quenching scenarios in which satellites are quenched but retain their disk morphologies. These results are also in qualitative agreement with SAM predictions, suggesting that environmental processes can quench low-mass galaxies at early times. However, SAMs produce $0.5-1$ dex more low-mass quiescent galaxies than observations.

    \item The stellar mass density of \logM~$>8$ quiescent galaxies rises rapidly and by $z\sim0.7$ overtakes that of the star-forming population, which shows a more gradual increase, consistent with previous results. Most of the SMD from quiescent galaxies is locked in bulge-dominated systems.

    \item For the number densities of massive \logM~$>10$ quiescent galaxies, we find lower values than recent literature measurements by $0.1-0.7$ dex. Compared to SAMs and hydrodynamical simulations, we find good agreement at $2<z<3$, but at $z>3$ the simulations increasingly underpredict the observed number densities of quiescent galaxies.

    \item We built a simple empirical model to describe the redshift evolution of the galaxy number densities that parametrizes the quenching rate of all and bulge-dominated galaxies, the baryon conversion efficiency for the total population and the bulge formation function. The model naturally links bulge growth in star-forming galaxies to the subsequent buildup of the quiescent bulge population.

    \item The quenching rate of bulge galaxies increases toward the present: as halos grow above the shock-heating threshold and the cosmological cold gas supply declines, star-forming bulges face a rising probability per unit time of becoming and remaining quenched. This trend is qualitatively consistent with theoretical predictions of halo-driven and morphology-driven quenching mechanisms.  

\end{itemize}

Taken together, our results support the scenarios that bulge-driven quenching in massive galaxies and environmentally driven quenching in low-mass systems act as complementary pathways that shape the buildup of the quiescent population across cosmic time. Complete spectroscopic surveys of both low- and high-mass galaxies with \JWST\ will be key to crystallizing the picture and to constraining the physical mechanisms that drive quenching. This work presents one of the most comprehensive measurements of the SMF and number densities of quiescent and star-forming galaxies selected from \JWST. As such, it provides key benchmark for comparing models and simulations.

\section*{Data availability} \label{sec:data-availability}

We provide all our measurements in tabulated form at \url{https://github.com/mShuntov/SMF-Q-SF-Morpho-paper-Shuntov25}.

\begin{acknowledgements}
We thank Gabriella De Lucia for providing us the SMF data from GAEA.
The Cosmic Dawn Center (DAWN) is funded by the Danish National Research Foundation under grant DNRF140. This work has received funding from the Swiss State Secretariat for Education, Research and Innovation (SERI) under contract number MB22.00072. 
This work was made possible by utilizing the CANDIDE cluster at the Institut d’Astrophysique de Paris, which was funded through grants from the PNCG, CNES, DIM-ACAV, and the Cosmic Dawn Center and maintained by S. Rouberol. French COSMOS team members are partly supported by the Centre National d’Etudes Spatiales (CNES). We acknowledge the funding of the French Agence Nationale de la Recherche for the project iMAGE (grant ANR-22-CE31-0007).  
GEM  acknowledges the Villum Fonden research grants 37440 and 13160.

\end{acknowledgements}

\bibliographystyle{aa}
\bibliography{biblio.bib}

@ARTICLE{Cimatti2006,
       author = {{Cimatti}, A. and {Daddi}, E. and {Renzini}, A.},
        title = "{Mass downsizing and ``top-down'' assembly of early-type galaxies}",
      journal = {\aap},
         year = 2006,
        month = jul,
       volume = {453},
       number = {2},
        pages = {L29-L33},
          doi = {10.1051/0004-6361:20065155},
archivePrefix = {arXiv},
       eprint = {astro-ph/0605353},
 primaryClass = {astro-ph},
       adsurl = {https://ui.adsabs.harvard.edu/abs/2006A&A...453L..29C}
}

@ARTICLE{YangQG2025,
       author = {{Yang}, Tiancheng and {Wang}, Tao and {Xu}, Ke and {Sun}, Hanwen and {Zhou}, Luwenjia and {Xie}, Lizhi and {De Lucia}, Gabriella and {Lagos}, Claudia del P. and {Wang}, Kai and {Fontanot}, Fabio and {Wu}, Yuxuan and {Lu}, Shiying and {Chen}, Longyue and {Hirschmann}, Michaela},
        title = "{A census of quiescent galaxies across $0.5 < z < 8$ with JWST/MIRI: Mass-dependent number density evolution of quiescent galaxies in the early Universe}",
      journal = {arXiv e-prints},
         year = 2025,
        month = oct,
          eid = {arXiv:2510.12235},
        pages = {arXiv:2510.12235},
          doi = {10.48550/arXiv.2510.12235},
archivePrefix = {arXiv},
       eprint = {2510.12235},
 primaryClass = {astro-ph.GA},
       adsurl = {https://ui.adsabs.harvard.edu/abs/2025arXiv251012235Y}
}

@ARTICLE{Lagos2008,
       author = {{Lagos}, Claudia Del P. and {Cora}, Sof{\'\i}a A. and {Padilla}, Nelson D.},
        title = "{Effects of AGN feedback on {\ensuremath{\Lambda}}CDM galaxies}",
      journal = {\mnras},
         year = 2008,
        month = aug,
       volume = {388},
       number = {2},
        pages = {587-602},
          doi = {10.1111/j.1365-2966.2008.13456.x},
archivePrefix = {arXiv},
       eprint = {0805.1930},
 primaryClass = {astro-ph},
       adsurl = {https://ui.adsabs.harvard.edu/abs/2008MNRAS.388..587L}
}

@ARTICLE{Fontana2004,
       author = {{Fontana}, A. and {Pozzetti}, L. and {Donnarumma}, I. and {Renzini}, A. and {Cimatti}, A. and {Zamorani}, G. and {Menci}, N. and {Daddi}, E. and {Giallongo}, E. and {Mignoli}, M. and {Perna}, C. and {Salimbeni}, S. and {Saracco}, P. and {Broadhurst}, T. and {Cristiani}, S. and {D'Odorico}, S. and {Gilmozzi}, R.},
        title = "{The K20 survey. VI. The distribution of the stellar masses in galaxies up to z ≃ 2}",
      journal = {\aap},
         year = 2004,
        month = sep,
       volume = {424},
        pages = {23-42},
          doi = {10.1051/0004-6361:20035626},
archivePrefix = {arXiv},
       eprint = {astro-ph/0405055},
 primaryClass = {astro-ph},
       adsurl = {https://ui.adsabs.harvard.edu/abs/2004A&A...424...23F}
}

@ARTICLE{Barnes1996,
       author = {{Barnes}, Joshua E. and {Hernquist}, Lars},
        title = "{Transformations of Galaxies. II. Gasdynamics in Merging Disk Galaxies}",
      journal = {\apj},
         year = 1996,
        month = nov,
       volume = {471},
        pages = {115},
          doi = {10.1086/177957},
       adsurl = {https://ui.adsabs.harvard.edu/abs/1996ApJ...471..115B}
}

@ARTICLE{Hopkins2010,
       author = {{Hopkins}, Philip F. and {Bundy}, Kevin and {Croton}, Darren and {Hernquist}, Lars and {Keres}, Dusan and {Khochfar}, Sadegh and {Stewart}, Kyle and {Wetzel}, Andrew and {Younger}, Joshua D.},
        title = "{Mergers and Bulge Formation in {\ensuremath{\Lambda}}CDM: Which Mergers Matter?}",
      journal = {\apj},
         year = 2010,
        month = may,
       volume = {715},
       number = {1},
        pages = {202-229},
          doi = {10.1088/0004-637X/715/1/202},
archivePrefix = {arXiv},
       eprint = {0906.5357},
 primaryClass = {astro-ph.CO},
       adsurl = {https://ui.adsabs.harvard.edu/abs/2010ApJ...715..202H}
}

@ARTICLE{Silverman2009,
       author = {{Silverman}, J.~D. and {Lamareille}, F. and {Maier}, C. and {Lilly}, S.~J. and {Mainieri}, V. and {Brusa}, M. and {Cappelluti}, N. and {Hasinger}, G. and {Zamorani}, G. and {Scodeggio}, M. and {Bolzonella}, M. and {Contini}, T. and {Carollo}, C.~M. and {Jahnke}, K. and {Kneib}, J.-P. and {Le F{\`e}vre}, O. and {Merloni}, A. and {Bardelli}, S. and {Bongiorno}, A. and {Brunner}, H. and {Caputi}, K. and {Civano}, F. and {Comastri}, A. and {Coppa}, G. and {Cucciati}, O. and {de la Torre}, S. and {de Ravel}, L. and {Elvis}, M. and {Finoguenov}, A. and {Fiore}, F. and {Franzetti}, P. and {Garilli}, B. and {Gilli}, R. and {Iovino}, A. and {Kampczyk}, P. and {Knobel}, C. and {Kova{\v{c}}}, K. and {Le Borgne}, J.-F. and {Le Brun}, V. and {Mignoli}, M. and {Pello}, R. and {Peng}, Y. and {Perez Montero}, E. and {Ricciardelli}, E. and {Tanaka}, M. and {Tasca}, L. and {Tresse}, L. and {Vergani}, D. and {Vignali}, C. and {Zucca}, E. and {Bottini}, D. and {Cappi}, A. and {Cassata}, P. and {Fumana}, M. and {Griffiths}, R. and {Kartaltepe}, J. and {Koekemoer}, A. and {Marinoni}, C. and {McCracken}, H.~J. and {Memeo}, P. and {Meneux}, B. and {Oesch}, P. and {Porciani}, C. and {Salvato}, M.},
        title = "{Ongoing and Co-Evolving Star Formation in zCOSMOS Galaxies Hosting Active Galactic Nuclei}",
      journal = {\apj},
         year = 2009,
        month = may,
       volume = {696},
       number = {1},
        pages = {396-410},
          doi = {10.1088/0004-637X/696/1/396},
archivePrefix = {arXiv},
       eprint = {0810.3653},
 primaryClass = {astro-ph},
       adsurl = {https://ui.adsabs.harvard.edu/abs/2009ApJ...696..396S}
}

@ARTICLE{Daddi2007b,
       author = {{Daddi}, E. and {Alexander}, D.~M. and {Dickinson}, M. and {Gilli}, R. and {Renzini}, A. and {Elbaz}, D. and {Cimatti}, A. and {Chary}, R. and {Frayer}, D. and {Bauer}, F.~E. and {Brandt}, W.~N. and {Giavalisco}, M. and {Grogin}, N.~A. and {Huynh}, M. and {Kurk}, J. and {Mignoli}, M. and {Morrison}, G. and {Pope}, A. and {Ravindranath}, S.},
        title = "{Multiwavelength Study of Massive Galaxies at z\raisebox{-0.5ex}\textasciitilde2. II. Widespread Compton-thick Active Galactic Nuclei and the Concurrent Growth of Black Holes and Bulges}",
      journal = {\apj},
         year = 2007,
        month = nov,
       volume = {670},
       number = {1},
        pages = {173-189},
          doi = {10.1086/521820},
archivePrefix = {arXiv},
       eprint = {0705.2832},
 primaryClass = {astro-ph},
       adsurl = {https://ui.adsabs.harvard.edu/abs/2007ApJ...670..173D}
}

@ARTICLE{Long2024,
       author = {{Long}, Arianna S. and {Antwi-Danso}, Jacqueline and {Lambrides}, Erini L. and {Lovell}, Christopher C. and {de la Vega}, Alexander and {Valentino}, Francesco and {Zavala}, Jorge A. and {Casey}, Caitlin M. and {Wilkins}, Stephen M. and {Yung}, L.~Y. Aaron and {Arrabal Haro}, Pablo and {Bagley}, Micaela B. and {Bisigello}, Laura and {Chworowsky}, Katherine and {Cooper}, M.~C. and {Cooper}, Olivia R. and {Cooray}, Asantha R. and {Croton}, Darren and {Dickinson}, Mark and {Finkelstein}, Steven L. and {Franco}, Maximilien and {Gould}, Katriona M.~L. and {Hirschmann}, Michaela and {Hutchison}, Taylor A. and {Kartaltepe}, Jeyhan S. and {Kocevski}, Dale D. and {Koekemoer}, Anton M. and {Lucas}, Ray A. and {McKinney}, Jed and {Nere}, Rachel and {Papovich}, Casey and {P{\'e}rez-Gonz{\'a}lez}, Pablo G. and {Pirzkal}, Nor and {Santini}, Paola},
        title = "{Efficient NIRCam Selection of Quiescent Galaxies at 3 < z < 6 in CEERS}",
      journal = {\apj},
         year = 2024,
        month = jul,
       volume = {970},
       number = {1},
          eid = {68},
        pages = {68},
          doi = {10.3847/1538-4357/ad4cea},
archivePrefix = {arXiv},
       eprint = {2305.04662},
 primaryClass = {astro-ph.GA},
       adsurl = {https://ui.adsabs.harvard.edu/abs/2024ApJ...970...68L}
}

@ARTICLE{Girelli2019,
       author = {{Girelli}, Giacomo and {Bolzonella}, Micol and {Cimatti}, Andrea},
        title = "{Massive and old quiescent galaxies at high redshift}",
      journal = {\aap},
         year = 2019,
        month = dec,
       volume = {632},
          eid = {A80},
        pages = {A80},
          doi = {10.1051/0004-6361/201834547},
archivePrefix = {arXiv},
       eprint = {1910.07544},
 primaryClass = {astro-ph.GA},
       adsurl = {https://ui.adsabs.harvard.edu/abs/2019A&A...632A..80G}
}

@ARTICLE{Gould2023,
       author = {{Gould}, Katriona M.~L. and {Brammer}, Gabriel and {Valentino}, Francesco and {Whitaker}, Katherine E. and {Weaver}, John. R. and {Lagos}, Claudia del P. and {Rizzo}, Francesca and {Franco}, Maximilien and {Hsieh}, Bau-Ching and {Ilbert}, Olivier and {Jin}, Shuowen and {Magdis}, Georgios and {McCracken}, Henry J. and {Mobasher}, Bahram and {Shuntov}, Marko and {Steinhardt}, Charles L. and {Strait}, Victoria and {Toft}, Sune},
        title = "{COSMOS2020: Exploring the Dawn of Quenching for Massive Galaxies at 3 < z < 5 with a New Color-selection Method}",
      journal = {\aj},
         year = 2023,
        month = jun,
       volume = {165},
       number = {6},
          eid = {248},
        pages = {248},
          doi = {10.3847/1538-3881/accadc},
archivePrefix = {arXiv},
       eprint = {2302.10934},
 primaryClass = {astro-ph.GA},
       adsurl = {https://ui.adsabs.harvard.edu/abs/2023AJ....165..248G}
}

@ARTICLE{FallEfstathiou1980,
       author = {{Fall}, S.~M. and {Efstathiou}, G.},
        title = "{Formation and rotation of disc galaxies with haloes.}",
      journal = {\mnras},
         year = 1980,
        month = oct,
       volume = {193},
        pages = {189-206},
          doi = {10.1093/mnras/193.2.189},
       adsurl = {https://ui.adsabs.harvard.edu/abs/1980MNRAS.193..189F}
}

@ARTICLE{Pan2025,
       author = {{Pan}, Richard and {Suess}, Katherine A. and {Marchesini}, Danilo and {Wang}, Bingjie and {Leja}, Joel and {Cutler}, Sam E. and {Whitaker}, Katherine E. and {Bezanson}, Rachel and {Price}, Sedona H. and {Furtak}, Lukas J. and {Weaver}, John R. and {Labb{\'e}}, Ivo and {Brammer}, Gabriel and {Zhang}, Yunchong and {Dayal}, Pratika and {Feldmann}, Robert and {Greene}, Jenny E. and {Glazebrook}, Karl and {Miller}, Tim B. and {Mitsuhashi}, Ikki and {Muzzin}, Adam and {Nanayakkara}, Themiya and {Nelson}, Erica J. and {Setton}, David J. and {Zitrin}, Adi},
        title = "{UNCOVER/MegaScience: No Evidence of Environmental Quenching in a z {\ensuremath{\sim}} 2.6 Protocluster}",
      journal = {\apjl},
         year = 2025,
        month = sep,
       volume = {990},
       number = {1},
          eid = {L24},
        pages = {L24},
          doi = {10.3847/2041-8213/adf7ab},
archivePrefix = {arXiv},
       eprint = {2504.06334},
 primaryClass = {astro-ph.GA},
       adsurl = {https://ui.adsabs.harvard.edu/abs/2025ApJ...990L..24P}
}

@ARTICLE{Toni2025,
       author = {{Toni}, Greta and {Maturi}, Matteo and {Castignani}, Gianluca and {Moscardini}, Lauro and {Gozaliasl}, Ghassem and {Finoguenov}, Alexis and {Taamoli}, Sina and {Hollis Akins}, B. and {Arango-Toro}, C. Rafael and {Casey}, M. Caitlin and {Drakos}, E. Nicole and {Faisst}, L. Andreas and {Flayhart}, Carter and {Franco}, Maximilien and {Gentile}, Fabrizio and {Hadi}, Ali and {Harish}, Santosh and {Hatamnia}, Hossein and {Ilbert}, Olivier and {Jin}, Shuowen and {Jeyhan Kartaltepe}, S. and {Khostovan}, Ali Ahmad and {Koekemoer}, M. Anton and {Leroy}, Gavin and {Georgios Magdis}, E. and {McCracken}, Henry Joy and {McKinney}, Jed and {Paquereau}, Louise and {Rhodes}, Jason and {Rich}, R. Michael and {Brant Robertson}, E. and {Rasha Samir}, M. and {Scognamiglio}, Diana and {Shamyati}, Samaneh and {Shuntov}, Marko and {Zavala}, A. Jorge},
        title = "{COSMOS-Web galaxy groups: Evolution of red sequence and quiescent galaxy fraction}",
      journal = {arXiv e-prints},
         year = 2025,
        month = sep,
          eid = {arXiv:2509.08040},
        pages = {arXiv:2509.08040},
          doi = {10.48550/arXiv.2509.08040},
archivePrefix = {arXiv},
       eprint = {2509.08040},
 primaryClass = {astro-ph.GA},
       adsurl = {https://ui.adsabs.harvard.edu/abs/2025arXiv250908040T}
}

@ARTICLE{Jin2024,
       author = {{Jin}, Shuowen and {Sillassen}, Nikolaj B. and {Magdis}, Georgios E. and {Brinch}, Malte and {Shuntov}, Marko and {Brammer}, Gabriel and {Gobat}, Raphael and {Valentino}, Francesco and {Carnall}, Adam C. and {Lee}, Minju and {Vijayan}, Aswin P. and {Gillman}, Steven and {Kokorev}, Vasily and {Le Bail}, Aur{\'e}lien and {Greve}, Thomas R. and {Gullberg}, Bitten and {Gould}, Katriona M.~L. and {Toft}, Sune},
        title = "{Cosmic Vine: A z = 3.44 large-scale structure hosting massive quiescent galaxies}",
      journal = {\aap},
         year = 2024,
        month = mar,
       volume = {683},
          eid = {L4},
        pages = {L4},
          doi = {10.1051/0004-6361/202348540},
archivePrefix = {arXiv},
       eprint = {2311.04867},
 primaryClass = {astro-ph.GA},
       adsurl = {https://ui.adsabs.harvard.edu/abs/2024A&A...683L...4J}
}

@ARTICLE{Bell2004,
       author = {{Bell}, Eric F. and {Wolf}, Christian and {Meisenheimer}, Klaus and {Rix}, Hans-Walter and {Borch}, Andrea and {Dye}, Simon and {Kleinheinrich}, Martina and {Wisotzki}, Lutz and {McIntosh}, Daniel H.},
        title = "{Nearly 5000 Distant Early-Type Galaxies in COMBO-17: A Red Sequence and Its Evolution since z\raisebox{-0.5ex}\textasciitilde1}",
      journal = {\apj},
         year = 2004,
        month = jun,
       volume = {608},
       number = {2},
        pages = {752-767},
          doi = {10.1086/420778},
archivePrefix = {arXiv},
       eprint = {astro-ph/0303394},
 primaryClass = {astro-ph},
       adsurl = {https://ui.adsabs.harvard.edu/abs/2004ApJ...608..752B}
}

@ARTICLE{Zhang2025,
       author = {{Zhang}, Yunchong and {de Graaff}, Anna and {Setton}, David J. and {Price}, Sedona H. and {Bezanson}, Rachel and {Lagos}, Claudia del P. and {Cutler}, Sam E. and {McConachie}, Ian and {Cleri}, Nikko J. and {Cooper}, Olivia R. and {Gottumukkala}, Rashmi and {Greene}, Jenny E. and {Hirschmann}, Michaela and {Khullar}, Gourav and {Labbe}, Ivo and {Leja}, Joel and {Maseda}, Michael V. and {Matthee}, Jorryt and {Miller}, Tim B. and {Nanayakkara}, Themiya and {Suess}, Katherine A. and {Wang}, Bingjie and {Whitaker}, Katherine E. and {Williams}, Christina C.},
        title = "{RUBIES spectroscopically confirms the high number density of quiescent galaxies from $\mathbf{2<z<5}$}",
      journal = {arXiv e-prints},
         year = 2025,
        month = aug,
          eid = {arXiv:2508.08577},
        pages = {arXiv:2508.08577},
          doi = {10.48550/arXiv.2508.08577},
archivePrefix = {arXiv},
       eprint = {2508.08577},
 primaryClass = {astro-ph.GA},
       adsurl = {https://ui.adsabs.harvard.edu/abs/2025arXiv250808577Z}
}

@ARTICLE{Stevenson2025,
       author = {{Stevenson}, Struan D. and {Carnall}, Adam C. and {Leung}, Ho-Hin and {Taylor}, Elizabeth and {Cullen}, Fergus and {Dunlop}, James S. and {McLeod}, Derek J. and {McLure}, Ross J. and {Begley}, Ryan and {Arellano-C{\'o}rdova}, Karla Z. and {Barrufet}, Laia and {Bondestam}, Cecilia and {Donnan}, Callum T. and {Ellis}, Richard S. and {Grogin}, Norman A. and {Liu}, Feng-Yuan and {Koekemoer}, Anton M. and {P{\'e}rez-Gonz{\'a}lez}, Pablo G. and {Rowlands}, Kate and {Sanders}, Ryan L. and {Scholte}, Dirk and {Shapley}, Alice E. and {Skarbinski}, Maya and {Stanton}, Thomas M. and {Wild}, Vivienne},
        title = "{PRIMER \& JADES reveal an abundance of massive quiescent galaxies at 2 < z < 5}",
      journal = {arXiv e-prints},
         year = 2025,
        month = sep,
          eid = {arXiv:2509.06913},
        pages = {arXiv:2509.06913},
archivePrefix = {arXiv},
       eprint = {2509.06913},
 primaryClass = {astro-ph.GA},
       adsurl = {https://ui.adsabs.harvard.edu/abs/2025arXiv250906913S}
}

@ARTICLE{Lilly2013,
       author = {{Lilly}, Simon J. and {Carollo}, C. Marcella and {Pipino}, Antonio and {Renzini}, Alvio and {Peng}, Yingjie},
        title = "{Gas Regulation of Galaxies: The Evolution of the Cosmic Specific Star Formation Rate, the Metallicity-Mass-Star-formation Rate Relation, and the Stellar Content of Halos}",
      journal = {\apj},
         year = 2013,
        month = aug,
       volume = {772},
       number = {2},
          eid = {119},
        pages = {119},
          doi = {10.1088/0004-637X/772/2/119},
archivePrefix = {arXiv},
       eprint = {1303.5059},
 primaryClass = {astro-ph.CO},
       adsurl = {https://ui.adsabs.harvard.edu/abs/2013ApJ...772..119L}
}

@ARTICLE{Chabrier03,
       author = {{Chabrier}, Gilles},
        title = "{Galactic Stellar and Substellar Initial Mass Function}",
      journal = {\pasp},
         year = 2003,
        month = jul,
       volume = {115},
       number = {809},
        pages = {763-795},
          doi = {10.1086/376392},
archivePrefix = {arXiv},
       eprint = {astro-ph/0304382},
 primaryClass = {astro-ph},
       adsurl = {https://ui.adsabs.harvard.edu/abs/2003PASP..115..763C}
}

@ARTICLE{Bundy2005,
       author = {{Bundy}, Kevin and {Ellis}, Richard S. and {Conselice}, Christopher J.},
        title = "{The Mass Assembly Histories of Galaxies of Various Morphologies in the GOODS Fields}",
      journal = {\apj},
         year = 2005,
        month = jun,
       volume = {625},
       number = {2},
        pages = {621-632},
          doi = {10.1086/429549},
archivePrefix = {arXiv},
       eprint = {astro-ph/0502204},
 primaryClass = {astro-ph},
       adsurl = {https://ui.adsabs.harvard.edu/abs/2005ApJ...625..621B}
}

@ARTICLE{Carnall2024,
       author = {{Carnall}, A.~C. and {Cullen}, F. and {McLure}, R.~J. and {McLeod}, D.~J. and {Begley}, R. and {Donnan}, C.~T. and {Dunlop}, J.~S. and {Shapley}, A.~E. and {Rowlands}, K. and {Almaini}, O. and {Arellano-C{\'o}rdova}, K.~Z. and {Barrufet}, L. and {Cimatti}, A. and {Ellis}, R.~S. and {Grogin}, N.~A. and {Hamadouche}, M.~L. and {Illingworth}, G.~D. and {Koekemoer}, A.~M. and {Leung}, H. -H. and {Lovell}, C.~C. and {P{\'e}rez-Gonz{\'a}lez}, P.~G. and {Santini}, P. and {Stanton}, T.~M. and {Wild}, V.},
        title = "{The JWST EXCELS survey: too much, too young, too fast? Ultra-massive quiescent galaxies at 3 < z < 5}",
      journal = {\mnras},
         year = 2024,
        month = oct,
       volume = {534},
       number = {1},
        pages = {325-348},
          doi = {10.1093/mnras/stae2092},
archivePrefix = {arXiv},
       eprint = {2405.02242},
 primaryClass = {astro-ph.GA},
       adsurl = {https://ui.adsabs.harvard.edu/abs/2024MNRAS.534..325C}
}

@ARTICLE{Moutard2016,
       author = {{Moutard}, T. and {Arnouts}, S. and {Ilbert}, O. and {Coupon}, J. and {Davidzon}, I. and {Guzzo}, L. and {Hudelot}, P. and {McCracken}, H.~J. and {Van Waerbeke}, L. and {Morrison}, G.~E. and {Le F{\`e}vre}, O. and {Comte}, V. and {Bolzonella}, M. and {Fritz}, A. and {Garilli}, B. and {Scodeggio}, M.},
        title = "{The VIPERS Multi-Lambda Survey. II. Diving with massive galaxies in 22 square degrees since z = 1.5}",
      journal = {\aap},
         year = 2016,
        month = may,
       volume = {590},
          eid = {A103},
        pages = {A103},
          doi = {10.1051/0004-6361/201527294},
archivePrefix = {arXiv},
       eprint = {1602.05917},
 primaryClass = {astro-ph.GA},
       adsurl = {https://ui.adsabs.harvard.edu/abs/2016A&A...590A.103M}
}

@ARTICLE{Mortlock2015,
       author = {{Mortlock}, Alice and {Conselice}, Christopher. J. and {Hartley}, William G. and {Duncan}, Ken and {Lani}, Caterina and {Ownsworth}, Jamie R. and {Almaini}, Omar and {Wel}, Arjen van der and {Huang}, Kuang-Han and {Ashby}, Matthew L.~N. and {Willner}, S.~P. and {Fontana}, Adriano and {Dekel}, Avishai and {Koekemoer}, Anton M. and {Ferguson}, Harry C. and {Faber}, Sandra M. and {Grogin}, Norman A. and {Kocevski}, Dale D.},
        title = "{Deconstructing the galaxy stellar mass function with UKIDSS and CANDELS: the impact of colour, structure and environment}",
      journal = {\mnras},
         year = 2015,
        month = feb,
       volume = {447},
       number = {1},
        pages = {2-24},
          doi = {10.1093/mnras/stu2403},
archivePrefix = {arXiv},
       eprint = {1411.3339},
 primaryClass = {astro-ph.GA},
       adsurl = {https://ui.adsabs.harvard.edu/abs/2015MNRAS.447....2M}
}

@ARTICLE{Muzzin2013SMF,
       author = {{Muzzin}, Adam and {Marchesini}, Danilo and {Stefanon}, Mauro and {Franx}, Marijn and {McCracken}, Henry J. and {Milvang-Jensen}, Bo and {Dunlop}, James S. and {Fynbo}, J.~P.~U. and {Brammer}, Gabriel and {Labb{\'e}}, Ivo and {van Dokkum}, Pieter G.},
        title = "{The Evolution of the Stellar Mass Functions of Star-forming and Quiescent Galaxies to z = 4 from the COSMOS/UltraVISTA Survey}",
      journal = {\apj},
         year = 2013,
        month = nov,
       volume = {777},
       number = {1},
          eid = {18},
        pages = {18},
          doi = {10.1088/0004-637X/777/1/18},
archivePrefix = {arXiv},
       eprint = {1303.4409},
 primaryClass = {astro-ph.CO},
       adsurl = {https://ui.adsabs.harvard.edu/abs/2013ApJ...777...18M}
}

@ARTICLE{Tomczak2014,
       author = {{Tomczak}, Adam R. and {Quadri}, Ryan F. and {Tran}, Kim-Vy H. and {Labb{\'e}}, Ivo and {Straatman}, Caroline M.~S. and {Papovich}, Casey and {Glazebrook}, Karl and {Allen}, Rebecca and {Brammer}, Gabriel B. and {Kacprzak}, Glenn G. and {Kawinwanichakij}, Lalitwadee and {Kelson}, Daniel D. and {McCarthy}, Patrick J. and {Mehrtens}, Nicola and {Monson}, Andrew J. and {Persson}, S. Eric and {Spitler}, Lee R. and {Tilvi}, Vithal and {van Dokkum}, Pieter},
        title = "{Galaxy Stellar Mass Functions from ZFOURGE/CANDELS: An Excess of Low-mass Galaxies since z = 2 and the Rapid Buildup of Quiescent Galaxies}",
      journal = {\apj},
         year = 2014,
        month = mar,
       volume = {783},
       number = {2},
          eid = {85},
        pages = {85},
          doi = {10.1088/0004-637X/783/2/85},
archivePrefix = {arXiv},
       eprint = {1309.5972},
 primaryClass = {astro-ph.CO},
       adsurl = {https://ui.adsabs.harvard.edu/abs/2014ApJ...783...85T}
}

@ARTICLE{Boselli2006,
       author = {{Boselli}, Alessandro and {Gavazzi}, Giuseppe},
        title = "{Environmental Effects on Late-Type Galaxies in Nearby Clusters}",
      journal = {\pasp},
         year = 2006,
        month = apr,
       volume = {118},
       number = {842},
        pages = {517-559},
          doi = {10.1086/500691},
archivePrefix = {arXiv},
       eprint = {astro-ph/0601108},
 primaryClass = {astro-ph},
       adsurl = {https://ui.adsabs.harvard.edu/abs/2006PASP..118..517B}
}

@ARTICLE{Chiang2013,
       author = {{Chiang}, Yi-Kuan and {Overzier}, Roderik and {Gebhardt}, Karl},
        title = "{Ancient Light from Young Cosmic Cities: Physical and Observational Signatures of Galaxy Proto-clusters}",
      journal = {\apj},
         year = 2013,
        month = dec,
       volume = {779},
       number = {2},
          eid = {127},
        pages = {127},
          doi = {10.1088/0004-637X/779/2/127},
archivePrefix = {arXiv},
       eprint = {1310.2938},
 primaryClass = {astro-ph.CO},
       adsurl = {https://ui.adsabs.harvard.edu/abs/2013ApJ...779..127C}
}

@ARTICLE{DeLucia2024,
       author = {{De Lucia}, Gabriella and {Fontanot}, Fabio and {Xie}, Lizhi and {Hirschmann}, Michaela},
        title = "{Tracing the quenching journey across cosmic time}",
      journal = {\aap},
         year = 2024,
        month = jul,
       volume = {687},
          eid = {A68},
        pages = {A68},
          doi = {10.1051/0004-6361/202349045},
archivePrefix = {arXiv},
       eprint = {2401.06211},
 primaryClass = {astro-ph.GA},
       adsurl = {https://ui.adsabs.harvard.edu/abs/2024A&A...687A..68D}
}

@ARTICLE{vanDokkum2001,
       author = {{van Dokkum}, Pieter G. and {Franx}, Marijn},
        title = "{Morphological Evolution and the Ages of Early-Type Galaxies in Clusters}",
      journal = {\apj},
         year = 2001,
        month = may,
       volume = {553},
       number = {1},
        pages = {90-102},
          doi = {10.1086/320645},
archivePrefix = {arXiv},
       eprint = {astro-ph/0501236},
 primaryClass = {astro-ph},
       adsurl = {https://ui.adsabs.harvard.edu/abs/2001ApJ...553...90V}
}

@ARTICLE{Carollo2013,
       author = {{Carollo}, C.~M. and {Bschorr}, T.~J. and {Renzini}, A. and {Lilly}, S.~J. and {Capak}, P. and {Cibinel}, A. and {Ilbert}, O. and {Onodera}, M. and {Scoville}, N. and {Cameron}, E. and {Mobasher}, B. and {Sanders}, D. and {Taniguchi}, Y.},
        title = "{Newly Quenched Galaxies as the Cause for the Apparent Evolution in Average Size of the Population}",
      journal = {\apj},
         year = 2013,
        month = aug,
       volume = {773},
       number = {2},
          eid = {112},
        pages = {112},
          doi = {10.1088/0004-637X/773/2/112},
archivePrefix = {arXiv},
       eprint = {1302.5115},
 primaryClass = {astro-ph.CO},
       adsurl = {https://ui.adsabs.harvard.edu/abs/2013ApJ...773..112C}
}

@ARTICLE{Zolotov2015,
       author = {{Zolotov}, Adi and {Dekel}, Avishai and {Mandelker}, Nir and {Tweed}, Dylan and {Inoue}, Shigeki and {DeGraf}, Colin and {Ceverino}, Daniel and {Primack}, Joel R. and {Barro}, Guillermo and {Faber}, Sandra M.},
        title = "{Compaction and quenching of high-z galaxies in cosmological simulations: blue and red nuggets}",
      journal = {\mnras},
         year = 2015,
        month = jul,
       volume = {450},
       number = {3},
        pages = {2327-2353},
          doi = {10.1093/mnras/stv740},
archivePrefix = {arXiv},
       eprint = {1412.4783},
 primaryClass = {astro-ph.GA},
       adsurl = {https://ui.adsabs.harvard.edu/abs/2015MNRAS.450.2327Z}
}

@ARTICLE{Martig2009,
       author = {{Martig}, Marie and {Bournaud}, Fr{\'e}d{\'e}ric and {Teyssier}, Romain and {Dekel}, Avishai},
        title = "{Morphological Quenching of Star Formation: Making Early-Type Galaxies Red}",
      journal = {\apj},
         year = 2009,
        month = dec,
       volume = {707},
       number = {1},
        pages = {250-267},
          doi = {10.1088/0004-637X/707/1/250},
archivePrefix = {arXiv},
       eprint = {0905.4669},
 primaryClass = {astro-ph.CO},
       adsurl = {https://ui.adsabs.harvard.edu/abs/2009ApJ...707..250M}
}

@ARTICLE{Shuntov2022,
       author = {{Shuntov}, M. and {McCracken}, H.~J. and {Gavazzi}, R. and {Laigle}, C. and {Weaver}, J.~R. and {Davidzon}, I. and {Ilbert}, O. and {Kauffmann}, O.~B. and {Faisst}, A. and {Dubois}, Y. and {Koekemoer}, A.~M. and {Moneti}, A. and {Milvang-Jensen}, B. and {Mobasher}, B. and {Sanders}, D.~B. and {Toft}, S.},
        title = "{COSMOS2020: Cosmic evolution of the stellar-to-halo mass relation for central and satellite galaxies up to z {\ensuremath{\sim}} 5}",
      journal = {\aap},
         year = 2022,
        month = aug,
       volume = {664},
          eid = {A61},
        pages = {A61},
          doi = {10.1051/0004-6361/202243136},
archivePrefix = {arXiv},
       eprint = {2203.10895},
 primaryClass = {astro-ph.GA},
       adsurl = {https://ui.adsabs.harvard.edu/abs/2022A&A...664A..61S}
}

@ARTICLE{diemer2020,
       author = {{Diemer}, Benedikt},
        title = "{Universal at Last? The Splashback Mass Function of Dark Matter Halos}",
      journal = {\apj},
         year = 2020,
        month = nov,
       volume = {903},
       number = {2},
          eid = {87},
        pages = {87},
          doi = {10.3847/1538-4357/abbf52},
archivePrefix = {arXiv},
       eprint = {2007.10346},
 primaryClass = {astro-ph.CO},
       adsurl = {https://ui.adsabs.harvard.edu/abs/2020ApJ...903...87D}
}

@ARTICLE{MHC2025,
       author = {{Huertas-Company}, M. and {Shuntov}, M. and {Dong}, Y. and {Walmsley}, M. and {Ilbert}, O. and {McCracken}, H.~J. and {Akins}, H.~B. and {Allen}, N. and {Casey}, C.~M. and {Costantin}, L. and {Daddi}, E. and {Dekel}, A. and {Franco}, M. and {Garland}, I.~L. and {G{\'e}ron}, T. and {Gozaliasl}, G. and {Hirschmann}, M. and {Kartaltepe}, J.~S. and {Koekemoer}, A.~M. and {Lintott}, C. and {Liu}, D. and {Lucas}, R. and {Masters}, K. and {Pacucci}, F. and {Paquereau}, L. and {P'erez-Gonz'alez}, P.~G. and {Rhodes}, J.~D. and {Robertson}, B.~E. and {Simmons}, B. and {Smethurst}, R. and {Toft}, S. and {Yang}, L.},
        title = "{COSMOS-Web: The emergence of the Hubble Sequence}",
      journal = {arXiv e-prints},
         year = 2025,
        month = feb,
          eid = {arXiv:2502.03532},
        pages = {arXiv:2502.03532},
          doi = {10.48550/arXiv.2502.03532},
archivePrefix = {arXiv},
       eprint = {2502.03532},
 primaryClass = {astro-ph.GA},
       adsurl = {https://ui.adsabs.harvard.edu/abs/2025arXiv250203532H}
}

@ARTICLE{MHC2016,
       author = {{Huertas-Company}, M. and {Bernardi}, M. and {P{\'e}rez-Gonz{\'a}lez}, P.~G. and {Ashby}, M.~L.~N. and {Barro}, G. and {Conselice}, C. and {Daddi}, E. and {Dekel}, A. and {Dimauro}, P. and {Faber}, S.~M. and {Grogin}, N.~A. and {Kartaltepe}, J.~S. and {Kocevski}, D.~D. and {Koekemoer}, A.~M. and {Koo}, D.~C. and {Mei}, S. and {Shankar}, F.},
        title = "{Mass assembly and morphological transformations since z {\ensuremath{\sim}} 3 from CANDELS}",
      journal = {\mnras},
         year = 2016,
        month = nov,
       volume = {462},
       number = {4},
        pages = {4495-4516},
          doi = {10.1093/mnras/stw1866},
archivePrefix = {arXiv},
       eprint = {1606.04952},
 primaryClass = {astro-ph.GA},
       adsurl = {https://ui.adsabs.harvard.edu/abs/2016MNRAS.462.4495H}
}

@ARTICLE{Gomez-Guijarro2019,
       author = {{G{\'o}mez-Guijarro}, C. and {Magdis}, G.~E. and {Valentino}, F. and {Toft}, S. and {Man}, A.~W.~S. and {Ivison}, R.~J. and {Tisani{\'c}}, K. and {van der Vlugt}, D. and {Stockmann}, M. and {Martin-Alvarez}, S. and {Brammer}, G.},
        title = "{Compact Star-forming Galaxies as Old Starbursts Becoming Quiescent}",
      journal = {\apj},
         year = 2019,
        month = dec,
       volume = {886},
       number = {2},
          eid = {88},
        pages = {88},
          doi = {10.3847/1538-4357/ab418b},
archivePrefix = {arXiv},
       eprint = {1909.02572},
 primaryClass = {astro-ph.GA},
       adsurl = {https://ui.adsabs.harvard.edu/abs/2019ApJ...886...88G}
}

@ARTICLE{Tacchella2015,
       author = {{Tacchella}, S. and {Carollo}, C.~M. and {Renzini}, A. and {F{\"o}rster Schreiber}, N.~M. and {Lang}, P. and {Wuyts}, S. and {Cresci}, G. and {Dekel}, A. and {Genzel}, R. and {Lilly}, S.~J. and {Mancini}, C. and {Newman}, S. and {Onodera}, M. and {Shapley}, A. and {Tacconi}, L. and {Woo}, J. and {Zamorani}, G.},
        title = "{Evidence for mature bulges and an inside-out quenching phase 3 billion years after the Big Bang}",
      journal = {Science},
         year = 2015,
        month = apr,
       volume = {348},
       number = {6232},
        pages = {314-317},
          doi = {10.1126/science.1261094},
archivePrefix = {arXiv},
       eprint = {1504.04021},
 primaryClass = {astro-ph.GA},
       adsurl = {https://ui.adsabs.harvard.edu/abs/2015Sci...348..314T}
}

@ARTICLE{DekelBurkert2014,
       author = {{Dekel}, A. and {Burkert}, A.},
        title = "{Wet disc contraction to galactic blue nuggets and quenching to red nuggets}",
      journal = {\mnras},
         year = 2014,
        month = feb,
       volume = {438},
       number = {2},
        pages = {1870-1879},
          doi = {10.1093/mnras/stt2331},
archivePrefix = {arXiv},
       eprint = {1310.1074},
 primaryClass = {astro-ph.CO},
       adsurl = {https://ui.adsabs.harvard.edu/abs/2014MNRAS.438.1870D}
}

@ARTICLE{Barro2014,
       author = {{Barro}, G. and {Faber}, S.~M. and {P{\'e}rez-Gonz{\'a}lez}, P.~G. and {Pacifici}, C. and {Trump}, J.~R. and {Koo}, D.~C. and {Wuyts}, S. and {Guo}, Y. and {Bell}, E. and {Dekel}, A. and {Porter}, L. and {Primack}, J. and {Ferguson}, H. and {Ashby}, M.~L.~N. and {Caputi}, K. and {Ceverino}, D. and {Croton}, D. and {Fazio}, G.~G. and {Giavalisco}, M. and {Hsu}, L. and {Kocevski}, D. and {Koekemoer}, A. and {Kurczynski}, P. and {Kollipara}, P. and {Lee}, J. and {McIntosh}, D.~H. and {McGrath}, E. and {Moody}, C. and {Somerville}, R. and {Papovich}, C. and {Salvato}, M. and {Santini}, P. and {Tal}, T. and {van der Wel}, A. and {Williams}, C.~C. and {Willner}, S.~P. and {Zolotov}, A.},
        title = "{CANDELS+3D-HST: Compact SFGs at z \raisebox{-0.5ex}\textasciitilde 2-3, the Progenitors of the First Quiescent Galaxies}",
      journal = {\apj},
         year = 2014,
        month = aug,
       volume = {791},
       number = {1},
          eid = {52},
        pages = {52},
          doi = {10.1088/0004-637X/791/1/52},
archivePrefix = {arXiv},
       eprint = {1311.5559},
 primaryClass = {astro-ph.CO},
       adsurl = {https://ui.adsabs.harvard.edu/abs/2014ApJ...791...52B}
}

@ARTICLE{Tacchella2016,
       author = {{Tacchella}, Sandro and {Dekel}, Avishai and {Carollo}, C. Marcella and {Ceverino}, Daniel and {DeGraf}, Colin and {Lapiner}, Sharon and {Mandelker}, Nir and {Primack}, Joel R.},
        title = "{Evolution of density profiles in high-z galaxies: compaction and quenching inside-out}",
      journal = {\mnras},
         year = 2016,
        month = may,
       volume = {458},
       number = {1},
        pages = {242-263},
          doi = {10.1093/mnras/stw303},
archivePrefix = {arXiv},
       eprint = {1509.00017},
 primaryClass = {astro-ph.GA},
       adsurl = {https://ui.adsabs.harvard.edu/abs/2016MNRAS.458..242T}
}

@ARTICLE{Slob2025,
       author = {{Slob}, Martje and {Kriek}, Mariska and {de Graaff}, Anna and {Cheng}, Chloe M. and {Beverage}, Aliza G. and {Bezanson}, Rachel and {Forster Schreiber}, Natascha M. and {Lorenz}, Brian and {Mancera Pi{\~n}a}, Pavel E. and {Marchesini}, Danilo and {Muzzin}, Adam and {Newman}, Andrew B. and {Price}, Sedona H. and {Suess}, Katherine A. and {van de Sande}, Jesse and {van Dokkum}, Pieter and {Weisz}, Daniel R.},
        title = "{Fast Rotators at Cosmic Noon: Stellar Kinematics for 15 Quiescent Galaxies from JWST-SUSPENSE}",
      journal = {arXiv e-prints},
         year = 2025,
        month = jun,
          eid = {arXiv:2506.04310},
        pages = {arXiv:2506.04310},
          doi = {10.48550/arXiv.2506.04310},
archivePrefix = {arXiv},
       eprint = {2506.04310},
 primaryClass = {astro-ph.GA},
       adsurl = {https://ui.adsabs.harvard.edu/abs/2025arXiv250604310S}
}

@ARTICLE{Lang2014,
       author = {{Lang}, Philipp and {Wuyts}, Stijn and {Somerville}, Rachel S. and {F{\"o}rster Schreiber}, Natascha M. and {Genzel}, Reinhard and {Bell}, Eric F. and {Brammer}, Gabe and {Dekel}, Avishai and {Faber}, Sandra M. and {Ferguson}, Henry C. and {Grogin}, Norman A. and {Kocevski}, Dale D. and {Koekemoer}, Anton M. and {Lutz}, Dieter and {McGrath}, Elizabeth J. and {Momcheva}, Ivelina and {Nelson}, Erica J. and {Primack}, Joel R. and {Rosario}, David J. and {Skelton}, Rosalind E. and {Tacconi}, Linda J. and {van Dokkum}, Pieter G. and {Whitaker}, Katherine E.},
        title = "{Bulge Growth and Quenching since z = 2.5 in CANDELS/3D-HST}",
      journal = {\apj},
         year = 2014,
        month = jun,
       volume = {788},
       number = {1},
          eid = {11},
        pages = {11},
          doi = {10.1088/0004-637X/788/1/11},
archivePrefix = {arXiv},
       eprint = {1402.0866},
 primaryClass = {astro-ph.GA},
       adsurl = {https://ui.adsabs.harvard.edu/abs/2014ApJ...788...11L}
}

@ARTICLE{Pacifici2016,
       author = {{Pacifici}, Camilla and {Kassin}, Susan A. and {Weiner}, Benjamin J. and {Holden}, Bradford and {Gardner}, Jonathan P. and {Faber}, Sandra M. and {Ferguson}, Henry C. and {Koo}, David C. and {Primack}, Joel R. and {Bell}, Eric F. and {Dekel}, Avishai and {Gawiser}, Eric and {Giavalisco}, Mauro and {Rafelski}, Marc and {Simons}, Raymond C. and {Barro}, Guillermo and {Croton}, Darren J. and {Dav{\'e}}, Romeel and {Fontana}, Adriano and {Grogin}, Norman A. and {Koekemoer}, Anton M. and {Lee}, Seong-Kook and {Salmon}, Brett and {Somerville}, Rachel and {Behroozi}, Peter},
        title = "{The Evolution of Star Formation Histories of Quiescent Galaxies}",
      journal = {\apj},
         year = 2016,
        month = nov,
       volume = {832},
       number = {1},
          eid = {79},
        pages = {79},
          doi = {10.3847/0004-637X/832/1/79},
archivePrefix = {arXiv},
       eprint = {1609.03572},
 primaryClass = {astro-ph.GA},
       adsurl = {https://ui.adsabs.harvard.edu/abs/2016ApJ...832...79P}
}

@ARTICLE{Carnall2018,
       author = {{Carnall}, A.~C. and {McLure}, R.~J. and {Dunlop}, J.~S. and {Dav{\'e}}, R.},
        title = "{Inferring the star formation histories of massive quiescent galaxies with BAGPIPES: evidence for multiple quenching mechanisms}",
      journal = {\mnras},
         year = 2018,
        month = nov,
       volume = {480},
       number = {4},
        pages = {4379-4401},
          doi = {10.1093/mnras/sty2169},
archivePrefix = {arXiv},
       eprint = {1712.04452},
 primaryClass = {astro-ph.GA},
       adsurl = {https://ui.adsabs.harvard.edu/abs/2018MNRAS.480.4379C}
}

@ARTICLE{Fontana2009,
       author = {{Fontana}, A. and {Santini}, P. and {Grazian}, A. and {Pentericci}, L. and {Fiore}, F. and {Castellano}, M. and {Giallongo}, E. and {Menci}, N. and {Salimbeni}, S. and {Cristiani}, S. and {Nonino}, M. and {Vanzella}, E.},
        title = "{The fraction of quiescent massive galaxies in the early Universe}",
      journal = {\aap},
         year = 2009,
        month = jul,
       volume = {501},
       number = {1},
        pages = {15-20},
          doi = {10.1051/0004-6361/200911650},
archivePrefix = {arXiv},
       eprint = {0901.2898},
 primaryClass = {astro-ph.GA},
       adsurl = {https://ui.adsabs.harvard.edu/abs/2009A&A...501...15F}
}

@ARTICLE{Williams2009,
       author = {{Williams}, Rik J. and {Quadri}, Ryan F. and {Franx}, Marijn and {van Dokkum}, Pieter and {Labb{\'e}}, Ivo},
        title = "{Detection of Quiescent Galaxies in a Bicolor Sequence from Z = 0-2}",
      journal = {\apj},
         year = 2009,
        month = feb,
       volume = {691},
       number = {2},
        pages = {1879-1895},
          doi = {10.1088/0004-637X/691/2/1879},
archivePrefix = {arXiv},
       eprint = {0806.0625},
 primaryClass = {astro-ph},
       adsurl = {https://ui.adsabs.harvard.edu/abs/2009ApJ...691.1879W}
}

@ARTICLE{Wuyts2008,
       author = {{Wuyts}, Stijn and {Labb{\'e}}, Ivo and {F{\"o}rster Schreiber}, Natascha M. and {Franx}, Marijn and {Rudnick}, Gregory and {Brammer}, Gabriel B. and {van Dokkum}, Pieter G.},
        title = "{FIREWORKS U$_{38}$-to-24 {\ensuremath{\mu}}m Photometry of the GOODS Chandra Deep Field-South: Multiwavelength Catalog and Total Infrared Properties of Distant K$_{s}$-selected Galaxies}",
      journal = {\apj},
         year = 2008,
        month = aug,
       volume = {682},
       number = {2},
        pages = {985-1003},
          doi = {10.1086/588749},
       adsurl = {https://ui.adsabs.harvard.edu/abs/2008ApJ...682..985W}
}

@ARTICLE{Lagos2025,
       author = {{Lagos}, Claudia del P. and {Valentino}, Francesco and {Wright}, Ruby J. and {de Graaff}, Anna and {Glazebrook}, Karl and {De Lucia}, Gabriella and {Robotham}, Aaron S.~G. and {Nanayakkara}, Themiya and {Chandro-Gomez}, Angel and {Bravo}, Mat{\'\i}as and {Baugh}, Carlton M. and {Harborne}, Katherine E. and {Hirschmann}, Michaela and {Fontanot}, Fabio and {Xie}, Lizhi and {Chittenden}, Harry},
        title = "{The diverse star formation histories of early massive, quenched galaxies in modern galaxy formation simulations}",
      journal = {\mnras},
         year = 2025,
        month = jan,
       volume = {536},
       number = {3},
        pages = {2324-2354},
          doi = {10.1093/mnras/stae2626},
archivePrefix = {arXiv},
       eprint = {2409.16916},
 primaryClass = {astro-ph.GA},
       adsurl = {https://ui.adsabs.harvard.edu/abs/2025MNRAS.536.2324L}
}

@ARTICLE{Vijayan2022,
       author = {{Vijayan}, Aswin P. and {Wilkins}, Stephen M. and {Lovell}, Christopher C. and {Thomas}, Peter A. and {Camps}, Peter and {Baes}, Maarten and {Trayford}, James and {Kuusisto}, Jussi and {Roper}, William J.},
        title = "{First Light And Reionisation Epoch Simulations (FLARES) - III. The properties of massive dusty galaxies at cosmic dawn}",
      journal = {\mnras},
         year = 2022,
        month = apr,
       volume = {511},
       number = {4},
        pages = {4999-5017},
          doi = {10.1093/mnras/stac338},
archivePrefix = {arXiv},
       eprint = {2108.00830},
 primaryClass = {astro-ph.GA},
       adsurl = {https://ui.adsabs.harvard.edu/abs/2022MNRAS.511.4999V}
}

@ARTICLE{Lovell2021,
       author = {{Lovell}, Christopher C. and {Vijayan}, Aswin P. and {Thomas}, Peter A. and {Wilkins}, Stephen M. and {Barnes}, David J. and {Irodotou}, Dimitrios and {Roper}, Will},
        title = "{First Light And Reionization Epoch Simulations (FLARES) - I. Environmental dependence of high-redshift galaxy evolution}",
      journal = {\mnras},
         year = 2021,
        month = jan,
       volume = {500},
       number = {2},
        pages = {2127-2145},
          doi = {10.1093/mnras/staa3360},
archivePrefix = {arXiv},
       eprint = {2004.07283},
 primaryClass = {astro-ph.GA},
       adsurl = {https://ui.adsabs.harvard.edu/abs/2021MNRAS.500.2127L}
}

@ARTICLE{Wilkins2023,
       author = {{Wilkins}, Stephen M. and {Vijayan}, Aswin P. and {Lovell}, Christopher C. and {Roper}, William J. and {Irodotou}, Dimitrios and {Caruana}, Joseph and {Seeyave}, Louise T.~C. and {Kuusisto}, Jussi K. and {Thomas}, Peter A. and {Parris}, Shedeur A.~K.},
        title = "{First light and reionization epoch simulations (FLARES) V: the redshift frontier}",
      journal = {\mnras},
         year = 2023,
        month = feb,
       volume = {519},
       number = {2},
        pages = {3118-3128},
          doi = {10.1093/mnras/stac3280},
archivePrefix = {arXiv},
       eprint = {2204.09431},
 primaryClass = {astro-ph.GA},
       adsurl = {https://ui.adsabs.harvard.edu/abs/2023MNRAS.519.3118W}
}

@ARTICLE{SIMBADave2019,
       author = {{Dav{\'e}}, Romeel and {Angl{\'e}s-Alc{\'a}zar}, Daniel and {Narayanan}, Desika and {Li}, Qi and {Rafieferantsoa}, Mika H. and {Appleby}, Sarah},
        title = "{SIMBA: Cosmological simulations with black hole growth and feedback}",
      journal = {\mnras},
         year = 2019,
        month = jun,
       volume = {486},
       number = {2},
        pages = {2827-2849},
          doi = {10.1093/mnras/stz937},
archivePrefix = {arXiv},
       eprint = {1901.10203},
 primaryClass = {astro-ph.GA},
       adsurl = {https://ui.adsabs.harvard.edu/abs/2019MNRAS.486.2827D}
}

@ARTICLE{Crain2015,
       author = {{Crain}, Robert A. and {Schaye}, Joop and {Bower}, Richard G. and {Furlong}, Michelle and {Schaller}, Matthieu and {Theuns}, Tom and {Dalla Vecchia}, Claudio and {Frenk}, Carlos S. and {McCarthy}, Ian G. and {Helly}, John C. and {Jenkins}, Adrian and {Rosas-Guevara}, Yetli M. and {White}, Simon D.~M. and {Trayford}, James W.},
        title = "{The EAGLE simulations of galaxy formation: calibration of subgrid physics and model variations}",
      journal = {\mnras},
         year = 2015,
        month = jun,
       volume = {450},
       number = {2},
        pages = {1937-1961},
          doi = {10.1093/mnras/stv725},
archivePrefix = {arXiv},
       eprint = {1501.01311},
 primaryClass = {astro-ph.GA},
       adsurl = {https://ui.adsabs.harvard.edu/abs/2015MNRAS.450.1937C}
}

@ARTICLE{Schaye2015,
       author = {{Schaye}, Joop and {Crain}, Robert A. and {Bower}, Richard G. and {Furlong}, Michelle and {Schaller}, Matthieu and {Theuns}, Tom and {Dalla Vecchia}, Claudio and {Frenk}, Carlos S. and {McCarthy}, I.~G. and {Helly}, John C. and {Jenkins}, Adrian and {Rosas-Guevara}, Y.~M. and {White}, Simon D.~M. and {Baes}, Maarten and {Booth}, C.~M. and {Camps}, Peter and {Navarro}, Julio F. and {Qu}, Yan and {Rahmati}, Alireza and {Sawala}, Till and {Thomas}, Peter A. and {Trayford}, James},
        title = "{The EAGLE project: simulating the evolution and assembly of galaxies and their environments}",
      journal = {\mnras},
         year = 2015,
        month = jan,
       volume = {446},
       number = {1},
        pages = {521-554},
          doi = {10.1093/mnras/stu2058},
archivePrefix = {arXiv},
       eprint = {1407.7040},
 primaryClass = {astro-ph.GA},
       adsurl = {https://ui.adsabs.harvard.edu/abs/2015MNRAS.446..521S}
}

@ARTICLE{GALFORM_Lacey2016,
       author = {{Lacey}, Cedric G. and {Baugh}, Carlton M. and {Frenk}, Carlos S. and {Benson}, Andrew J. and {Bower}, Richard G. and {Cole}, Shaun and {Gonzalez-Perez}, Violeta and {Helly}, John C. and {Lagos}, Claudia D.~P. and {Mitchell}, Peter D.},
        title = "{A unified multiwavelength model of galaxy formation}",
      journal = {\mnras},
         year = 2016,
        month = nov,
       volume = {462},
       number = {4},
        pages = {3854-3911},
          doi = {10.1093/mnras/stw1888},
archivePrefix = {arXiv},
       eprint = {1509.08473},
 primaryClass = {astro-ph.GA},
       adsurl = {https://ui.adsabs.harvard.edu/abs/2016MNRAS.462.3854L}
}

@ARTICLE{Lagos2024,
       author = {{Lagos}, Claudia del P. and {Bravo}, Mat{\'\i}as and {Tobar}, Rodrigo and {Obreschkow}, Danail and {Power}, Chris and {Robotham}, Aaron S.~G. and {Proctor}, Katy L. and {Hansen}, Samuel and {Chandro-G{\'o}mez}, {\'A}ngel and {Carrivick}, Julian},
        title = "{Quenching massive galaxies across cosmic time with the semi-analytic model SHARK V2.0}",
      journal = {\mnras},
         year = 2024,
        month = jul,
       volume = {531},
       number = {3},
        pages = {3551-3578},
          doi = {10.1093/mnras/stae1024},
archivePrefix = {arXiv},
       eprint = {2309.02310},
 primaryClass = {astro-ph.GA},
       adsurl = {https://ui.adsabs.harvard.edu/abs/2024MNRAS.531.3551L}
}

@ARTICLE{Lagos2018,
       author = {{Lagos}, Claudia del P. and {Tobar}, Rodrigo J. and {Robotham}, Aaron S.~G. and {Obreschkow}, Danail and {Mitchell}, Peter D. and {Power}, Chris and {Elahi}, Pascal J.},
        title = "{Shark: introducing an open source, free, and flexible semi-analytic model of galaxy formation}",
      journal = {\mnras},
         year = 2018,
        month = dec,
       volume = {481},
       number = {3},
        pages = {3573-3603},
          doi = {10.1093/mnras/sty2440},
archivePrefix = {arXiv},
       eprint = {1807.11180},
 primaryClass = {astro-ph.GA},
       adsurl = {https://ui.adsabs.harvard.edu/abs/2018MNRAS.481.3573L}
}

@ARTICLE{Ferrara2000,
       author = {{Ferrara}, Andrea and {Tolstoy}, Eline},
        title = "{The role of stellar feedback and dark matter in the evolution of dwarf galaxies}",
      journal = {\mnras},
         year = 2000,
        month = apr,
       volume = {313},
       number = {2},
        pages = {291-309},
          doi = {10.1046/j.1365-8711.2000.03209.x},
archivePrefix = {arXiv},
       eprint = {astro-ph/9905280},
 primaryClass = {astro-ph},
       adsurl = {https://ui.adsabs.harvard.edu/abs/2000MNRAS.313..291F}
}

@ARTICLE{Gelli2024,
       author = {{Gelli}, Viola and {Salvadori}, Stefania and {Ferrara}, Andrea and {Pallottini}, Andrea},
        title = "{Can Supernovae Quench Star Formation in High-z Galaxies?}",
      journal = {\apj},
         year = 2024,
        month = mar,
       volume = {964},
       number = {1},
          eid = {76},
        pages = {76},
          doi = {10.3847/1538-4357/ad23ec},
archivePrefix = {arXiv},
       eprint = {2310.03065},
 primaryClass = {astro-ph.GA},
       adsurl = {https://ui.adsabs.harvard.edu/abs/2024ApJ...964...76G}
}

@ARTICLE{Moore1996,
       author = {{Moore}, Ben and {Katz}, Neal and {Lake}, George and {Dressler}, Alan and {Oemler}, Augustus},
        title = "{Galaxy harassment and the evolution of clusters of galaxies}",
      journal = {\nat},
         year = 1996,
        month = feb,
       volume = {379},
       number = {6566},
        pages = {613-616},
          doi = {10.1038/379613a0},
archivePrefix = {arXiv},
       eprint = {astro-ph/9510034},
 primaryClass = {astro-ph},
       adsurl = {https://ui.adsabs.harvard.edu/abs/1996Natur.379..613M}
}

@ARTICLE{Gunn1972,
       author = {{Gunn}, James E. and {Gott}, III, J. Richard},
        title = "{On the Infall of Matter Into Clusters of Galaxies and Some Effects on Their Evolution}",
      journal = {\apj},
         year = 1972,
        month = aug,
       volume = {176},
        pages = {1},
          doi = {10.1086/151605},
       adsurl = {https://ui.adsabs.harvard.edu/abs/1972ApJ...176....1G}
}

@ARTICLE{Moran2007,
       author = {{Moran}, Sean M. and {Ellis}, Richard S. and {Treu}, Tommaso and {Smith}, Graham P. and {Rich}, R. Michael and {Smail}, Ian},
        title = "{A Wide-Field Survey of Two z \raisebox{-0.5ex}\textasciitilde 0.5 Galaxy Clusters: Identifying the Physical Processes Responsible for the Observed Transformation of Spirals into S0s}",
      journal = {\apj},
         year = 2007,
        month = dec,
       volume = {671},
       number = {2},
        pages = {1503-1522},
          doi = {10.1086/522303},
archivePrefix = {arXiv},
       eprint = {0707.4173},
 primaryClass = {astro-ph},
       adsurl = {https://ui.adsabs.harvard.edu/abs/2007ApJ...671.1503M}
}

@ARTICLE{Cortese2021,
       author = {{Cortese}, L. and {Catinella}, B. and {Smith}, R.},
        title = "{The Dawes Review 9: The role of cold gas stripping on the star formation quenching of satellite galaxies}",
      journal = {\pasa},
         year = 2021,
        month = aug,
       volume = {38},
          eid = {e035},
        pages = {e035},
          doi = {10.1017/pasa.2021.18},
archivePrefix = {arXiv},
       eprint = {2104.02193},
 primaryClass = {astro-ph.GA},
       adsurl = {https://ui.adsabs.harvard.edu/abs/2021PASA...38...35C}
}

@ARTICLE{Larson1980,
       author = {{Larson}, R.~B. and {Tinsley}, B.~M. and {Caldwell}, C.~N.},
        title = "{The evolution of disk galaxies and the origin of S0 galaxies}",
      journal = {\apj},
         year = 1980,
        month = may,
        pages = {692-707},
          doi = {10.1086/157917},
       adsurl = {https://ui.adsabs.harvard.edu/abs/1980ApJ...237..692L}
}

@ARTICLE{Schreiber2015,
       author = {{Schreiber}, C. and {Pannella}, M. and {Elbaz}, D. and {B{\'e}thermin}, M. and {Inami}, H. and {Dickinson}, M. and {Magnelli}, B. and {Wang}, T. and {Aussel}, H. and {Daddi}, E. and {Juneau}, S. and {Shu}, X. and {Sargent}, M.~T. and {Buat}, V. and {Faber}, S.~M. and {Ferguson}, H.~C. and {Giavalisco}, M. and {Koekemoer}, A.~M. and {Magdis}, G. and {Morrison}, G.~E. and {Papovich}, C. and {Santini}, P. and {Scott}, D.},
        title = "{The Herschel view of the dominant mode of galaxy growth from z = 4 to the present day}",
      journal = {\aap},
         year = 2015,
        month = mar,
       volume = {575},
          eid = {A74},
        pages = {A74},
          doi = {10.1051/0004-6361/201425017},
archivePrefix = {arXiv},
       eprint = {1409.5433},
 primaryClass = {astro-ph.GA},
       adsurl = {https://ui.adsabs.harvard.edu/abs/2015A&A...575A..74S}
}

@ARTICLE{Porras-Valverde2025,
       author = {{Porras-Valverde}, Antonio J. and {Forbes}, John C.},
        title = "{On the Signature of Black Holes on the Quenched Stellar Mass Function}",
      journal = {\apj},
         year = 2025,
        month = may,
       volume = {984},
       number = {2},
          eid = {96},
        pages = {96},
          doi = {10.3847/1538-4357/adcc2d},
archivePrefix = {arXiv},
       eprint = {2412.04553},
 primaryClass = {astro-ph.GA},
       adsurl = {https://ui.adsabs.harvard.edu/abs/2025ApJ...984...96P}
}

@ARTICLE{Nanayakkara2024,
       author = {{Nanayakkara}, Themiya and {Glazebrook}, Karl and {Jacobs}, Colin and {Kawinwanichakij}, Lalitwadee and {Schreiber}, Corentin and {Brammer}, Gabriel and {Esdaile}, James and {Kacprzak}, Glenn G. and {Labbe}, Ivo and {Lagos}, Claudia and {Marchesini}, Danilo and {Marsan}, Z. Cemile and {Oesch}, Pascal A. and {Papovich}, Casey and {Remus}, Rhea-Silvia and {Tran}, Kim-Vy H.},
        title = "{A population of faint, old, and massive quiescent galaxies at 3 <z <4 revealed by JWST NIRSpec Spectroscopy}",
      journal = {Scientific Reports},
         year = 2024,
        month = feb,
       volume = {14},
          eid = {3724},
        pages = {3724},
          doi = {10.1038/s41598-024-52585-4},
archivePrefix = {arXiv},
       eprint = {2212.11638},
 primaryClass = {astro-ph.GA},
       adsurl = {https://ui.adsabs.harvard.edu/abs/2024NatSR..14.3724N}
}

@ARTICLE{Valentino2023,
       author = {{Valentino}, Francesco and {Brammer}, Gabriel and {Gould}, Katriona M.~L. and {Kokorev}, Vasily and {Fujimoto}, Seiji and {Jespersen}, Christian Kragh and {Vijayan}, Aswin P. and {Weaver}, John R. and {Ito}, Kei and {Tanaka}, Masayuki and {Ilbert}, Olivier and {Magdis}, Georgios E. and {Whitaker}, Katherine E. and {Faisst}, Andreas L. and {Gallazzi}, Anna and {Gillman}, Steven and {Gim{\'e}nez-Arteaga}, Clara and {G{\'o}mez-Guijarro}, Carlos and {Kubo}, Mariko and {Heintz}, Kasper E. and {Hirschmann}, Michaela and {Oesch}, Pascal and {Onodera}, Masato and {Rizzo}, Francesca and {Lee}, Minju and {Strait}, Victoria and {Toft}, Sune},
        title = "{An Atlas of Color-selected Quiescent Galaxies at z > 3 in Public JWST Fields}",
      journal = {\apj},
         year = 2023,
        month = apr,
       volume = {947},
       number = {1},
          eid = {20},
        pages = {20},
          doi = {10.3847/1538-4357/acbefa},
archivePrefix = {arXiv},
       eprint = {2302.10936},
 primaryClass = {astro-ph.GA},
       adsurl = {https://ui.adsabs.harvard.edu/abs/2023ApJ...947...20V}
}

@ARTICLE{Weibel2025,
       author = {{Weibel}, Andrea and {de Graaff}, Anna and {Setton}, David J. and {Miller}, Tim B. and {Oesch}, Pascal A. and {Brammer}, Gabriel and {Lagos}, Claudia D.~P. and {Whitaker}, Katherine E. and {Williams}, Christina C. and {Baggen}, Josephine F.~W. and {Bezanson}, Rachel and {Boogaard}, Leindert A. and {Cleri}, Nikko J. and {Greene}, Jenny E. and {Hirschmann}, Michaela and {Hviding}, Raphael E. and {Kuruvanthodi}, Adarsh and {Labb{\'e}}, Ivo and {Leja}, Joel and {Maseda}, Michael V. and {Matthee}, Jorryt and {McConachie}, Ian and {Naidu}, Rohan P. and {Roberts-Borsani}, Guido and {Schaerer}, Daniel and {Suess}, Katherine A. and {Valentino}, Francesco and {van Dokkum}, Pieter and {Wang}, Bingjie},
        title = "{RUBIES Reveals a Massive Quiescent Galaxy at z = 7.3}",
      journal = {\apj},
         year = 2025,
        month = apr,
       volume = {983},
       number = {1},
          eid = {11},
        pages = {11},
          doi = {10.3847/1538-4357/adab7a},
archivePrefix = {arXiv},
       eprint = {2409.03829},
 primaryClass = {astro-ph.GA},
       adsurl = {https://ui.adsabs.harvard.edu/abs/2025ApJ...983...11W}
}

@ARTICLE{Alberts2024,
       author = {{Alberts}, Stacey and {Williams}, Christina C. and {Helton}, Jakob M. and {Suess}, Katherine A. and {Ji}, Zhiyuan and {Shivaei}, Irene and {Lyu}, Jianwei and {Rieke}, George and {Baker}, William M. and {Bonaventura}, Nina and {Bunker}, Andrew J. and {Carniani}, Stefano and {Charlot}, Stephane and {Curtis-Lake}, Emma and {D'Eugenio}, Francesco and {Eisenstein}, Daniel J. and {de Graaff}, Anna and {Hainline}, Kevin N. and {Hausen}, Ryan and {Johnson}, Benjamin D. and {Maiolino}, Roberto and {Parlanti}, Eleonora and {Rieke}, Marcia J. and {Robertson}, Brant E. and {Sun}, Yang and {Tacchella}, Sandro and {Willmer}, Christopher N.~A. and {Willott}, Chris J.},
        title = "{To High Redshift and Low Mass: Exploring the Emergence of Quenched Galaxies and Their Environments at 3 < z < 6 in the Ultra-deep JADES MIRI F770W Parallel}",
      journal = {\apj},
         year = 2024,
        month = nov,
       volume = {975},
       number = {1},
          eid = {85},
        pages = {85},
          doi = {10.3847/1538-4357/ad66cc},
archivePrefix = {arXiv},
       eprint = {2312.12207},
 primaryClass = {astro-ph.GA},
       adsurl = {https://ui.adsabs.harvard.edu/abs/2024ApJ...975...85A}
}

@ARTICLE{Carnall2020,
       author = {{Carnall}, A.~C. and {Walker}, S. and {McLure}, R.~J. and {Dunlop}, J.~S. and {McLeod}, D.~J. and {Cullen}, F. and {Wild}, V. and {Amorin}, R. and {Bolzonella}, M. and {Castellano}, M. and {Cimatti}, A. and {Cucciati}, O. and {Fontana}, A. and {Gargiulo}, A. and {Garilli}, B. and {Jarvis}, M.~J. and {Pentericci}, L. and {Pozzetti}, L. and {Zamorani}, G. and {Calabro}, A. and {Hathi}, N.~P. and {Koekemoer}, A.~M.},
        title = "{Timing the earliest quenching events with a robust sample of massive quiescent galaxies at 2 < z < 5}",
      journal = {\mnras},
         year = 2020,
        month = jul,
       volume = {496},
       number = {1},
        pages = {695-707},
          doi = {10.1093/mnras/staa1535},
archivePrefix = {arXiv},
       eprint = {2001.11975},
 primaryClass = {astro-ph.GA},
       adsurl = {https://ui.adsabs.harvard.edu/abs/2020MNRAS.496..695C}
}

@ARTICLE{Carnall2023,
       author = {{Carnall}, A.~C. and {McLeod}, D.~J. and {McLure}, R.~J. and {Dunlop}, J.~S. and {Begley}, R. and {Cullen}, F. and {Donnan}, C.~T. and {Hamadouche}, M.~L. and {Jewell}, S.~M. and {Jones}, E.~W. and {Pollock}, C.~L. and {Wild}, V.},
        title = "{A surprising abundance of massive quiescent galaxies at 3 < z < 5 in the first data from JWST CEERS}",
      journal = {\mnras},
         year = 2023,
        month = apr,
       volume = {520},
       number = {3},
        pages = {3974-3985},
          doi = {10.1093/mnras/stad369},
archivePrefix = {arXiv},
       eprint = {2208.00986},
 primaryClass = {astro-ph.GA},
       adsurl = {https://ui.adsabs.harvard.edu/abs/2023MNRAS.520.3974C}
}

@ARTICLE{Toft2017,
       author = {{Toft}, Sune and {Zabl}, Johannes and {Richard}, Johan and {Gallazzi}, Anna and {Zibetti}, Stefano and {Prescott}, Moire and {Grillo}, Claudio and {Man}, Allison W.~S. and {Lee}, Nicholas Y. and {G{\'o}mez-Guijarro}, Carlos and {Stockmann}, Mikkel and {Magdis}, Georgios and {Steinhardt}, Charles L.},
        title = "{A massive, dead disk galaxy in the early Universe}",
      journal = {\nat},
         year = 2017,
        month = jun,
       volume = {546},
       number = {7659},
        pages = {510-513},
          doi = {10.1038/nature22388},
archivePrefix = {arXiv},
       eprint = {1706.07030},
 primaryClass = {astro-ph.GA},
       adsurl = {https://ui.adsabs.harvard.edu/abs/2017Natur.546..510T}
}

@ARTICLE{DEugenio2024,
       author = {{D'Eugenio}, Francesco and {P{\'e}rez-Gonz{\'a}lez}, Pablo G. and {Maiolino}, Roberto and {Scholtz}, Jan and {Perna}, Michele and {Circosta}, Chiara and {{\"U}bler}, Hannah and {Arribas}, Santiago and {B{\"o}ker}, Torsten and {Bunker}, Andrew J. and {Carniani}, Stefano and {Charlot}, Stephane and {Chevallard}, Jacopo and {Cresci}, Giovanni and {Curtis-Lake}, Emma and {Jones}, Gareth C. and {Kumari}, Nimisha and {Lamperti}, Isabella and {Looser}, Tobias J. and {Parlanti}, Eleonora and {Rix}, Hans-Walter and {Robertson}, Brant and {Rodr{\'\i}guez Del Pino}, Bruno and {Tacchella}, Sandro and {Venturi}, Giacomo and {Willott}, Chris J.},
        title = "{A fast-rotator post-starburst galaxy quenched by supermassive black-hole feedback at z = 3}",
      journal = {Nature Astronomy},
         year = 2024,
        month = nov,
       volume = {8},
        pages = {1443-1456},
          doi = {10.1038/s41550-024-02345-1},
archivePrefix = {arXiv},
       eprint = {2308.06317},
 primaryClass = {astro-ph.GA},
       adsurl = {https://ui.adsabs.harvard.edu/abs/2024NatAs...8.1443D}
}

@ARTICLE{Hamadouche2025,
       author = {{Hamadouche}, M.~L. and {McLure}, R.~J. and {Carnall}, A.~C. and {McLeod}, D.~J. and {Dunlop}, J.~S. and {Whitaker}, K.~E. and {Donnan}, C.~T. and {Begley}, R. and {Stanton}, T.~M. and {Almaini}, O. and {Aird}, J. and {Cullen}, F. and {Cutler}, S. and {Grogin}, N.~A. and {Koekemoer}, A.~M.},
        title = "{JWST PRIMER: strong evidence for the environmental quenching of low-mass galaxies out to z ≃ 2}",
      journal = {\mnras},
         year = 2025,
        month = jun,
          doi = {10.1093/mnras/staf971},
archivePrefix = {arXiv},
       eprint = {2412.09592},
 primaryClass = {astro-ph.GA},
       adsurl = {https://ui.adsabs.harvard.edu/abs/2025MNRAS.tmp..939H}
}

@ARTICLE{Baker25b,
       author = {{Baker}, William M. and {Valentino}, Francesco and {Lagos}, Claudia del P. and {Ito}, Kei and {Jespersen}, Christian Kragh and {Gottumukkala}, Rashmi and {Hjorth}, Jens and {Langeroodi}, Danial and {Sedgewick}, Aidan},
        title = "{Exploring over 700 massive quiescent galaxies at z = 2-7: Demographics and stellar mass functions}",
      journal = {arXiv e-prints},
         year = 2025,
        month = jun,
          eid = {arXiv:2506.04119},
        pages = {arXiv:2506.04119},
          doi = {10.48550/arXiv.2506.04119},
archivePrefix = {arXiv},
       eprint = {2506.04119},
 primaryClass = {astro-ph.GA},
       adsurl = {https://ui.adsabs.harvard.edu/abs/2025arXiv250604119B}
}

@ARTICLE{Eddington1913,
       author = {{Eddington}, A.~S.},
        title = "{On a formula for correcting statistics for the effects of a known error of observation}",
      journal = {\mnras},
         year = 1913,
        month = mar,
       volume = {73},
        pages = {359-360},
          doi = {10.1093/mnras/73.5.359},
       adsurl = {https://ui.adsabs.harvard.edu/abs/1913MNRAS..73..359E}
}

@ARTICLE{Matthee2024,
       author = {{Matthee}, Jorryt and {Naidu}, Rohan P. and {Brammer}, Gabriel and {Chisholm}, John and {Eilers}, Anna-Christina and {Goulding}, Andy and {Greene}, Jenny and {Kashino}, Daichi and {Labbe}, Ivo and {Lilly}, Simon J. and {Mackenzie}, Ruari and {Oesch}, Pascal A. and {Weibel}, Andrea and {Wuyts}, Stijn and {Xiao}, Mengyuan and {Bordoloi}, Rongmon and {Bouwens}, Rychard and {van Dokkum}, Pieter and {Illingworth}, Garth and {Kramarenko}, Ivan and {Maseda}, Michael V. and {Mason}, Charlotte and {Meyer}, Romain A. and {Nelson}, Erica J. and {Reddy}, Naveen A. and {Shivaei}, Irene and {Simcoe}, Robert A. and {Yue}, Minghao},
        title = "{Little Red Dots: An Abundant Population of Faint Active Galactic Nuclei at z {\ensuremath{\sim}} 5 Revealed by the EIGER and FRESCO JWST Surveys}",
      journal = {\apj},
         year = 2024,
        month = mar,
       volume = {963},
       number = {2},
          eid = {129},
        pages = {129},
          doi = {10.3847/1538-4357/ad2345},
archivePrefix = {arXiv},
       eprint = {2306.05448},
 primaryClass = {astro-ph.GA},
       adsurl = {https://ui.adsabs.harvard.edu/abs/2024ApJ...963..129M}
}

@article{Santini2022,
   title={The Stellar Mass Function in CANDELS and Frontier Fields: The Buildup of Low-mass Passive Galaxies since z ∼ 3},
   volume={940},
   ISSN={1538-4357},
   url={http://dx.doi.org/10.3847/1538-4357/ac9a48},
   DOI={10.3847/1538-4357/ac9a48},
   number={2},
   journal={The Astrophysical Journal},
   publisher={American Astronomical Society},
   author={Santini, Paola and Castellano, Marco and Fontana, Adriano and Fortuni, Flaminia and Menci, Nicola and Merlin, Emiliano and Pagul, Amanda and Testa, Vincenzo and Calabrò, Antonello and Paris, Diego and Pentericci, Laura},
   year={2022},
   month=nov, pages={135} }

@ARTICLE{Jespersen2025,
       author = {{Jespersen}, Christian Kragh and {Steinhardt}, Charles L. and {Somerville}, Rachel S. and {Lovell}, Christopher C.},
        title = "{On the Significance of Rare Objects at High Redshift: The Impact of Cosmic Variance}",
      journal = {\apj},
         year = 2025,
        month = mar,
       volume = {982},
       number = {1},
          eid = {23},
        pages = {23},
          doi = {10.3847/1538-4357/adb422},
archivePrefix = {arXiv},
       eprint = {2403.00050},
 primaryClass = {astro-ph.GA},
       adsurl = {https://ui.adsabs.harvard.edu/abs/2025ApJ...982...23J}
}

@ARTICLE{Malmquist1922,
       author = {{Malmquist}, K.~G.},
        title = "{On some relations in stellar statistics}",
      journal = {Lund Medd Serie I},
         year = 1922,
        month = mar,
       volume = {100},
        pages = {1-52},
       adsurl = {https://ui.adsabs.harvard.edu/abs/1922MeLuF.100....1M}
}

@ARTICLE{McLeod2021,
       author = {{McLeod}, D.~J. and {McLure}, R.~J. and {Dunlop}, J.~S. and {Cullen}, F. and {Carnall}, A.~C. and {Duncan}, K.},
        title = "{The evolution of the galaxy stellar-mass function over the last 12 billion years from a combination of ground-based and HST surveys}",
      journal = {\mnras},
         year = 2021,
        month = may,
       volume = {503},
       number = {3},
        pages = {4413-4435},
          doi = {10.1093/mnras/stab731},
archivePrefix = {arXiv},
       eprint = {2009.03176},
 primaryClass = {astro-ph.GA},
       adsurl = {https://ui.adsabs.harvard.edu/abs/2021MNRAS.503.4413M}
}

@ARTICLE{Schechter76,
       author = {{Schechter}, P.},
        title = "{An analytic expression for the luminosity function for galaxies.}",
      journal = {\apj},
         year = 1976,
        month = jan,
       volume = {203},
        pages = {297-306},
          doi = {10.1086/154079},
       adsurl = {https://ui.adsabs.harvard.edu/abs/1976ApJ...203..297S}
}

@article{Civano_2016,
doi = {10.3847/0004-637X/819/1/62},
url = {https://dx.doi.org/10.3847/0004-637X/819/1/62},
year = {2016},
month = {feb},
publisher = {The American Astronomical Society},
volume = {819},
number = {1},
pages = {62},
author = {F. Civano and S. Marchesi and A. Comastri and M. C. Urry and M. Elvis and N. Cappelluti and S. Puccetti and M. Brusa and G. Zamorani and G. Hasinger and T. Aldcroft and D. M. Alexander and V. Allevato and H. Brunner and P. Capak and A. Finoguenov and F. Fiore and A. Fruscione and R. Gilli and K. Glotfelty and R. E. Griffiths and H. Hao and F. A. Harrison and K. Jahnke and J. Kartaltepe and A. Karim and S. M. LaMassa and G. Lanzuisi and T. Miyaji and P. Ranalli and M. Salvato and M. Sargent and N. J. Scoville and K. Schawinski and E. Schinnerer and J. Silverman and V. Smolcic and D. Stern and S. Toft and B. Trakhenbrot and E. Treister and C. Vignali},
title = {THE CHANDRA  COSMOS LEGACY SURVEY: OVERVIEW AND POINT SOURCE CATALOG},
journal = {The Astrophysical Journal}
}

@ARTICLE{ShuntovCW2025,
       author = {{Shuntov}, Marko and {Akins}, Hollis B. and {Paquereau}, Louise and {Casey}, Caitlin M. and {Ilbert}, Olivier and {Arango-Toro}, Rafael C. and {McCracken}, Henry Joy and {Franco}, Maximilien and {Harish}, Santosh and {Kartaltepe}, Jeyhan S. and {Koekemoer}, Anton M. and {Yang}, Lilan and {Huertas-Company}, Marc and {Berman}, Edward M. and {McCleary}, Jacqueline E. and {Toft}, Sune and {Gavazzi}, Rapha{\"e}l and {Achenbach}, Mark J. and {Bertin}, Emmanuel and {Brinch}, Malte and {Champagne}, Jackie and {Chartab}, Nima and {Drakos}, Nicole E. and {Egami}, Eiichi and {Endsley}, Ryan and {Faisst}, Andreas L. and {Fan}, Xiaohui and {Flayhart}, Carter and {Hartley}, William G. and {Hatamnia}, Hossein and {Gozaliasl}, Ghassem and {Gentile}, Fabrizio and {Jermann}, Iris and {Jin}, Shuowen and {Kakiichi}, Koki and {Khostovan}, Ali Ahmad and {K{\"u}mmel}, Martin and {Laigle}, Clotilde and {Laishram}, Ronaldo and {Lambrides}, Erini and {Liu}, Daizhong and {Lyu}, Jianwei and {Magdis}, Georgios and {Mobasher}, Bahram and {Moutard}, Thibaud and {Renzini}, Alvio and {Robertson}, Brant E. and {Schefer}, Marc and {Scognamiglio}, Diana and {Scoville}, Nick and {Sattari}, Zahra and {Sanders}, David B. and {Taamoli}, Sina and {Trakhtenbrot}, Benny and {Valentino}, Francesco and {Wang}, Feige and {Weaver}, John R. and {Yang}, Jinyl},
        title = "{COSMOS2025: The COSMOS-Web galaxy catalog of photometry, morphology, redshifts, and physical parameters from JWST, HST, and ground-based imaging}",
      journal = {arXiv e-prints},
         year = 2025,
        month = jun,
          eid = {arXiv:2506.03243},
        pages = {arXiv:2506.03243},
          doi = {10.48550/arXiv.2506.03243},
archivePrefix = {arXiv},
       eprint = {2506.03243},
 primaryClass = {astro-ph.GA},
       adsurl = {https://ui.adsabs.harvard.edu/abs/2025arXiv250603243S}
}

@ARTICLE{ShuntovSMF2025,
       author = {{Shuntov}, M. and {Ilbert}, O. and {Toft}, S. and {Arango-Toro}, R.~C. and {Akins}, H.~B. and {Casey}, C.~M. and {Franco}, M. and {Harish}, S. and {Kartaltepe}, J.~S. and {Koekemoer}, A.~M. and {McCracken}, H.~J. and {Paquereau}, L. and {Laigle}, C. and {Bethermin}, M. and {Dubois}, Y. and {Drakos}, N.~E. and {Faisst}, A. and {Gozaliasl}, G. and {Gillman}, S. and {Hayward}, C.~C. and {Hirschmann}, M. and {Huertas-Company}, M. and {Jespersen}, C.~K. and {Jin}, S. and {Kokorev}, V. and {Lambrides}, E. and {Le Borgne}, D. and {Liu}, D. and {Magdis}, G. and {Massey}, R. and {McPartland}, C.~J.~R. and {Mercier}, W. and {McCleary}, J.~E. and {McKinney}, J. and {Oesch}, P.~A. and {Renzini}, A. and {Rhodes}, J.~D. and {Rich}, R.~M. and {Robertson}, B.~E. and {Sanders}, D. and {Trebitsch}, M. and {Tresse}, L. and {Valentino}, F. and {Vijayan}, A.~P. and {Weaver}, J.~R. and {Weibel}, A. and {Wilkins}, S.~M. and {Yang}, L.},
        title = "{COSMOS-Web: Stellar mass assembly in relation to dark matter halos across 0.2 < z < 12 of cosmic history}",
      journal = {\aap},
         year = 2025,
        month = mar,
       volume = {695},
          eid = {A20},
        pages = {A20},
          doi = {10.1051/0004-6361/202452570},
archivePrefix = {arXiv},
       eprint = {2410.08290},
 primaryClass = {astro-ph.GA},
       adsurl = {https://ui.adsabs.harvard.edu/abs/2025A&A...695A..20S}
}

@ARTICLE{Barro2017,
       author = {{Barro}, Guillermo and {Faber}, S.~M. and {Koo}, David C. and {Dekel}, Avishai and {Fang}, Jerome J. and {Trump}, Jonathan R. and {P{\'e}rez-Gonz{\'a}lez}, Pablo G. and {Pacifici}, Camilla and {Primack}, Joel R. and {Somerville}, Rachel S. and {Yan}, Haojing and {Guo}, Yicheng and {Liu}, Fengshan and {Ceverino}, Daniel and {Kocevski}, Dale D. and {McGrath}, Elizabeth},
        title = "{Structural and Star-forming Relations since z {\ensuremath{\sim}} 3: Connecting Compact Star-forming and Quiescent Galaxies}",
      journal = {\apj},
         year = 2017,
        month = may,
       volume = {840},
       number = {1},
          eid = {47},
        pages = {47},
          doi = {10.3847/1538-4357/aa6b05},
archivePrefix = {arXiv},
       eprint = {1509.00469},
 primaryClass = {astro-ph.GA},
       adsurl = {https://ui.adsabs.harvard.edu/abs/2017ApJ...840...47B}
}

@ARTICLE{Barro2013,
       author = {{Barro}, Guillermo and {Faber}, S.~M. and {P{\'e}rez-Gonz{\'a}lez}, Pablo G. and {Koo}, David C. and {Williams}, Christina C. and {Kocevski}, Dale D. and {Trump}, Jonathan R. and {Mozena}, Mark and {McGrath}, Elizabeth and {van der Wel}, Arjen and {Wuyts}, Stijn and {Bell}, Eric F. and {Croton}, Darren J. and {Ceverino}, Daniel and {Dekel}, Avishai and {Ashby}, M.~L.~N. and {Cheung}, Edmond and {Ferguson}, Henry C. and {Fontana}, Adriano and {Fang}, Jerome and {Giavalisco}, Mauro and {Grogin}, Norman A. and {Guo}, Yicheng and {Hathi}, Nimish P. and {Hopkins}, Philip F. and {Huang}, Kuang-Han and {Koekemoer}, Anton M. and {Kartaltepe}, Jeyhan S. and {Lee}, Kyoung-Soo and {Newman}, Jeffrey A. and {P@ARTICLE{Barro2013,
       author = {{Barro}, Guillermo and {Faber}, S.~M. and {P{\'e}rez-Gonz{\'a}lez}, Pablo G. and {Koo}, David C. and {Williams}, Christina C. and {Kocevski}, Dale D. and {Trump}, Jonathan R. and {Mozena}, Mark and {McGrath}, Elizabeth and {van der Wel}, Arjen and {Wuyts}, Stijn and {Bell}, Eric F. and {Croton}, Darren J. and {Ceverino}, Daniel and {Dekel}, Avishai and {Ashby}, M.~L.~N. and {Cheung}, Edmond and {Ferguson}, Henry C. and {Fontana}, Adriano and {Fang}, Jerome and {Giavalisco}, Mauro and {Grogin}, Norman A. and {Guo}, Yicheng and {Hathi}, Nimish P. and {Hopkins}, Philip F. and {Huang}, Kuang-Han and {Koekemoer}, Anton M. and {Kartaltepe}, Jeyhan S. and {Lee}, Kyoung-Soo and {Newman}, Jeffrey A. and {Porter}, Lauren A. and {Primack}, Joel R. and {Ryan}, Russell E. and {Rosario}, David and {Somerville}, Rachel S. and {Salvato}, Mara and {Hsu}, Li-Ting},
        title = "{CANDELS: The Progenitors of Compact Quiescent Galaxies at z \raisebox{-0.5ex}\textasciitilde 2}",
      journal = {\apj},
     keywords = {galaxies: high-redshift, galaxies: photometry, galaxies: starburst, Astrophysics - Cosmology and Nongalactic Astrophysics},
         year = 2013,
        month = mar,
       volume = {765},
       number = {2},
          eid = {104},
        pages = {104},
          doi = {10.1088/0004-637X/765/2/104},
archivePrefix = {arXiv},
       eprint = {1206.5000},
 primaryClass = {astro-ph.CO},
       adsurl = {https://ui.adsabs.harvard.edu/abs/2013ApJ...765..104B},
      adsnote = {Provided by the SAO/NASA Astrophysics Data System}
}orter}, Lauren A. and {Primack}, Joel R. and {Ryan}, Russell E. and {Rosario}, David and {Somerville}, Rachel S. and {Salvato}, Mara and {Hsu}, Li-Ting},
        title = "{CANDELS: The Progenitors of Compact Quiescent Galaxies at z \raisebox{-0.5ex}\textasciitilde 2}",
      journal = {\apj},
         year = 2013,
        month = mar,
       volume = {765},
       number = {2},
          eid = {104},
        pages = {104},
          doi = {10.1088/0004-637X/765/2/104},
archivePrefix = {arXiv},
       eprint = {1206.5000},
 primaryClass = {astro-ph.CO},
       adsurl = {https://ui.adsabs.harvard.edu/abs/2013ApJ...765..104B}
}

@ARTICLE{Sersic1963,
       author = {{S{\'e}rsic}, J.~L.},
        title = "{Influence of the atmospheric and instrumental dispersion on the brightness distribution in a galaxy}",
      journal = {Boletin de la Asociacion Argentina de Astronomia La Plata Argentina},
         year = 1963,
        month = feb,
       volume = {6},
        pages = {41-43},
       adsurl = {https://ui.adsabs.harvard.edu/abs/1963BAAA....6...41S}
}

@ARTICLE{Akins2024,
       author = {{Akins}, Hollis B. and {Casey}, Caitlin M. and {Lambrides}, Erini and {Allen}, Natalie and {Andika}, Irham T. and {Brinch}, Malte and {Champagne}, Jaclyn B. and {Cooper}, Olivia and {Ding}, Xuheng and {Drakos}, Nicole E. and {Faisst}, Andreas and {Finkelstein}, Steven L. and {Franco}, Maximilien and {Fujimoto}, Seiji and {Gentile}, Fabrizio and {Gillman}, Steven and {Gozaliasl}, Ghassem and {Harish}, Santosh and {Hayward}, Christopher C. and {Hirschmann}, Michaela and {Ilbert}, Olivier and {Kartaltepe}, Jeyhan S. and {Kocevski}, Dale D. and {Koekemoer}, Anton M. and {Kokorev}, Vasily and {Liu}, Daizhong and {Long}, Arianna S. and {McCracken}, Henry Joy and {McKinney}, Jed and {Onoue}, Masafusa and {Paquereau}, Louise and {Renzini}, Alvio and {Rhodes}, Jason and {Robertson}, Brant E. and {Shuntov}, Marko and {Silverman}, John D. and {Tanaka}, Takumi S. and {Toft}, Sune and {Trakhtenbrot}, Benny and {Valentino}, Francesco and {Zavala}, Jorge},
        title = "{COSMOS-Web: The Overabundance and Physical Nature of ``Little Red Dots''{\textemdash}Implications for Early Galaxy and SMBH Assembly}",
      journal = {\apj},
         year = 2025,
        month = sep,
       volume = {991},
       number = {1},
          eid = {37},
        pages = {37},
          doi = {10.3847/1538-4357/ade984},
archivePrefix = {arXiv},
       eprint = {2406.10341},
 primaryClass = {astro-ph.GA},
       adsurl = {https://ui.adsabs.harvard.edu/abs/2025ApJ...991...37A}
}

@ARTICLE{Paquereau2025,
       author = {{Paquereau}, L. and {Laigle}, C. and {McCracken}, H.~J. and {Shuntov}, M. and {Ilbert}, O. and {Akins}, H.~B. and {Allen}, N. and {Arango-Togo}, R. and {Berman}, E.~M. and {B{\'e}thermin}, M. and {Casey}, C.~M. and {McCleary}, J. and {Dubois}, Y. and {Drakos}, N.~E. and {Faisst}, A.~L. and {Franco}, M. and {Harish}, S. and {Jespersen}, C.~K. and {Kartaltepe}, J.~S. and {Koekemoer}, A.~M. and {Kokorev}, V. and {Lambrides}, E. and {Larson}, R. and {Liu}, D. and {Le Borgne}, D. and {Lewis}, J.~S.~W. and {McKinney}, J. and {Mercier}, W. and {Rhodes}, J.~D. and {Robertson}, B.~E. and {Toft}, S. and {Trebitsch}, M. and {Tresse}, L. and {Weaver}, J.~R.},
        title = "{Tracing the galaxy-halo connection with galaxy clustering in COSMOS-Web from z = 0.1 to z {\ensuremath{\sim}} 12}",
      journal = {\aap},
         year = 2025,
        month = oct,
       volume = {702},
          eid = {A163},
        pages = {A163},
          doi = {10.1051/0004-6361/202553828},
archivePrefix = {arXiv},
       eprint = {2501.11674},
 primaryClass = {astro-ph.GA},
       adsurl = {https://ui.adsabs.harvard.edu/abs/2025A&A...702A.163P}
}

@ARTICLE{Weaver2023,
       author = {{Weaver}, J.~R. and {Davidzon}, I. and {Toft}, S. and {Ilbert}, O. and {McCracken}, H.~J. and {Gould}, K.~M.~L. and {Jespersen}, C.~K. and {Steinhardt}, C. and {Lagos}, C.~D.~P. and {Capak}, P.~L. and {Casey}, C.~M. and {Chartab}, N. and {Faisst}, A.~L. and {Hayward}, C.~C. and {Kartaltepe}, J.~S. and {Kauffmann}, O.~B. and {Koekemoer}, A.~M. and {Kokorev}, V. and {Laigle}, C. and {Liu}, D. and {Long}, A. and {Magdis}, G.~E. and {McPartland}, C.~J.~R. and {Milvang-Jensen}, B. and {Mobasher}, B. and {Moneti}, A. and {Peng}, Y. and {Sanders}, D.~B. and {Shuntov}, M. and {Sneppen}, A. and {Valentino}, F. and {Zalesky}, L. and {Zamorani}, G.},
        title = "{COSMOS2020: The galaxy stellar mass function. The assembly and star formation cessation of galaxies at 0.2< z {\ensuremath{\leq}} 7.5}",
      journal = {\aap},
         year = 2023,
        month = sep,
       volume = {677},
          eid = {A184},
        pages = {A184},
          doi = {10.1051/0004-6361/202245581},
archivePrefix = {arXiv},
       eprint = {2212.02512},
 primaryClass = {astro-ph.GA},
       adsurl = {https://ui.adsabs.harvard.edu/abs/2023A&A...677A.184W}
}

@ARTICLE{Ilbert2010,
       author = {{Ilbert}, O. and {Salvato}, M. and {Le Floc'h}, E. and {Aussel}, H. and {Capak}, P. and {McCracken}, H.~J. and {Mobasher}, B. and {Kartaltepe}, J. and {Scoville}, N. and {Sanders}, D.~B. and {Arnouts}, S. and {Bundy}, K. and {Cassata}, P. and {Kneib}, J. -P. and {Koekemoer}, A. and {Le F{\`e}vre}, O. and {Lilly}, S. and {Surace}, J. and {Taniguchi}, Y. and {Tasca}, L. and {Thompson}, D. and {Tresse}, L. and {Zamojski}, M. and {Zamorani}, G. and {Zucca}, E.},
        title = "{Galaxy Stellar Mass Assembly Between 0.2 < z < 2 from the S-COSMOS Survey}",
      journal = {\apj},
         year = 2010,
        month = feb,
       volume = {709},
       number = {2},
        pages = {644-663},
          doi = {10.1088/0004-637X/709/2/644},
archivePrefix = {arXiv},
       eprint = {0903.0102},
 primaryClass = {astro-ph.CO},
       adsurl = {https://ui.adsabs.harvard.edu/abs/2010ApJ...709..644I}
}

@ARTICLE{Casey2023CW,
       author = {{Casey}, Caitlin M. and {Kartaltepe}, Jeyhan S. and {Drakos}, Nicole E. and {Franco}, Maximilien and {Harish}, Santosh and {Paquereau}, Louise and {Ilbert}, Olivier and {Rose}, Caitlin and {Cox}, Isabella G. and {Nightingale}, James W. and {Robertson}, Brant E. and {Silverman}, John D. and {Koekemoer}, Anton M. and {Massey}, Richard and {McCracken}, Henry Joy and {Rhodes}, Jason and {Akins}, Hollis B. and {Allen}, Natalie and {Amvrosiadis}, Aristeidis and {Arango-Toro}, Rafael C. and {Bagley}, Micaela B. and {Bongiorno}, Angela and {Capak}, Peter L. and {Champagne}, Jaclyn B. and {Chartab}, Nima and {Ch{\'a}vez Ortiz}, {\'O}scar A. and {Chworowsky}, Katherine and {Cooke}, Kevin C. and {Cooper}, Olivia R. and {Darvish}, Behnam and {Ding}, Xuheng and {Faisst}, Andreas L. and {Finkelstein}, Steven L. and {Fujimoto}, Seiji and {Gentile}, Fabrizio and {Gillman}, Steven and {Gould}, Katriona M.~L. and {Gozaliasl}, Ghassem and {Hayward}, Christopher C. and {He}, Qiuhan and {Hemmati}, Shoubaneh and {Hirschmann}, Michaela and {Jahnke}, Knud and {Jin}, Shuowen and {Khostovan}, Ali Ahmad and {Kokorev}, Vasily and {Lambrides}, Erini and {Laigle}, Clotilde and {Larson}, Rebecca L. and {Leung}, Gene C.~K. and {Liu}, Daizhong and {Liaudat}, Tobias and {Long}, Arianna S. and {Magdis}, Georgios and {Mahler}, Guillaume and {Mainieri}, Vincenzo and {Manning}, Sinclaire M. and {Maraston}, Claudia and {Martin}, Crystal L. and {McCleary}, Jacqueline E. and {McKinney}, Jed and {McPartland}, Conor J.~R. and {Mobasher}, Bahram and {Pattnaik}, Rohan and {Renzini}, Alvio and {Rich}, R. Michael and {Sanders}, David B. and {Sattari}, Zahra and {Scognamiglio}, Diana and {Scoville}, Nick and {Sheth}, Kartik and {Shuntov}, Marko and {Sparre}, Martin and {Suzuki}, Tomoko L. and {Talia}, Margherita and {Toft}, Sune and {Trakhtenbrot}, Benny and {Urry}, C. Megan and {Valentino}, Francesco and {Vanderhoof}, Brittany N. and {Vardoulaki}, Eleni and {Weaver}, John R. and {Whitaker}, Katherine E. and {Wilkins}, Stephen M. and {Yang}, Lilan and {Zavala}, Jorge A.},
        title = "{COSMOS-Web: An Overview of the JWST Cosmic Origins Survey}",
      journal = {\apj},
         year = 2023,
        month = sep,
       volume = {954},
       number = {1},
          eid = {31},
        pages = {31},
          doi = {10.3847/1538-4357/acc2bc},
archivePrefix = {arXiv},
       eprint = {2211.07865},
 primaryClass = {astro-ph.GA},
       adsurl = {https://ui.adsabs.harvard.edu/abs/2023ApJ...954...31C}
}

@ARTICLE{Arango-Toro2024,
     author = {{Arango-Toro}, R.~C. and {Ilbert}, O. and {Ciesla}, L. and {Shuntov}, M. and {Aufort}, G. and {Mercier}, W. and {Laigle}, C. and {Franco}, M. and {Bethermin}, M. and {Le Borgne}, D. and {Dubois}, Y. and {McCracken}, H.~J. and {Paquereau}, L. and {Huertas-Company}, M. and {Kartaltepe}, J. and {Casey}, C.~M. and {Akins}, H. and {Allen}, N. and {Andika}, I. and {Brinch}, M. and {Drakos}, N.~E. and {Faisst}, A. and {Gozaliasl}, G. and {Harish}, S. and {Kaminsky}, A. and {Koekemoer}, A. and {Kokorev}, V. and {Liu}, D. and {Magdis}, G. and {Martin}, C.~L. and {Moutard}, T. and {Rhodes}, J. and {Rich}, R.~M. and {Robertson}, B. and {Sanders}, D.~B. and {Sheth}, K. and {Talia}, M. and {Toft}, S. and {Tresse}, L. and {Valentino}, F. and {Vijayan}, A. and {Weaver}, J.},
        title = "{COSMOS-Web: A history of galaxy migrations over the stellar mass{\textendash}star formation rate plane}",
      journal = {\aap},
         year = 2025,
        month = apr,
       volume = {696},
          eid = {A159},
        pages = {A159},
          doi = {10.1051/0004-6361/202452519},
archivePrefix = {arXiv},
       eprint = {2410.05375},
 primaryClass = {astro-ph.GA},}

@ARTICLE{Boquien19,
       author = {{Boquien}, M. and {Burgarella}, D. and {Roehlly}, Y. and {Buat}, V. and {Ciesla}, L. and {Corre}, D. and {Inoue}, A.~K. and {Salas}, H.},
        title = "{CIGALE: a python Code Investigating GALaxy Emission}",
      journal = {\aap},
         year = 2019,
        month = feb,
       volume = {622},
          eid = {A103},
        pages = {A103},
          doi = {10.1051/0004-6361/201834156},
archivePrefix = {arXiv},
       eprint = {1811.03094},
 primaryClass = {astro-ph.GA},
       adsurl = {https://ui.adsabs.harvard.edu/abs/2019A&A...622A.103B}
}

@ARTICLE{Grogin2011,
       author = {{Grogin}, Norman A. and {Kocevski}, Dale D. and {Faber}, S.~M. and
         {Ferguson}, Henry C. and {Koekemoer}, Anton M. and {Riess}, Adam G. and
         {Acquaviva}, Viviana and {Alexander}, David M. and {Almaini}, Omar and
         {Ashby}, Matthew L.~N. and {Barden}, Marco and {Bell}, Eric F. and
         {Bournaud}, Fr{\'e}d{\'e}ric and {Brown}, Thomas M. and
         {Caputi}, Karina I. and {Casertano}, Stefano and {Cassata}, Paolo and
         {Castellano}, Marco and {Challis}, Peter and {Chary}, Ranga-Ram and
         {Cheung}, Edmond and {Cirasuolo}, Michele and
         {Conselice}, Christopher J. and {Roshan Cooray}, Asantha and
         {Croton}, Darren J. and {Daddi}, Emanuele and {Dahlen}, Tomas and
         {Dav{\'e}}, Romeel and {de Mello}, Du{\'\i}lia F. and {Dekel}, Avishai and
         {Dickinson}, Mark and {Dolch}, Timothy and {Donley}, Jennifer L. and
         {Dunlop}, James S. and {Dutton}, Aaron A. and {Elbaz}, David and
         {Fazio}, Giovanni G. and {Filippenko}, Alexei V. and
         {Finkelstein}, Steven L. and {Fontana}, Adriano and
         {Gardner}, Jonathan P. and {Garnavich}, Peter M. and {Gawiser}, Eric and
         {Giavalisco}, Mauro and {Grazian}, Andrea and {Guo}, Yicheng and
         {Hathi}, Nimish P. and {H{\"a}ussler}, Boris and {Hopkins}, Philip F. and
         {Huang}, Jia-Sheng and {Huang}, Kuang-Han and {Jha}, Saurabh W. and
         {Kartaltepe}, Jeyhan S. and {Kirshner}, Robert P. and {Koo}, David C. and
         {Lai}, Kamson and {Lee}, Kyoung-Soo and {Li}, Weidong and
         {Lotz}, Jennifer M. and {Lucas}, Ray A. and {Madau}, Piero and
         {McCarthy}, Patrick J. and {McGrath}, Elizabeth J. and
         {McIntosh}, Daniel H. and {McLure}, Ross J. and {Mobasher}, Bahram and
         {Moustakas}, Leonidas A. and {Mozena}, Mark and {Nandra}, Kirpal and
         {Newman}, Jeffrey A. and {Niemi}, Sami-Matias and {Noeske}, Kai G. and
         {Papovich}, Casey J. and {Pentericci}, Laura and {Pope}, Alexandra and
         {Primack}, Joel R. and {Rajan}, Abhijith and {Ravindranath}, Swara and
         {Reddy}, Naveen A. and {Renzini}, Alvio and {Rix}, Hans-Walter and
         {Robaina}, Aday R. and {Rodney}, Steven A. and {Rosario}, David J. and
         {Rosati}, Piero and {Salimbeni}, Sara and {Scarlata}, Claudia and
         {Siana}, Brian and {Simard}, Luc and {Smidt}, Joseph and
         {Somerville}, Rachel S. and {Spinrad}, Hyron and {Straughn}, Amber N. and
         {Strolger}, Louis-Gregory and {Telford}, Olivia and
         {Teplitz}, Harry I. and {Trump}, Jonathan R. and {van der Wel}, Arjen and
         {Villforth}, Carolin and {Wechsler}, Risa H. and {Weiner}, Benjamin J. and
         {Wiklind}, Tommy and {Wild}, Vivienne and {Wilson}, Grant and
         {Wuyts}, Stijn and {Yan}, Hao-Jing and {Yun}, Min S.},
        title = "{CANDELS: The Cosmic Assembly Near-infrared Deep Extragalactic Legacy Survey}",
      journal = {\apjs},
         year = 2011,
        month = dec,
       volume = {197},
       number = {2},
          eid = {35},
        pages = {35},
          doi = {10.1088/0067-0049/197/2/35},
archivePrefix = {arXiv},
       eprint = {1105.3753},
 primaryClass = {astro-ph.CO},
       adsurl = {https://ui.adsabs.harvard.edu/abs/2011ApJS..197...35G}
}

@ARTICLE{Koekemoer2011,
       author = {{Koekemoer}, Anton M. and {Faber}, S.~M. and {Ferguson}, Henry C. and
         {Grogin}, Norman A. and {Kocevski}, Dale D. and {Koo}, David C. and
         {Lai}, Kamson and {Lotz}, Jennifer M. and {Lucas}, Ray A. and
         {McGrath}, Elizabeth J. and {Ogaz}, Sara and {Rajan}, Abhijith and
         {Riess}, Adam G. and {Rodney}, Steve A. and {Strolger}, Louis and
         {Casertano}, Stefano and {Castellano}, Marco and {Dahlen}, Tomas and
         {Dickinson}, Mark and {Dolch}, Timothy and {Fontana}, Adriano and
         {Giavalisco}, Mauro and {Grazian}, Andrea and {Guo}, Yicheng and
         {Hathi}, Nimish P. and {Huang}, Kuang-Han and {van der Wel}, Arjen and
         {Yan}, Hao-Jing and {Acquaviva}, Viviana and {Alexander}, David M. and
         {Almaini}, Omar and {Ashby}, Matthew L.~N. and {Barden}, Marco and
         {Bell}, Eric F. and {Bournaud}, Fr{\'e}d{\'e}ric and
         {Brown}, Thomas M. and {Caputi}, Karina I. and {Cassata}, Paolo and
         {Challis}, Peter J. and {Chary}, Ranga-Ram and {Cheung}, Edmond and
         {Cirasuolo}, Michele and {Conselice}, Christopher J. and
         {Roshan Cooray}, Asantha and {Croton}, Darren J. and {Daddi}, Emanuele and
         {Dav{\'e}}, Romeel and {de Mello}, Duilia F. and {de Ravel}, Loic and
         {Dekel}, Avishai and {Donley}, Jennifer L. and {Dunlop}, James S. and
         {Dutton}, Aaron A. and {Elbaz}, David and {Fazio}, Giovanni G. and
         {Filippenko}, Alexei V. and {Finkelstein}, Steven L. and
         {Frazer}, Chris and {Gardner}, Jonathan P. and {Garnavich}, Peter M. and
         {Gawiser}, Eric and {Gruetzbauch}, Ruth and {Hartley}, Will G. and
         {H{\"a}ussler}, Boris and {Herrington}, Jessica and
         {Hopkins}, Philip F. and {Huang}, Jia-Sheng and {Jha}, Saurabh W. and
         {Johnson}, Andrew and {Kartaltepe}, Jeyhan S. and {Khostovan}, Ali A. and
         {Kirshner}, Robert P. and {Lani}, Caterina and {Lee}, Kyoung-Soo and
         {Li}, Weidong and {Madau}, Piero and {McCarthy}, Patrick J. and
         {McIntosh}, Daniel H. and {McLure}, Ross J. and {McPartland}, Conor and
         {Mobasher}, Bahram and {Moreira}, Heidi and {Mortlock}, Alice and
         {Moustakas}, Leonidas A. and {Mozena}, Mark and {Nandra}, Kirpal and
         {Newman}, Jeffrey A. and {Nielsen}, Jennifer L. and {Niemi}, Sami and
         {Noeske}, Kai G. and {Papovich}, Casey J. and {Pentericci}, Laura and
         {Pope}, Alexandra and {Primack}, Joel R. and {Ravindranath}, Swara and
         {Reddy}, Naveen A. and {Renzini}, Alvio and {Rix}, Hans-Walter and
         {Robaina}, Aday R. and {Rosario}, David J. and {Rosati}, Piero and
         {Salimbeni}, Sara and {Scarlata}, Claudia and {Siana}, Brian and
         {Simard}, Luc and {Smidt}, Joseph and {Snyder}, Diana and
         {Somerville}, Rachel S. and {Spinrad}, Hyron and {Straughn}, Amber N. and
         {Telford}, Olivia and {Teplitz}, Harry I. and {Trump}, Jonathan R. and
         {Vargas}, Carlos and {Villforth}, Carolin and {Wagner}, Cory R. and {Wand
        ro}, Pat and {Wechsler}, Risa H. and {Weiner}, Benjamin J. and
         {Wiklind}, Tommy and {Wild}, Vivienne and {Wilson}, Grant and
         {Wuyts}, Stijn and {Yun}, Min S.},
        title = "{CANDELS: The Cosmic Assembly Near-infrared Deep Extragalactic Legacy Survey{\textemdash}The Hubble Space Telescope Observations, Imaging Data Products, and Mosaics}",
      journal = {\apjs},
         year = 2011,
        month = dec,
       volume = {197},
       number = {2},
          eid = {36},
        pages = {36},
          doi = {10.1088/0067-0049/197/2/36},
archivePrefix = {arXiv},
       eprint = {1105.3754},
 primaryClass = {astro-ph.CO},
       adsurl = {https://ui.adsabs.harvard.edu/abs/2011ApJS..197...36K}
}

@ARTICLE{Arnouts2007,
       author = {{Arnouts}, S. and {Walcher}, C.~J. and {Le F{\`e}vre}, O. and {Zamorani}, G. and {Ilbert}, O. and {Le Brun}, V. and {Pozzetti}, L. and {Bardelli}, S. and {Tresse}, L. and {Zucca}, E. and {Charlot}, S. and {Lamareille}, F. and {McCracken}, H.~J. and {Bolzonella}, M. and {Iovino}, A. and {Lonsdale}, C. and {Polletta}, M. and {Surace}, J. and {Bottini}, D. and {Garilli}, B. and {Maccagni}, D. and {Picat}, J.~P. and {Scaramella}, R. and {Scodeggio}, M. and {Vettolani}, G. and {Zanichelli}, A. and {Adami}, C. and {Cappi}, A. and {Ciliegi}, P. and {Contini}, T. and {de la Torre}, S. and {Foucaud}, S. and {Franzetti}, P. and {Gavignaud}, I. and {Guzzo}, L. and {Marano}, B. and {Marinoni}, C. and {Mazure}, A. and {Meneux}, B. and {Merighi}, R. and {Paltani}, S. and {Pell{\`o}}, R. and {Pollo}, A. and {Radovich}, M. and {Temporin}, S. and {Vergani}, D.},
        title = "{The SWIRE-VVDS-CFHTLS surveys: stellar mass assembly over the last 10 Gyr. Evidence for a major build up of the red sequence between z = 2 and z = 1}",
      journal = {\aap},
         year = 2007,
        month = dec,
       volume = {476},
       number = {1},
        pages = {137-150},
          doi = {10.1051/0004-6361:20077632},
archivePrefix = {arXiv},
       eprint = {0705.2438},
 primaryClass = {astro-ph},
       adsurl = {https://ui.adsabs.harvard.edu/abs/2007A&A...476..137A}
}

@ARTICLE{Keres2005,
       author = {{Kere{\v{s}}}, Du{\v{s}}an and {Katz}, Neal and {Weinberg}, David H. and {Dav{\'e}}, Romeel},
        title = "{How do galaxies get their gas?}",
      journal = {\mnras},
         year = 2005,
        month = oct,
       volume = {363},
       number = {1},
        pages = {2-28},
          doi = {10.1111/j.1365-2966.2005.09451.x},
archivePrefix = {arXiv},
       eprint = {astro-ph/0407095},
 primaryClass = {astro-ph},
       adsurl = {https://ui.adsabs.harvard.edu/abs/2005MNRAS.363....2K}
}

@ARTICLE{Wuyts2011,
       author = {{Wuyts}, Stijn and {F{\"o}rster Schreiber}, Natascha M. and {van der Wel}, Arjen and {Magnelli}, Benjamin and {Guo}, Yicheng and {Genzel}, Reinhard and {Lutz}, Dieter and {Aussel}, Herv{\'e} and {Barro}, Guillermo and {Berta}, Stefano and {Cava}, Antonio and {Graci{\'a}-Carpio}, Javier and {Hathi}, Nimish P. and {Huang}, Kuang-Han and {Kocevski}, Dale D. and {Koekemoer}, Anton M. and {Lee}, Kyoung-Soo and {Le Floc'h}, Emeric and {McGrath}, Elizabeth J. and {Nordon}, Raanan and {Popesso}, Paola and {Pozzi}, Francesca and {Riguccini}, Laurie and {Rodighiero}, Giulia and {Saintonge}, Amelie and {Tacconi}, Linda},
        title = "{Galaxy Structure and Mode of Star Formation in the SFR-Mass Plane from z \raisebox{-0.5ex}\textasciitilde 2.5 to z \raisebox{-0.5ex}\textasciitilde 0.1}",
      journal = {\apj},
         year = 2011,
        month = dec,
       volume = {742},
       number = {2},
          eid = {96},
        pages = {96},
          doi = {10.1088/0004-637X/742/2/96},
archivePrefix = {arXiv},
       eprint = {1107.0317},
 primaryClass = {astro-ph.CO},
       adsurl = {https://ui.adsabs.harvard.edu/abs/2011ApJ...742...96W}
}

@ARTICLE{Kennicutt2012,
       author = {{Kennicutt}, Robert C. and {Evans}, Neal J.},
        title = "{Star Formation in the Milky Way and Nearby Galaxies}",
      journal = {\araa},
         year = 2012,
        month = sep,
       volume = {50},
        pages = {531-608},
          doi = {10.1146/annurev-astro-081811-125610},
archivePrefix = {arXiv},
       eprint = {1204.3552},
 primaryClass = {astro-ph.GA},
       adsurl = {https://ui.adsabs.harvard.edu/abs/2012ARA&A..50..531K}
}

@ARTICLE{Diemer2018,
       author = {{Diemer}, Benedikt},
        title = "{COLOSSUS: A Python Toolkit for Cosmology, Large-scale Structure, and Dark Matter Halos}",
      journal = {\apjs},
         year = 2018,
        month = dec,
       volume = {239},
       number = {2},
          eid = {35},
        pages = {35},
          doi = {10.3847/1538-4365/aaee8c},
archivePrefix = {arXiv},
       eprint = {1712.04512},
 primaryClass = {astro-ph.CO},
       adsurl = {https://ui.adsabs.harvard.edu/abs/2018ApJS..239...35D}
}

@ARTICLE{Bower2006,
       author = {{Bower}, R.~G. and {Benson}, A.~J. and {Malbon}, R. and {Helly}, J.~C. and {Frenk}, C.~S. and {Baugh}, C.~M. and {Cole}, S. and {Lacey}, C.~G.},
        title = "{Breaking the hierarchy of galaxy formation}",
      journal = {\mnras},
         year = 2006,
        month = aug,
       volume = {370},
       number = {2},
        pages = {645-655},
          doi = {10.1111/j.1365-2966.2006.10519.x},
archivePrefix = {arXiv},
       eprint = {astro-ph/0511338},
 primaryClass = {astro-ph},
       adsurl = {https://ui.adsabs.harvard.edu/abs/2006MNRAS.370..645B}
}

@ARTICLE{dubois_2013,
       author = {{Dubois}, Yohan and {Gavazzi}, Rapha{\"e}l and {Peirani}, S{\'e}bastien and {Silk}, Joseph},
        title = "{AGN-driven quenching of star formation: morphological and dynamical implications for early-type galaxies}",
      journal = {\mnras},
         year = 2013,
        month = aug,
       volume = {433},
       number = {4},
        pages = {3297-3313},
          doi = {10.1093/mnras/stt997},
archivePrefix = {arXiv},
       eprint = {1301.3092},
 primaryClass = {astro-ph.CO},
       adsurl = {https://ui.adsabs.harvard.edu/abs/2013MNRAS.433.3297D}
}

@article{behroozi_most_2018,
	title = {The {Most} {Massive} {Galaxies} and {Black} {Holes} {Allowed} by $\Lambda$CDM},
	volume = {477},
	issn = {0035-8711, 1365-2966},
	url = {http://arxiv.org/abs/1609.04402},
	doi = {10.1093/mnras/sty945},
	number = {4},
	urldate = {2021-12-03},
	journal = {\mnras},
	author = {Behroozi, Peter and Silk, Joseph},
	month = jul,
	year = {2018},
	note = {arXiv: 1609.04402},
	pages = {5382--5387},
}

@article{birnboim_virial_2003,
	title = {Virial shocks in galactic haloes?},
	volume = {345},
	issn = {0035-8711},
	url = {https://ui.adsabs.harvard.edu/abs/2003MNRAS.345..349B},
	doi = {10.1046/j.1365-8711.2003.06955.x},
	urldate = {2021-10-29},
	journal = {\mnras},
	author = {Birnboim, Yuval and Dekel, Avishai},
	month = oct,
	year = {2003},
	pages = {349--364},
}

@article{dekel_galaxy_2006,
	title = {Galaxy bimodality due to cold flows and shock heating},
	volume = {368},
	issn = {0035-8711, 1365-2966},
	url = {https://academic.oup.com/mnras/article-lookup/doi/10.1111/j.1365-2966.2006.10145.x},
	doi = {10.1111/j.1365-2966.2006.10145.x},
	language = {en},
	number = {1},
	urldate = {2021-10-29},
	journal = {\mnras},
	author = {Dekel, A. and Birnboim, Y.},
	month = may,
	year = {2006},
	pages = {2--20},
}

@ARTICLE{weaver_cosmos2020_2022,
       author = {{Weaver}, J.~R. and {Kauffmann}, O.~B. and {Ilbert}, O. and {McCracken}, H.~J. and {Moneti}, A. and {Toft}, S. and {Brammer}, G. and {Shuntov}, M. and {Davidzon}, I. and {Hsieh}, B.~C. and {Laigle}, C. and {Anastasiou}, A. and {Jespersen}, C.~K. and {Vinther}, J. and {Capak}, P. and {Casey}, C.~M. and {McPartland}, C.~J.~R. and {Milvang-Jensen}, B. and {Mobasher}, B. and {Sanders}, D.~B. and {Zalesky}, L. and {Arnouts}, S. and {Aussel}, H. and {Dunlop}, J.~S. and {Faisst}, A. and {Franx}, M. and {Furtak}, L.~J. and {Fynbo}, J.~P.~U. and {Gould}, K.~M.~L. and {Greve}, T.~R. and {Gwyn}, S. and {Kartaltepe}, J.~S. and {Kashino}, D. and {Koekemoer}, A.~M. and {Kokorev}, V. and {Le F{\`e}vre}, O. and {Lilly}, S. and {Masters}, D. and {Magdis}, G. and {Mehta}, V. and {Peng}, Y. and {Riechers}, D.~A. and {Salvato}, M. and {Sawicki}, M. and {Scarlata}, C. and {Scoville}, N. and {Shirley}, R. and {Silverman}, J.~D. and {Sneppen}, A. and {Smolc̆i{\'c}}, V. and {Steinhardt}, C. and {Stern}, D. and {Tanaka}, M. and {Taniguchi}, Y. and {Teplitz}, H.~I. and {Vaccari}, M. and {Wang}, W. -H. and {Zamorani}, G.},
        title = "{COSMOS2020: A Panchromatic View of the Universe to z 10 from Two Complementary Catalogs}",
      journal = {\apjs},
         year = 2022,
        month = jan,
       volume = {258},
       number = {1},
          eid = {11},
        pages = {11},
          doi = {10.3847/1538-4365/ac3078},
archivePrefix = {arXiv},
       eprint = {2110.13923},
 primaryClass = {astro-ph.GA},
       adsurl = {https://ui.adsabs.harvard.edu/abs/2022ApJS..258...11W}
}

@article{1983ApJ...266..713O,
	title = {Secondary standard stars for absolute spectrophotometry.},
	volume = {266},
	issn = {0004-637X},
	url = {https://ui.adsabs.harvard.edu/abs/1983ApJ...266..713O},
	doi = {10.1086/160817},
	urldate = {2021-10-20},
	journal = {\apj},
	author = {Oke, J. B. and Gunn, J. E.},
	month = mar,
	year = {1983},
	pages = {713--717},
}

@article{1996AJ....112..839C,
	title = {New {Insight} on {Galaxy} {Formation} and {Evolution} {From} {Keck} {Spectroscopy} of the {Hawaii} {Deep} {Fields}},
	volume = {112},
	issn = {0004-6256},
	url = {https://ui.adsabs.harvard.edu/abs/1996AJ....112..839C},
	doi = {10.1086/118058},
	urldate = {2021-10-06},
	journal = {\aj},
	author = {Cowie, Lennox L. and Songaila, Antoinette and Hu, Esther M. and Cohen, J. G.},
	month = sep,
	year = {1996},
	pages = {839},
}

@article{croton_many_2006,
	title = {The many lives of active galactic nuclei: cooling flows, black holes and the luminosities and colours of galaxies},
	volume = {365},
	issn = {00358711, 13652966},
	shorttitle = {The many lives of active galactic nuclei},
	url = {http://arxiv.org/abs/astro-ph/0508046},
	doi = {10.1111/j.1365-2966.2005.09675.x},
	language = {en},
	number = {1},
	urldate = {2021-09-30},
	journal = {\mnras},
	author = {Croton, Darren J. and Springel, Volker and White, Simon D. M. and De Lucia, G. and Frenk, C. S. and Gao, L. and Jenkins, A. and Kauffmann, G. and Navarro, J. F. and Yoshida, N.},
	month = jan,
	year = {2006},
	note = {arXiv: astro-ph/0508046},
	pages = {11--28},
}

@article{gabor_hot_2015,
	title = {Hot gas in massive haloes drives both mass quenching and environment quenching},
	volume = {447},
	issn = {1365-2966, 0035-8711},
	url = {http://academic.oup.com/mnras/article/447/1/374/988785/Hot-gas-in-massive-haloes-drives-both-mass},
	doi = {10.1093/mnras/stu2399},
	language = {en},
	number = {1},
	urldate = {2021-09-27},
	journal = {\mnras},
	author = {Gabor, J. M. and Davé, R.},
	month = feb,
	year = {2015},
	pages = {374--391},
}

@article{thomas_environment_2010,
	title = {Environment and self-regulation in galaxy formation},
	issn = {00358711, 13652966},
	url = {http://arxiv.org/abs/0912.0259},
	doi = {10.1111/j.1365-2966.2010.16427.x},
	urldate = {2021-09-27},
	journal = {\mnras},
	author = {Thomas, Daniel and Maraston, Claudia and Schawinski, Kevin and Sarzi, Marc and Silk, Joseph},
	month = mar,
	year = {2010},
	note = {arXiv: 0912.0259},
}

@article{de_lucia_formation_2006,
	title = {The formation history of elliptical galaxies},
	volume = {366},
	issn = {0035-8711},
	url = {https://doi.org/10.1111/j.1365-2966.2005.09879.x},
	doi = {10.1111/j.1365-2966.2005.09879.x},
	number = {2},
	urldate = {2021-09-27},
	journal = {\mnras},
	author = {De Lucia, Gabriella and Springel, Volker and White, Simon D. M. and Croton, Darren and Kauffmann, Guinevere},
	month = feb,
	year = {2006},
	pages = {499--509},
}

@article{peng_mass_2010,
	title = {Mass and environment as drivers of galaxy evolution in {SDSS} and {zCOSMOS} and the origin of the {Schechter} function},
	volume = {721},
	issn = {0004-637X, 1538-4357},
	url = {http://arxiv.org/abs/1003.4747},
	doi = {10.1088/0004-637X/721/1/193},
	number = {1},
	urldate = {2021-09-26},
	journal = {\apj},
	author = {Peng, Y. and Lilly, S. J. and Kovac, K. and Bolzonella, M. and Pozzetti, L. and Renzini, A. and Zamorani, G. and Ilbert, O. and Knobel, C. and Iovino, A. and Maier, C. and Cucciati, O. and Tasca, L. and Carollo, C. M. and Silverman, J. and Kampczyk, P. and de Ravel, L. and Sanders, D. and Scoville, N. and Contini, T. and Mainieri, V. and Scodeggio, M. and Kneib, J.-P. and Fevre, O. Le and Bardelli, S. and Bongiorno, A. and Caputi, K. and Coppa, G. and de la Torre, S. and Franzetti, P. and Garilli, B. and Lamareille, F. and Borgne, J.-F. Le and Brun, V. Le and Mignoli, M. and Montero, E. Perez and Pello, R. and Ricciardelli, E. and Tanaka, M. and Tresse, L. and Vergani, D. and Welikala, N. and Zucca, E. and Oesch, P. and Abbas, U. and Barnes, L. and Bordoloi, R. and Bottini, D. and Cappi, A. and Cassata, P. and Cimatti, A. and Fumana, M. and Hasinger, G. and Koekemoer, A. M. and Leauthaud, A. and Maccagni, D. and Marinoni, C. and McCracken, H. J. and Memeo, P. and Meneux, B. and Nair, P. and Porciani, C. and Presotto, V. and Scaramella, R.},
	month = sep,
	year = {2010},
	note = {arXiv: 1003.4747},
	pages = {193--221},
}

@article{pillepich_first_2018,
	title = {First results from the {IllustrisTNG} simulations: the stellar mass content of groups and clusters of galaxies},
	volume = {475},
	issn = {0035-8711, 1365-2966},
	shorttitle = {First results from the {IllustrisTNG} simulations},
	url = {http://arxiv.org/abs/1707.03406},
	doi = {10.1093/mnras/stx3112},
	number = {1},
	urldate = {2021-09-23},
	journal = {\mnras},
	author = {Pillepich, Annalisa and Nelson, Dylan and Hernquist, Lars and Springel, Volker and Pakmor, Rüdiger and Torrey, Paul and Weinberger, Rainer and Genel, Shy and Naiman, Jill and Marinacci, Federico and Vogelsberger, Mark},
	month = mar,
	year = {2018},
	note = {arXiv: 1707.03406},
	pages = {648--675},
}

@article{springel_first_2018,
	title = {First results from the {IllustrisTNG} simulations: matter and galaxy clustering},
	volume = {475},
	doi = {10.1093/mnras/stx3304},
	number = {1},
	journal = {\mnras},
	author = {Springel, Volker and Pakmor, Rüdiger and Pillepich, Annalisa and Weinberger, Rainer and Nelson, Dylan and Hernquist, Lars and Vogelsberger, Mark and Genel, Shy and Torrey, Paul and Marinacci, Federico and Naiman, Jill},
	month = mar,
	year = {2018},
	pages = {676--698},
}

@article{ilbert_mass_2013,
	title = {Mass assembly in quiescent and star-forming galaxies since z ≃ 4 from {UltraVISTA}},
	volume = {556},
	issn = {0004-6361},
	url = {https://ui.adsabs.harvard.edu/abs/2013A&A...556A..55I/abstract},
	doi = {10.1051/0004-6361/201321100},
	language = {en},
	urldate = {2021-07-21},
	journal = {\aap},
	author = {Ilbert, O. and McCracken, H. J. and Le Fèvre, O. and Capak, P. and Dunlop, J. and Karim, A. and Renzini, M. A. and Caputi, K. and Boissier, S. and Arnouts, S. and Aussel, H. and Comparat, J. and Guo, Q. and Hudelot, P. and Kartaltepe, J. and Kneib, J. P. and Krogager, J. K. and Le Floc'h, E. and Lilly, S. and Mellier, Y. and Milvang-Jensen, B. and Moutard, T. and Onodera, M. and Richard, J. and Salvato, M. and Sanders, D. B. and Scoville, N. and Silverman, J. D. and Taniguchi, Y. and Tasca, L. and Thomas, R. and Toft, S. and Tresse, L. and Vergani, D. and Wolk, M. and Zirm, A.},
	month = aug,
	year = {2013},
	pages = {A55},
}

@article{foreman-mackey_emcee_2013,
	title = {emcee: {The} {MCMC} {Hammer}},
	volume = {125},
	issn = {00046280, 15383873},
	shorttitle = {emcee},
	url = {http://arxiv.org/abs/1202.3665},
	doi = {10.1086/670067},
	number = {925},
	urldate = {2021-07-03},
	journal = {\pasp},
	author = {Foreman-Mackey, Daniel and Hogg, David W. and Lang, Dustin and Goodman, Jonathan},
	month = mar,
	year = {2013},
	note = {arXiv: 1202.3665},
	pages = {306--312},
}

\appendix
  \onecolumn 

\section{Impact of quiescent classification criteria on the SMF}\label{sec:quiescent-classification}

\begin{figure}[th!]
\centering
\includegraphics[width=0.95\columnwidth]{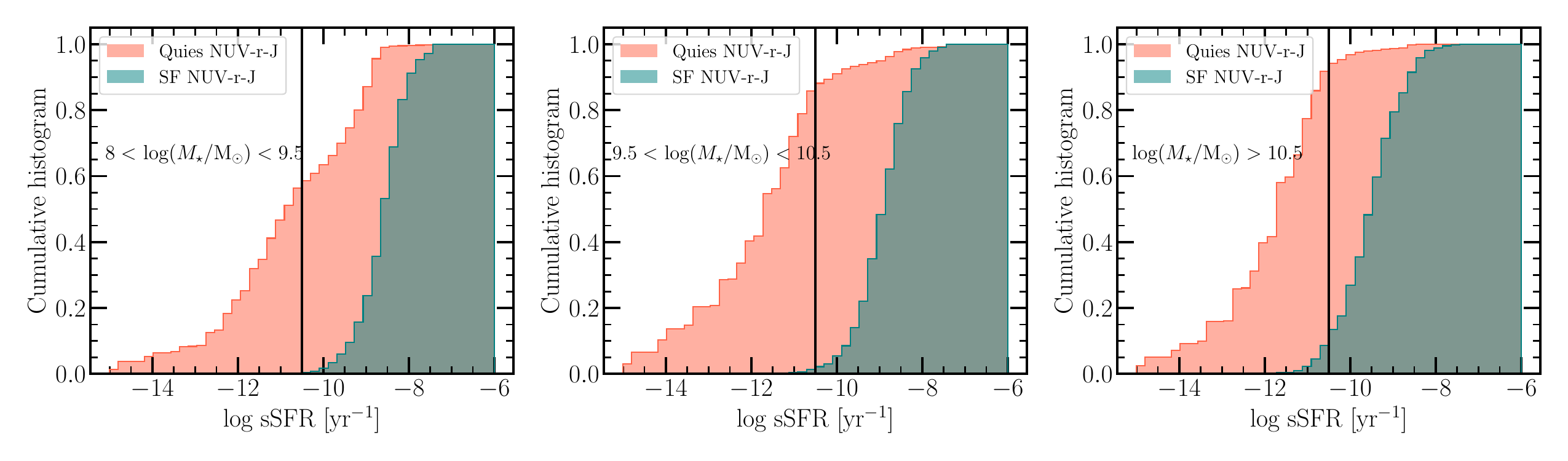}
\vspace{-5mm}
\caption{Cumulative histogram of the sSFR for $NUVrJ$ selected quiescent and star-forming galaxies. The different panels show the distributions in three different stellar mass bins. The vertical black line marks ${\rm log(sSFR)} = -10.5\, \si{\year}^{-1}$, typically used to separate quiescent and star-forming galaxies.}
\label{fig:ssfr-histogram}
\end{figure}

\begin{figure}[th!]
\centering
\includegraphics[width=0.95\columnwidth]{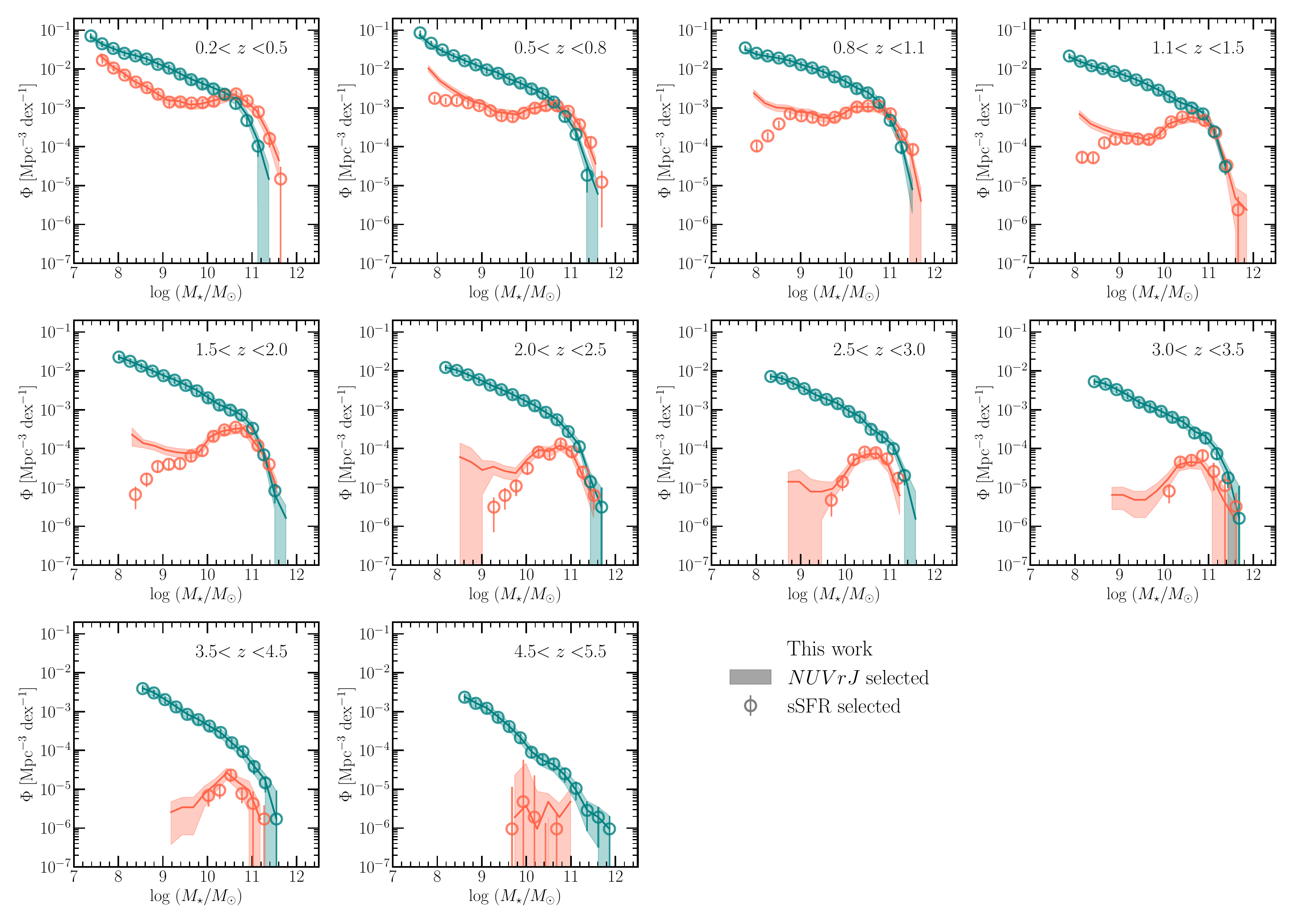}
\vspace{-5mm}
\caption{SMF for quiescent and star forming galaxies computed using the nominal $NUVrJ$ rest-frame color diagram (solid line and envelope) and ${\rm sSFR}/{\rm yr }^{-1} < 0.2/t_{\rm H}(z)$ (circles with error bars). While the high mass end of the SMF is robust with respect to the selection criterion, the low-mass end is highly sensitive.}
\label{fig:SMF-ssfr}
\end{figure}

Several methods exist in the literature to separate quiescent and star-forming galaxies, that include rest-frame colors diagrams such as the traditional $UVJ$ \citep{Wuyts2008, Williams2009}, $NUVrK$ and $NUVrJ$ \citep{arnouts_NRK_2013, ilbert_mass_2013}, as well as sSFR cuts that are either fixed (e.g., ${\rm log(sSFR/yr^{-1})} = -10.5$) or scale with the Hubble time as ${\rm sSFR}/{\rm yr }^{-1} = 0.2/t_{\rm H}(z)$ \citep[e.g.,][]{Fontana2009, Pacifici2016, Carnall2018}. Adopting different criteria can result in different selections of quiescent galaxies, which can be mass dependent and affect the shape of the SMF. In this section, we discuss how our classification is impacted by different selection methods, namely the $NUVrJ$ diagram versus a sSFR-based selection.

In Fig.~\ref{fig:ssfr-histogram} we show the cumulative histogram of the sSFR for the quiescent and star-forming samples selected using the $NUVrJ$ color diagram. To demonstrate the dependence on stellar mass, we separated the samples in three stellar mass bins $8<$~\logM~$<9.5$ (top), $9.5<$~\logM~$<10.5$ (middle), and \logM~$>10.5$ (bottom panel). This shows that the low-mass quiescent galaxies are most affected by the classification criterion --- only about $60\%$ of $8<$~\logM~$<9.5$, $NUVrJ$-selected quiescent galaxies at all redshifts have ${\rm log(sSFR/yr^{-1})}<-10.5$ and would be classified as quiescent by their sSFR too. More massive samples are more robust --- over $90\%$ ($95\%$) of $9.5<$~\logM~$<10.5$ (\logM~$>10.5$) satisfy both the $NUVrJ$ and the ${\rm log(sSFR/yr^{-1})}<-10.5$ criteria. The star-forming sample is more robust --- almost 100\%  of the low and intermediate mass $NUVrJ$-selected samples have ${\rm log(sSFR/yr^{-1})}>-10.5$, while this drops to $\sim 90\%$ for the \logM~$>10.5$ sample.

We also investigated how the choice of quiescent selection criterion affects the stellar mass function (SMF). In Fig.~\ref{fig:SMF-ssfr}, we compare the quiescent and star-forming SMFs derived using the $NUVrJ$ and an ${\rm sSFR}/{\rm yr }^{-1} < 0.2/t_{\rm H}(z)$ selection. The high-mass end (\logM~$>10$) is largely robust to the adopted criterion, with differences smaller than the statistical uncertainties of the SMF. At lower masses (\logM~$<10$), however, the results diverge more substantially. For example, at $z\sim1$, the quiescent SMF exhibits a double-peaked shape and declines at \logM~$<8.5$, while at $z>1.5$ the low-mass upturn seen in the $NUVrJ$-based SMF largely disappears. These differences highlight that the inferred abundance and shape of the low-mass quiescent population depend sensitively on the adopted selection, which has implications for interpreting the physical mechanisms responsible for quenching in low-mass galaxies.

\section{Fitting results} \label{appdx:fitting-results}
Tables \ref{tab:smf_parameters} and \ref{tab:smf_parameters_q} show the best fit values for the Schechter model parameters, SMD, and galaxy number densities.

\begin{table*}
\centering
\caption{The best-fitting single and double Schechter function parameters to the observed star-forming SMFs, where ${M}^* \equiv \log (M_\star/{\rm M}_\odot)$. The units of $ \Phi^{*}_1$ and $ \Phi^{*}_2$ are dex$^{-1}$Mpc$^{-3}$, $\rho_{\star}$ is in M$_{\sun} \, {\rm Mpc}^{-3}$ and $n_{\rm gal}$ is in ${\rm Mpc}^{-3}$ }
\label{tab:smf_parameters}
\begin{tabular}{lccccccc}
\hline\hline
 \multicolumn{8}{c}{Star-forming, all} \\
\hline
Redshift & ${M}^*$ & $\log \Phi^{*}_1$ & $\alpha_1$ & $\log \Phi^{*}_2$ & $\alpha_2$ & $\log \rho_{\star}$ & $\log n_{\rm gal}$ \\
\hline
$0.2 < z < 0.5$ & $10.76^{+0.06}_{-0.07}$ & $-3.01^{+0.08}_{-0.09}$ & $-1.42^{+0.03}_{-0.03}$ &  &  & $7.92^{+0.05}_{-0.05}$ & $-2.68^{+0.02}_{-0.02}$ \\
$0.5 < z < 0.8$ & $10.82^{+0.06}_{-0.06}$ & $-3.13^{+0.09}_{-0.10}$ & $-1.46^{+0.03}_{-0.03}$ &  &  & $7.89^{+0.05}_{-0.05}$ & $-2.71^{+0.01}_{-0.01}$ \\
$0.8 < z < 1.1$ & $10.73^{+0.04}_{-0.05}$ & $-2.79^{+0.07}_{-0.08}$ & $-1.31^{+0.03}_{-0.03}$ &  &  & $8.05^{+0.04}_{-0.05}$ & $-2.59^{+0.01}_{-0.01}$ \\
$1.1 < z < 1.5$ & $10.85^{+0.05}_{-0.06}$ & $-3.21^{+0.08}_{-0.09}$ & $-1.38^{+0.03}_{-0.03}$ &  &  & $7.78^{+0.04}_{-0.05}$ & $-2.87^{+0.01}_{-0.01}$ \\
$1.5 < z < 2.0$ & $10.84^{+0.05}_{-0.05}$ & $-3.32^{+0.08}_{-0.09}$ & $-1.46^{+0.04}_{-0.04}$ &  &  & $7.72^{+0.04}_{-0.05}$ & $-2.94^{+0.01}_{-0.01}$ \\
$2.0 < z < 2.5$ & $10.90^{+0.07}_{-0.07}$ & $-3.48^{+0.11}_{-0.10}$ & $-1.45^{+0.04}_{-0.04}$ &  &  & $7.61^{+0.05}_{-0.06}$ & $-3.05^{+0.01}_{-0.01}$ \\
$2.5 < z < 3.0$ & $10.90^{+0.09}_{-0.10}$ & $-3.73^{+0.12}_{-0.13}$ & $-1.49^{+0.05}_{-0.05}$ &  &  & $7.39^{+0.05}_{-0.06}$ & $-3.29^{+0.01}_{-0.01}$ \\
$3.0 < z < 3.5$ & $11.11^{+0.11}_{-0.11}$ & $-3.97^{+0.14}_{-0.15}$ & $-1.50^{+0.05}_{-0.05}$ &  &  & $7.37^{+0.06}_{-0.07}$ & $-3.32^{+0.01}_{-0.01}$ \\
$3.5 < z < 4.5$ & $11.06^{+0.21}_{-0.18}$ & $-4.39^{+0.23}_{-0.26}$ & $-1.65^{+0.07}_{-0.07}$ &  &  & $7.03^{+0.07}_{-0.08}$ & $-3.67^{+0.01}_{-0.01}$ \\
$4.5 < z < 5.5$ & $11.46^{+0.48}_{-0.34}$ & $-5.57^{+0.44}_{-0.59}$ & $-1.92^{+0.07}_{-0.06}$ &  &  & $6.62^{+0.09}_{-0.09}$ & $-4.25^{+0.03}_{-0.03}$ \\
\hline
 \multicolumn{8}{c}{$B/T<0.2$} \\
\hline
$0.2 < z < 0.5$ & $10.64^{+0.07}_{-0.07}$ & $-3.20^{+0.10}_{-0.10}$ & $-1.47^{+0.03}_{-0.03}$ &  &  & $7.64^{+0.05}_{-0.05}$ & $-2.94^{+0.02}_{-0.03}$ \\
$0.5 < z < 0.8$ & $10.63^{+0.07}_{-0.07}$ & $-3.23^{+0.11}_{-0.12}$ & $-1.50^{+0.04}_{-0.04}$ &  &  & $7.61^{+0.05}_{-0.05}$ & $-2.93^{+0.02}_{-0.02}$ \\
$0.8 < z < 1.1$ & $10.46^{+0.05}_{-0.06}$ & $-2.86^{+0.08}_{-0.08}$ & $-1.33^{+0.04}_{-0.04}$ &  &  & $7.71^{+0.04}_{-0.05}$ & $-2.91^{+0.01}_{-0.01}$ \\
$1.1 < z < 1.5$ & $10.60^{+0.06}_{-0.06}$ & $-3.30^{+0.09}_{-0.09}$ & $-1.41^{+0.04}_{-0.04}$ &  &  & $7.46^{+0.04}_{-0.05}$ & $-3.17^{+0.01}_{-0.01}$ \\
$1.5 < z < 2.0$ & $10.75^{+0.07}_{-0.07}$ & $-3.52^{+0.11}_{-0.11}$ & $-1.51^{+0.04}_{-0.04}$ &  &  & $7.46^{+0.05}_{-0.05}$ & $-3.18^{+0.01}_{-0.01}$ \\
$2.0 < z < 2.5$ & $10.77^{+0.06}_{-0.07}$ & $-3.64^{+0.10}_{-0.10}$ & $-1.48^{+0.04}_{-0.04}$ &  &  & $7.35^{+0.04}_{-0.05}$ & $-3.29^{+0.01}_{-0.01}$ \\
$2.5 < z < 3.0$ & $10.87^{+0.10}_{-0.10}$ & $-3.97^{+0.14}_{-0.14}$ & $-1.54^{+0.05}_{-0.05}$ &  &  & $7.16^{+0.06}_{-0.06}$ & $-3.50^{+0.01}_{-0.02}$ \\
$3.0 < z < 3.5$ & $11.04^{+0.12}_{-0.12}$ & $-4.36^{+0.15}_{-0.17}$ & $-1.60^{+0.06}_{-0.05}$ &  &  & $6.99^{+0.06}_{-0.07}$ & $-3.65^{+0.02}_{-0.02}$ \\
$3.5 < z < 4.5$ & $11.24^{+0.53}_{-0.27}$ & $-5.03^{+0.34}_{-0.58}$ & $-1.78^{+0.07}_{-0.07}$ &  &  & $6.72^{+0.11}_{-0.08}$ & $-4.01^{+0.02}_{-0.02}$ \\
$4.5 < z < 5.5$ & $11.32^{+0.44}_{-0.38}$ & $-5.79^{+0.47}_{-0.56}$ & $-1.97^{+0.08}_{-0.06}$ &  &  & $6.32^{+0.09}_{-0.09}$ & $-4.52^{+0.04}_{-0.04}$ \\
\hline
 \multicolumn{8}{c}{$0.2<B/T<0.6$} \\
\hline
$0.2 < z < 0.5$ & $10.75^{+0.08}_{-0.09}$ & $-3.37^{+0.08}_{-0.09}$ & $-1.25^{+0.03}_{-0.03}$ &  &  & $7.46^{+0.06}_{-0.07}$ & $-3.14^{+0.03}_{-0.03}$ \\
$0.5 < z < 0.8$ & $10.86^{+0.07}_{-0.08}$ & $-3.64^{+0.09}_{-0.10}$ & $-1.35^{+0.03}_{-0.03}$ &  &  & $7.35^{+0.06}_{-0.06}$ & $-3.24^{+0.02}_{-0.02}$ \\
$0.8 < z < 1.1$ & $10.77^{+0.08}_{-0.08}$ & $-3.43^{+0.08}_{-0.10}$ & $-1.25^{+0.03}_{-0.03}$ &  &  & $7.43^{+0.06}_{-0.06}$ & $-3.20^{+0.02}_{-0.02}$ \\
$1.1 < z < 1.5$ & $10.69^{+0.07}_{-0.08}$ & $-3.75^{+0.10}_{-0.10}$ & $-1.32^{+0.04}_{-0.04}$ &  &  & $7.06^{+0.05}_{-0.06}$ & $-3.59^{+0.02}_{-0.02}$ \\
$1.5 < z < 2.0$ & $10.77^{+0.08}_{-0.10}$ & $-3.97^{+0.12}_{-0.11}$ & $-1.45^{+0.04}_{-0.04}$ &  &  & $6.99^{+0.05}_{-0.06}$ & $-3.72^{+0.02}_{-0.02}$ \\
$2.0 < z < 2.5$ & $10.76^{+0.07}_{-0.08}$ & $-3.97^{+0.10}_{-0.10}$ & $-1.38^{+0.04}_{-0.04}$ &  &  & $6.94^{+0.05}_{-0.06}$ & $-3.72^{+0.02}_{-0.02}$ \\
$2.5 < z < 3.0$ & $10.76^{+0.11}_{-0.13}$ & $-4.22^{+0.14}_{-0.15}$ & $-1.42^{+0.06}_{-0.05}$ &  &  & $6.71^{+0.06}_{-0.07}$ & $-3.96^{+0.03}_{-0.03}$ \\
$3.0 < z < 3.5$ & $11.02^{+0.16}_{-0.15}$ & $-4.34^{+0.15}_{-0.18}$ & $-1.43^{+0.06}_{-0.05}$ &  &  & $6.86^{+0.07}_{-0.08}$ & $-3.82^{+0.02}_{-0.02}$ \\
$3.5 < z < 4.5$ & $10.84^{+0.21}_{-0.17}$ & $-4.64^{+0.21}_{-0.27}$ & $-1.56^{+0.08}_{-0.08}$ &  &  & $6.47^{+0.08}_{-0.09}$ & $-4.17^{+0.02}_{-0.03}$ \\
$4.5 < z < 5.5$ & $10.49^{+0.32}_{-0.26}$ & $-4.95^{+0.37}_{-0.46}$ & $-1.75^{+0.13}_{-0.13}$ &  &  & $5.96^{+0.09}_{-0.10}$ & $-4.83^{+0.05}_{-0.06}$ \\
\hline
 \multicolumn{8}{c}{$B/T>0.6$} \\
\hline
$0.2 < z < 0.5$ & $11.38^{+0.56}_{-0.36}$ & $-4.39^{+0.21}_{-0.29}$ & $-1.36^{+0.03}_{-0.03}$ &  &  & $7.13^{+0.30}_{-0.21}$ & $-3.64^{+0.05}_{-0.06}$ \\
$0.5 < z < 0.8$ & $10.70^{+0.21}_{-0.26}$ & $-4.32^{+0.21}_{-0.30}$ & $-1.44^{+0.05}_{-0.09}$ & $-3.91^{+0.18}_{-0.22}$ & $-0.26^{+0.82}_{-0.59}$ & $7.00^{+0.11}_{-0.12}$ & $-3.67^{+0.04}_{-0.04}$ \\
$0.8 < z < 1.1$ & $10.52^{+0.17}_{-0.13}$ & $-3.68^{+0.12}_{-0.25}$ & $-1.19^{+0.06}_{-0.09}$ & $-3.30^{+0.12}_{-0.15}$ & $0.42^{+0.64}_{-0.67}$ & $7.47^{+0.09}_{-0.11}$ & $-3.16^{+0.02}_{-0.02}$ \\
$1.1 < z < 1.5$ & $10.39^{+0.13}_{-0.07}$ & $-3.95^{+0.09}_{-0.24}$ & $-1.24^{+0.05}_{-0.08}$ & $-3.66^{+0.16}_{-0.18}$ & $1.66^{+0.56}_{-0.58}$ & $7.36^{+0.09}_{-0.17}$ & $-3.37^{+0.02}_{-0.02}$ \\
$1.5 < z < 2.0$ & $10.48^{+0.12}_{-0.09}$ & $-3.99^{+0.09}_{-0.14}$ & $-1.27^{+0.05}_{-0.06}$ & $-3.72^{+0.12}_{-0.12}$ & $0.98^{+0.50}_{-0.66}$ & $7.20^{+0.08}_{-0.10}$ & $-3.52^{+0.02}_{-0.02}$ \\
$2.0 < z < 2.5$ & $10.74^{+0.19}_{-0.16}$ & $-4.30^{+0.14}_{-0.28}$ & $-1.32^{+0.06}_{-0.09}$ & $-4.02^{+0.13}_{-0.14}$ & $0.07^{+0.71}_{-0.77}$ & $6.97^{+0.09}_{-0.09}$ & $-3.75^{+0.02}_{-0.02}$ \\
$2.5 < z < 3.0$ & $11.03^{+0.14}_{-0.14}$ & $-4.49^{+0.13}_{-0.14}$ & $-1.27^{+0.05}_{-0.05}$ &  &  & $6.64^{+0.08}_{-0.09}$ & $-4.04^{+0.03}_{-0.03}$ \\
$3.0 < z < 3.5$ & $11.13^{+0.14}_{-0.13}$ & $-4.48^{+0.13}_{-0.15}$ & $-1.30^{+0.05}_{-0.05}$ &  &  & $6.76^{+0.08}_{-0.09}$ & $-3.96^{+0.03}_{-0.03}$ \\
$3.5 < z < 4.5$ & $10.98^{+0.20}_{-0.15}$ & $-4.77^{+0.19}_{-0.20}$ & $-1.40^{+0.08}_{-0.07}$ &  &  & $6.34^{+0.10}_{-0.10}$ & $-4.29^{+0.03}_{-0.03}$ \\
$4.5 < z < 5.5$ & $10.92^{+0.73}_{-0.43}$ & $-5.60^{+0.52}_{-0.78}$ & $-1.73^{+0.14}_{-0.10}$ &  &  & $5.74^{+0.17}_{-0.13}$ & $-4.95^{+0.06}_{-0.07}$ \\
\hline
\end{tabular}
\end{table*}

\begin{table*}
\centering
\caption{The best-fitting single and double Schechter function parameters to the observed quiescent SMFs, where ${M}^* \equiv \log (M_\star/{\rm M}_\odot)$. The units of $ \Phi^{*}_1$ and $ \Phi^{*}_2$ are dex$^{-1}$Mpc$^{-3}$, $\rho_{\star}$ is in M$_{\sun} \, {\rm Mpc}^{-3}$ and $n_{\rm gal}$ is in ${\rm Mpc}^{-3}$} 
\label{tab:smf_parameters_q}
\begin{tabular}{lccccccc}
\hline\hline
 \multicolumn{8}{c}{Quiescent, all} \\
Redshift & ${M}^*$ & $\log \Phi^{*}_1$ & $\alpha_1$ & $\log \Phi^{*}_2$ & $\alpha_2$ & $\log \rho_{\star}$ & $\log n_{\rm gal}$ \\
\hline
$0.2 < z < 0.5$ & $10.78^{+0.10}_{-0.10}$ & $-4.61^{+0.29}_{-0.37}$ & $-1.80^{+0.09}_{-0.11}$ & $-2.73^{+0.09}_{-0.12}$ & $-0.39^{+0.26}_{-0.21}$ & $8.03^{+0.08}_{-0.09}$ & $-2.70^{+0.02}_{-0.02}$ \\
$0.5 < z < 0.8$ & $10.74^{+0.10}_{-0.10}$ & $-4.84^{+0.34}_{-0.43}$ & $-1.82^{+0.12}_{-0.14}$ & $-2.91^{+0.08}_{-0.11}$ & $-0.30^{+0.27}_{-0.23}$ & $7.80^{+0.08}_{-0.09}$ & $-2.91^{+0.02}_{-0.02}$ \\
$0.8 < z < 1.1$ & $10.72^{+0.08}_{-0.09}$ & $-4.34^{+0.33}_{-0.51}$ & $-1.45^{+0.13}_{-0.19}$ & $-2.95^{+0.07}_{-0.09}$ & $-0.16^{+0.34}_{-0.26}$ & $7.78^{+0.07}_{-0.08}$ & $-2.95^{+0.01}_{-0.01}$ \\
$1.1 < z < 1.5$ & $10.58^{+0.08}_{-0.08}$ & $-4.98^{+0.32}_{-0.40}$ & $-1.55^{+0.15}_{-0.17}$ & $-3.19^{+0.06}_{-0.07}$ & $0.25^{+0.30}_{-0.25}$ & $7.46^{+0.07}_{-0.08}$ & $-3.25^{+0.01}_{-0.01}$ \\
$1.5 < z < 2.0$ & $10.60^{+0.11}_{-0.11}$ & $-5.30^{+0.41}_{-0.64}$ & $-1.54^{+0.23}_{-0.30}$ & $-3.47^{+0.06}_{-0.09}$ & $0.29^{+0.42}_{-0.31}$ & $7.21^{+0.09}_{-0.11}$ & $-3.49^{+0.02}_{-0.02}$ \\
$2.0 < z < 2.5$ & $10.63^{+0.11}_{-0.12}$ & $-7.46^{+1.88}_{-3.14}$ & $-1.84^{+0.57}_{-1.00}$ & $-3.93^{+0.07}_{-0.10}$ & $0.09^{+0.43}_{-0.28}$ & $6.74^{+0.10}_{-0.11}$ & $-3.96^{+0.03}_{-0.03}$ \\
$2.5 < z < 3.0$ & $10.29^{+0.12}_{-0.12}$ & $-8.92^{+2.34}_{-1.99}$ & $-2.17^{+0.83}_{-1.15}$ & $-4.22^{+0.08}_{-0.15}$ & $1.04^{+0.56}_{-0.44}$ & $6.40^{+0.11}_{-0.13}$ & $-4.21^{+0.03}_{-0.04}$ \\
$3.0 < z < 3.5$ & $10.43^{+0.19}_{-0.21}$ & $-8.22^{+1.90}_{-2.43}$ & $-2.03^{+0.74}_{-1.17}$ & $-4.39^{+0.10}_{-0.19}$ & $0.69^{+0.82}_{-0.50}$ & $6.26^{+0.17}_{-0.21}$ & $-4.39^{+0.04}_{-0.05}$ \\
$3.5 < z < 4.5$ & $10.27^{+0.21}_{-0.19}$ & $-4.76^{+0.11}_{-0.16}$ & $0.67^{+0.74}_{-0.57}$ &  &  & $5.72^{+0.18}_{-0.20}$ & $-4.77^{+0.05}_{-0.05}$ \\
$4.5 < z < 5.5$ & $10.77^{+1.10}_{-0.97}$ & $-5.56^{+0.53}_{-0.56}$ & $0.37^{+1.07}_{-0.92}$ &  &  & $4.79^{+1.71}_{-2.60}$ & $-5.45^{+0.10}_{-0.13}$ \\
\hline
 \multicolumn{8}{c}{$B/T<0.2$} \\
\hline
$0.2 < z < 0.5$ & $11.84^{+0.38}_{-0.29}$ & $-4.37^{+0.29}_{-0.31}$ & $-1.50^{+0.06}_{-0.05}$ &  &  & $6.47^{+0.13}_{-0.15}$ & $-4.15^{+0.09}_{-0.11}$  \\
$0.5 < z < 0.8$ & $11.65^{+0.42}_{-0.28}$ & $-4.33^{+0.31}_{-0.36}$ & $-1.44^{+0.08}_{-0.07}$ &  &  & $6.40^{+0.14}_{-0.14}$ & $-4.16^{+0.06}_{-0.07}$  \\
$0.8 < z < 1.1$ & $11.05^{+0.06}_{-0.06}$ & $-3.54^{+0.11}_{-0.13}$ & $-1.06^{+0.06}_{-0.06}$ &  &  & $6.28^{+0.12}_{-0.11}$ & $-4.25^{+0.05}_{-0.06}$  \\
$1.1 < z < 1.5$ & $11.06^{+0.10}_{-0.09}$ & $-4.01^{+0.19}_{-0.20}$ & $-1.01^{+0.12}_{-0.10}$ &  &  & $6.21^{+0.25}_{-0.22}$ & $-4.59^{+0.06}_{-0.07}$  \\
$1.5 < z < 2.0$ & $10.96^{+0.11}_{-0.10}$ & $-3.84^{+0.16}_{-0.19}$ & $-0.67^{+0.15}_{-0.15}$ &  &  & $6.50^{+0.27}_{-0.31}$ & $-4.52^{+0.05}_{-0.05}$  \\
$2.0 < z < 2.5$ & $10.67^{+0.10}_{-0.10}$ & $-3.94^{+0.08}_{-0.10}$ & $-0.07^{+0.24}_{-0.21}$ &  &  & $6.03^{+0.46}_{-0.46}$ & $-4.96^{+0.08}_{-0.09}$  \\
$2.5 < z < 3.0$ & $10.31^{+0.13}_{-0.12}$ & $-4.20^{+0.09}_{-0.12}$ & $1.00^{+0.44}_{-0.45}$ &  &  & $5.39^{+0.28}_{-0.46}$ & $-4.95^{+0.07}_{-0.09}$  \\
$3.0 < z < 3.5$ & $10.52^{+0.18}_{-0.19}$ & $-4.37^{+0.09}_{-0.13}$ & $0.38^{+0.60}_{-0.40}$ &  &  & $5.16^{+0.63}_{-0.73}$ & $-5.19^{+0.10}_{-0.12}$  \\
$3.5 < z < 4.5$ & $10.27^{+0.21}_{-0.19}$ & $-4.76^{+0.11}_{-0.16}$ & $0.67^{+0.74}_{-0.57}$ &  &  & $5.13^{+0.48}_{-0.59}$ & $-5.39^{+0.09}_{-0.11}$  \\
$4.5 < z < 5.5$ & $10.77^{+1.10}_{-0.97}$ & $-5.56^{+0.53}_{-0.56}$ & $0.37^{+1.07}_{-0.92}$ &  &  & $4.66^{+1.98}_{-3.41}$ & $-5.72^{+0.13}_{-0.19}$  \\
\hline
 \multicolumn{8}{c}{$0.2<B/T<0.6$} \\
\hline
$0.2 < z < 0.5$ & $10.78^{+0.10}_{-0.10}$ & $-4.61^{+0.29}_{-0.37}$ & $-1.80^{+0.09}_{-0.11}$ & $-2.73^{+0.09}_{-0.12}$ & $-0.39^{+0.26}_{-0.21}$  & $7.65^{+0.11}_{-0.12}$ & $-3.02^{+0.03}_{-0.03}$ \\
$0.5 < z < 0.8$ & $10.74^{+0.10}_{-0.10}$ & $-4.84^{+0.34}_{-0.43}$ & $-1.82^{+0.12}_{-0.14}$ & $-2.91^{+0.08}_{-0.11}$ & $-0.30^{+0.27}_{-0.23}$  & $7.42^{+0.09}_{-0.11}$ & $-3.23^{+0.02}_{-0.02}$ \\
$0.8 < z < 1.1$ & $10.72^{+0.08}_{-0.09}$ & $-4.34^{+0.33}_{-0.51}$ & $-1.45^{+0.13}_{-0.19}$ & $-2.95^{+0.07}_{-0.09}$ & $-0.16^{+0.34}_{-0.26}$  & $7.34^{+0.09}_{-0.11}$ & $-3.30^{+0.02}_{-0.02}$ \\
$1.1 < z < 1.5$ & $10.58^{+0.08}_{-0.08}$ & $-4.98^{+0.32}_{-0.40}$ & $-1.55^{+0.15}_{-0.17}$ & $-3.19^{+0.06}_{-0.07}$ & $0.25^{+0.30}_{-0.25}$  & $6.85^{+0.10}_{-0.12}$ & $-3.74^{+0.02}_{-0.03}$ \\
$1.5 < z < 2.0$ & $10.60^{+0.11}_{-0.11}$ & $-5.30^{+0.41}_{-0.64}$ & $-1.54^{+0.23}_{-0.30}$ & $-3.47^{+0.06}_{-0.09}$ & $0.29^{+0.42}_{-0.31}$  & $6.54^{+0.13}_{-0.16}$ & $-4.02^{+0.03}_{-0.03}$ \\
$2.0 < z < 2.5$ & $10.63^{+0.11}_{-0.12}$ & $-7.46^{+1.88}_{-3.14}$ & $-1.84^{+0.57}_{-1.00}$ & $-3.93^{+0.07}_{-0.10}$ & $0.09^{+0.43}_{-0.28}$  & $6.23^{+0.12}_{-0.15}$ & $-4.40^{+0.04}_{-0.05}$ \\
$2.5 < z < 3.0$ & $10.29^{+0.12}_{-0.12}$ & $-8.92^{+2.34}_{-1.99}$ & $-2.17^{+0.83}_{-1.15}$ & $-4.22^{+0.08}_{-0.15}$ & $1.04^{+0.56}_{-0.44}$  & $5.94^{+0.15}_{-0.21}$ & $-4.58^{+0.05}_{-0.06}$ \\
$3.0 < z < 3.5$ & $10.43^{+0.19}_{-0.21}$ & $-8.22^{+1.90}_{-2.43}$ & $-2.03^{+0.74}_{-1.17}$ & $-4.39^{+0.10}_{-0.19}$ & $0.69^{+0.82}_{-0.50}$  & $5.68^{+0.22}_{-0.31}$ & $-4.85^{+0.07}_{-0.08}$ \\
$3.5 < z < 4.5$ & $10.27^{+0.21}_{-0.19}$ & $-4.76^{+0.11}_{-0.16}$ & $0.67^{+0.74}_{-0.57}$ &  &  & $5.00^{+0.35}_{-0.57}$ & $-5.22^{+0.08}_{-0.09}$  \\
$4.5 < z < 5.5$ & $10.77^{+1.10}_{-0.97}$ & $-5.56^{+0.53}_{-0.56}$ & $0.37^{+1.07}_{-0.92}$ &  &  & $3.63^{+2.89}_{-2.70}$ & $-6.15^{+0.20}_{-0.37}$  \\
\hline
 \multicolumn{8}{c}{$B/T>0.6$} \\
\hline
$0.2 < z < 0.5$ & $10.78^{+0.10}_{-0.10}$ & $-4.61^{+0.29}_{-0.37}$ & $-1.80^{+0.09}_{-0.11}$ & $-2.73^{+0.09}_{-0.12}$ & $-0.39^{+0.26}_{-0.21}$ & $7.73^{+0.10}_{-0.13}$ & $-3.02^{+0.03}_{-0.03}$  \\
$0.5 < z < 0.8$ & $10.74^{+0.10}_{-0.10}$ & $-4.84^{+0.34}_{-0.43}$ & $-1.82^{+0.12}_{-0.14}$ & $-2.91^{+0.08}_{-0.11}$ & $-0.30^{+0.27}_{-0.23}$ & $7.51^{+0.09}_{-0.10}$ & $-3.26^{+0.02}_{-0.02}$  \\
$0.8 < z < 1.1$ & $10.72^{+0.08}_{-0.09}$ & $-4.34^{+0.33}_{-0.51}$ & $-1.45^{+0.13}_{-0.19}$ & $-2.95^{+0.07}_{-0.09}$ & $-0.16^{+0.34}_{-0.26}$ & $7.50^{+0.07}_{-0.08}$ & $-3.24^{+0.02}_{-0.02}$  \\
$1.1 < z < 1.5$ & $10.58^{+0.08}_{-0.08}$ & $-4.98^{+0.32}_{-0.40}$ & $-1.55^{+0.15}_{-0.17}$ & $-3.19^{+0.06}_{-0.07}$ & $0.25^{+0.30}_{-0.25}$ & $7.29^{+0.08}_{-0.09}$ & $-3.45^{+0.02}_{-0.02}$  \\
$1.5 < z < 2.0$ & $10.60^{+0.11}_{-0.11}$ & $-5.30^{+0.41}_{-0.64}$ & $-1.54^{+0.23}_{-0.30}$ & $-3.47^{+0.06}_{-0.09}$ & $0.29^{+0.42}_{-0.31}$ & $7.04^{+0.08}_{-0.10}$ & $-3.71^{+0.02}_{-0.02}$  \\
$2.0 < z < 2.5$ & $10.63^{+0.11}_{-0.12}$ & $-7.46^{+1.88}_{-3.14}$ & $-1.84^{+0.57}_{-1.00}$ & $-3.93^{+0.07}_{-0.10}$ & $0.09^{+0.43}_{-0.28}$ & $6.46^{+0.11}_{-0.14}$ & $-4.23^{+0.03}_{-0.04}$  \\
$2.5 < z < 3.0$ & $10.29^{+0.12}_{-0.12}$ & $-8.92^{+2.34}_{-1.99}$ & $-2.17^{+0.83}_{-1.15}$ & $-4.22^{+0.08}_{-0.15}$ & $1.04^{+0.56}_{-0.44}$ & $6.03^{+0.15}_{-0.22}$ & $-4.61^{+0.05}_{-0.06}$  \\
$3.0 < z < 3.5$ & $10.43^{+0.19}_{-0.21}$ & $-8.22^{+1.90}_{-2.43}$ & $-2.03^{+0.74}_{-1.17}$ & $-4.39^{+0.10}_{-0.19}$ & $0.69^{+0.82}_{-0.50}$ & $5.93^{+0.20}_{-0.23}$ & $-4.69^{+0.06}_{-0.07}$  \\
$3.5 < z < 4.5$ & $10.27^{+0.21}_{-0.19}$ & $-4.76^{+0.11}_{-0.16}$ & $0.67^{+0.74}_{-0.57}$ &  &  & $5.16^{+0.32}_{-0.43}$ & $-5.16^{+0.07}_{-0.08}$  \\
$4.5 < z < 5.5$ & $10.77^{+1.10}_{-0.97}$ & $-5.56^{+0.53}_{-0.56}$ & $0.37^{+1.07}_{-0.92}$ &  &  & $4.45^{+2.00}_{-3.13}$ & $-6.02^{+0.18}_{-0.30}$  \\
\hline
\end{tabular}
\end{table*}

\end{document}